\documentclass[apj,iop,appendixfloats,twocolappendix]{emulateapj}
\usepackage[caption=false]{subfig}
\usepackage{amsmath}

\begin{document}

\title{Massive Compact Galaxies with High-Velocity Outflows: Morphological Analysis and Constraints on AGN Activity}

\author{P. H. Sell\altaffilmark{1,2}}
\author{C. A. Tremonti\altaffilmark{1}}
\author{R. C. Hickox\altaffilmark{3}}
\author{A. M. Diamond-Stanic\altaffilmark{1}}
\author{J. Moustakas\altaffilmark{4}}
\author{A. Coil\altaffilmark{5}}
\author{A. Williams\altaffilmark{1}}
\author{G. Rudnick\altaffilmark{6}}
\author{A. Robaina\altaffilmark{7}}
\author{J. E. Geach\altaffilmark{8}}
\author{S. Heinz\altaffilmark{1}}
\author{E. M. Wilcots\altaffilmark{1}}
\altaffiltext{1}{Department of Astronomy, University of Wisconsin-Madison, Madison, WI, USA; paul.sell@ttu.edu}
\altaffiltext{2}{Department of Physics, Texas Tech University, Lubbock, TX, USA}
\altaffiltext{3}{Department of Physics \& Astronomy, Dartmouth College, Hanover, NH, USA}
\altaffiltext{4}{Department of Physics and Astronomy, Siena College, 515 Loudon Road, Loudonville, NY 12211-1462, USA}
\altaffiltext{5}{Center for Astrophysics and Space Sciences, University of California, San Diego, La Jolla, CA, USA}
\altaffiltext{6}{Department of Physics and Astronomy, 1251 Wescoe Hall Drive, University of Kansas, Lawrence, KS, USA}
\altaffiltext{7}{Department of Astronomy, University of Michigan, 550 Church Street, Ann Arbor, MI, USA}
\altaffiltext{8}{Centre for Astrophysics Research, University of Hertfordshire, Hatfield, Hertfordshire, AL10 9AB, UK}

\begin{abstract}

We investigate the process of rapid star formation quenching in a sample
of 12 massive galaxies at intermediate redshift ($z\sim0.6$) that host
high-velocity ionized gas outflows ($v > 1000$~km~s$^{-1}$).  We
conclude that these fast outflows are most likely driven by feedback
from star formation rather than active galactic nuclei (AGN).
We use multiwavelength survey and targeted observations of the
galaxies to assess their star formation, AGN activity, and
morphology.  Common attributes include diffuse tidal
features indicative of recent mergers accompanied by bright, unresolved
cores with effective radii less than a few hundred parsecs.  The
galaxies are extraordinarily compact for their stellar mass, even when
compared with galaxies at $z\sim2$~--~3. For 9/12 galaxies, we rule out
an AGN contribution to the nuclear light and hypothesize that the
unresolved core comes from a compact central starburst triggered by the
dissipative collapse of very gas-rich progenitor merging disks. We find
evidence of AGN activity in half the sample but we argue that it
accounts for only a small fraction ($\lesssim10$\%) of the total
bolometric luminosity.  We find no correlation between AGN activity and
outflow velocity and we conclude that the fast outflows in our galaxies
are not powered by on-going AGN activity, but rather by recent,
extremely compact starbursts.

\end{abstract}


\section{Introduction}
\label{section:intro}

Cosmological simulations based on a $\Lambda$CDM framework overpredict by an order of magnitude the
fraction of baryons that will form stars by the present day \citep[e.g.,][]{keres09}.  This
``overcooling" problem is manifested at the massive end ($M_{\ast} \sim 10^{11} M_{\sun}$) by
simulated galaxies that are too luminous and blue to match observations
\citep[e.g.,][]{croton06,gabor11}.  The preferred solution is that feedback from massive stars and
accreting supermassive black holes (SMBHs) regulates the cold gas supply for star formation by
ejecting cold gas from galaxies and preventing hot gas from cooling.

The global effect of this feedback can be tuned to match the observed stellar mass function by
reducing the efficiency of star formation in both low-mass and high-mass dark matter halos
\citep[e.g.,][]{somerville09,behroozi13}.  However, the relevant gas physics
(e.g., shocks, dissipation, heating, cooling) occurs on scales that are unresolved by modern
simulations.  At the massive end, the central problem is how to quench star formation to form the
population of elliptical galaxies found in the local universe and the ``red sequence" galaxies
observed out to $z\ge2$.  The most massive ellipticals have high $\alpha$/Fe abundance ratios
implying very short formation times ($\Delta t  \lesssim 1$~Gyr;  \citealt{thomas05,thomas10}).

The ejective feedback that is necessary to quench star formation quickly in simulations is predicted
to be most effective in major mergers of massive gas-rich galaxies \citep[e.g.,][]{wuyts10}.  Under
the assumption that such mergers form dynamically hot spheroids and quench subsequent star formation
\citep[e.g.,][]{springel05}, this merger-driven model has been proposed to be the dominant formation
mechanism for red elliptical galaxies \citep{hopkins08a}.  Feedback from an active galactic nucleus
(AGN) is commonly invoked as the mechanism for heating up and driving out large fractions of cold
gas, effectively quenching star formation \citep[e.g.,][]{booth13,granato04,hopkins06,menci06}.
These models have been successful in reproducing a number of empirical trends, including the
color-magnitude relation and the correlation between supermassive black hole (SMBH) mass and bulge
stellar velocity dispersion.  AGN feedback models accomplish this by assuming that $\sim5$\% of the
radiated quasar luminosity can couple thermally and isotropically to the surrounding gas.  However,
linking galaxy-wide outflows to feedback processes (e.g., radiation pressure, jets) from a SMBH that
originate on parsec scales remains a challenging problem because momentum coupling is not fully
understood.  Therefore, it is crucial to search for direct evidence of merger-induced quasar
feedback.  Unfortunately, even recent evidence for AGN feedback is still largely circumstantial
\citep[e.g.,][]{cano-diaz12,veilleux13}.  Only in a limited number of cases in very low-redshift
galaxies where powerful, kiloparsec-scale outflows can be resolved and examined in detail can quasar
feedback be most clearly traced back to the SMBH
\citep[e.g.,][]{lipari09,rupke11,greene11,greene12,hainline13}.

Several observational studies have found that massive, quiescent
galaxies at $z \sim 2$~--~3 are remarkably compact, with sizes a
factor of 4~--~6 smaller than local galaxies
\citep{zirm07,trujillo07,vandokkum08,buitrago08}.  Highly dissipational
mergers between gas-rich progenitors, which are more common at high
redshift, have been invoked to explain these super-compact massive
galaxies \citep{covington11}.  It has been suggested that these massive
galaxies evolve ``inside out'' in order to arrive on the local size-mass
relation \citep{hopkins09b,fan13,vandesande13}.
There have been considerable recent efforts to identify the $z\sim3$
star-forming progenitors of massive, compact, quiescent galaxies
\citep{patel12,barro13,stefanon13}.  Studying such faint systems in
sufficient detail to gain insight into the physical mechanisms
responsible for shutting down their star formation is very difficult
at these redshifts.  By identifying and studying lower redshift
analogues, we may be able to more readily learn about higher redshift
massive galaxy evolution.

With the preceding ideas in mind, we have been studying a sample of
massive galaxies ($log(M_*/M_{\sun}) = 10.5 - 11.5$) at
$z=0.40$~--~$0.75$ selected to be in the midst of star formation
quenching.  They have very blue B- and A-star dominated stellar continua
but relatively weak nebular emission lines (H$\beta$ EW$ < 12$~\AA).
\citet{tremonti07} inferred that the star formation rate in the last
10~Myr was significantly lower than it was in the past 100~Myr, and
labeled them young post-starburst galaxies.  Subsequent restframe
mid-infrared (IR) measurements revealed large luminosities, which might be
explained if these galaxies are unusually compact young post-starbursts
\citep{groves08}.  However, modeling of the ultraviolet (UV) to near IR
spectral energy distribution (SED) suggests a high level of heavily
obscured star formation \citep{diamond-stanic12b}.  In either case, the
galaxies are very different from classic post-starburst galaxies
\citep[i.e., E+A or K+A galaxies][]{dressler83,zabludoff96,poggianti99},
which do not exhibit such unusual properties and have been shown
to have very little obscured star formation \citep{nielsen12}.
Therefore, the galaxies in our sample are likely to be very close to
their peak star formation rate, when quenching processes are expected to
be the most active.  Notably, two-thirds of the galaxies exhibit $\ge
1000$~km~s$^{-1}$ outflows \citep{tremonti07}, the largest outflow
velocities observed in star-forming galaxies at any redshift,
suggesting that feedback may play a significant role in quenching.

We present detailed multiwavelength analysis of a small sub-sample of these galaxies.  The 12
galaxies in this study are a subset of the 29 galaxies initially considered by
\cite{diamond-stanic12b}.  They presented the basic result that many of these galaxies have compact
morphologies (as small as $r_e \sim 100$~pc).   The compact morphologies of these galaxies suggested
that their high-velocity outflows could have been driven by extreme star-formation feedback
\citep{heckman11}.

\cite{diamond-stanic12b} highlighted the UV though IR SEDs, Hubble Space Telescope (HST) images, and
optical spectra for three galaxies with extraordinarily high star-formation rate surface densities
(up to $\Sigma_{\textnormal{\scriptsize{SFR}}}\approx3000$~M$_{\odot}$~yr$^{-1}$~kpc$^{-2}$) that
approach the theoretical Eddington limit \citep{lehnert96,meurer97,murray05,thompson05}.  One of
these galaxies, J1506+54, which is one of the 12 galaxies in our sub-sample, has also been
investigated by \cite{geach13}.  Their recent CO observations of this galaxy indicate that it
contains $\sim10^{10}~M_{\odot}$ of cold gas.  However, the very high $L_{IR}$ / $L_{CO}$ ratio
implies that it is being consumed with near 100\% efficiency, and will be exhausted in a few tens of
Myr.  Thus, we surmise that our galaxies are in the midst of starburst quenching.

An important issue is whether feedback from an AGN contributes to these outflows and whether the
presence of an AGN could have affected the size measurements from {\em HST}.  We consider whether
the galaxies' recent activity is related to a merger, and we examine whether there is any evidence
that the black hole plays a role in driving the fast outflows we observe and in quenching the
starburst.  For our investigation, we use multi-wavelength diagnostics to build a comprehensive view
of these galaxies.  We combine targeted MMT UV-optical spectroscopy, \emph{Chandra} X-ray
observations, \emph{HST} optical imaging, Jansky Very Large Array (JVLA) radio observations, and
{\em Spitzer Space Telescope} \citep{werner04} near IR imaging with survey imaging from the
{\it Galaxy Evolution Explorer} \citep[GALEX;][]{martin05a}, Sloan Digital Sky Survey
\citep[SDSS;][]{york02}, and {\em Wide-field Infrared Survey Explorer} \citep[WISE;][]{wright10}.

This paper is organized as follows.  Our sample selection is discussed
in \S~\ref{section:sample}; data reduction and basic analysis of our
multiwavelength observations is presented in \S~\ref{section:data_red}
and in the Appendix.  Readers wishing to go straight to the results are
encouraged to begin reading in \S~\ref{section:analysis_agn}, which
provides a summary of the preceding analysis.  In this section, we
estimate the accretion rate of the three broad-line AGN and consider the
available evidence for AGN activity in the other nine galaxies.  We also
include a case study of a galaxy in our sample with one of most extreme
starbursts currently known (\S~\ref{section:j1506}).  In
\S~\ref{section:discussion}, we summarize the HST morphological analysis
and highlight the very high star formation surface densities implied by
the compact sizes of the galaxies.  We assess whether AGN feedback is
responsible for starburst quenching and the ultra-fast outflows in our
sample.  Finally, we summarize this work and state our most important
conclusions in \S~\ref{section:conclusions}.

Throughout this paper, we adopt the AB magnitude system, unless otherwise noted, and standard
cosmological parameters:  $H_0 = 70$~km~s$^{-1}$~Mpc$^{-1}$, $\Omega_M = 0.3$, and
$\Omega_\Lambda = 0.7$.

\section{Sample Selection}
\label{section:sample}

\renewcommand{\thefootnote}{\alph{footnote}}
\begin{deluxetable*}{cccccccc}
\tablecaption{General Sample Information}
\tablehead{
	\colhead{ID} &
	\colhead{Galaxy} &
	\colhead{RA} &
	\colhead{Dec} &
	\colhead{z} &
	\colhead{Log($M_*$)} &
	\colhead{$v_{avg}$} &
	\colhead{$v_{max}$} \\
	\colhead{} &
	\colhead{} &
	\multicolumn{2}{c}{(J2000)} &
	\colhead{} &
	\colhead{(M$_{\odot}$)} &
	\multicolumn{2}{c}{(km~s$^{-1}$)} \\
	\colhead{(1)} &
	\colhead{(2)} &
	\colhead{(3)} &
	\colhead{(4)} &
	\colhead{(5)} &
	\colhead{(6)} &
	\colhead{(7)} &
	\colhead{(8)}
	}
\startdata
1  & J0826+43 & 126.66004 & 43.09151 & 0.604 & 10.8 & $-$1230 & $-$1500 \\
2  & J0944+09 & 146.07438 & 9.50538  & 0.514 & 10.5 & $-$1330 & $-$1860 \\
3  & J1104+59 & 166.15608 & 59.77767 & 0.573 & 10.6 & $-$1040 & $-$1490 \\
4  & J1506+54 & 226.65125 & 54.03914 & 0.609 & 10.7 & $-$1480 & $-$2200 \\
5  & J1506+61 & 226.51533 & 61.53003 & 0.437 & 10.2 & $-$1000$^a$ &  $-$800 \\
6  & J1558+39 & 239.54683 & 39.95579 & 0.403 & 10.6 & $-$1000 & $-$1220 \\
7  & J1613+28 & 243.38554 & 28.57077 & 0.450 & 11.2 & $-$1520 & $-$2440 \\
8  & J1713+28 & 258.25163 & 28.28562 & 0.577 & 10.8 &  $-$930 & $-$1190 \\
9  & J2118+00 & 319.60025 & 0.29150  & 0.460 & 11.1 &   ---   &   ---   \\ \hline
10 & J1359+51 & 209.83742 & 51.62748 & 0.413 & 10.5 &   ---   &   ---   \\
11 & J1634+46 & 248.69371 & 46.32965 & 0.576 & 11.8 &   ---   &   ---   \\
12 & J2140+12 & 325.00204 & 12.15406 & 0.752 & 10.9 &  $-$490 &  $-$950 
\enddata
\tablecomments{
Column 1:  Galaxy identification number used in some figures for brevity.
Column 2:  Abbreviated IAU designation used throughout the text.
Columns 3 and 4:  RA and Dec in decimal degrees.
Column 5:  Redshift.
Column 6:  Stellar mass (uncertainties are approximately a factor of two).
Column 6 is based on SED fits assuming a \citet{chabrier03} IMF from 0.1~--~100~M$_\odot$.
Columns 7 and 8:  The average and maximum outflow velocities (see \S~\ref{section:data_red_MMT} for
details).  No Mg~II $\lambda\lambda2796,2804$ outflows were detected in 3/12 of the galaxies in this
sample.  Galaxies are ordered by their short name except the three galaxies with broad Mg~II
$\lambda\lambda2796,2804$ emission lines.  These are placed at the end of the list because they are
analyzed separately.
\vspace{0.1in}
\\
$^a$ In J1506+61, strong emission line in-filling of the Mg~II absorption line causes the average
velocity to be biased to larger negative values; the maximum velocity is less affected.}
\label{table:general_info}
\end{deluxetable*}
\renewcommand{\thefootnote}{\arabic{footnote}}

The parent sample from which our targets are drawn was selected from the SDSS Data Release 4
\citep{adelman-mccarthy06}.  It is composed of $i < 20.5$ mag objects that were targeted for SDSS
spectroscopy as low-redshift quasar candidates but were subsequently classified as galaxies at
$z = 0.4$~--~1 by the SDSS spectroscopic pipeline.  We used the low signal-to-noise
(S/N $\sim 2$~--~4 per pixel) SDSS spectra to select 159 galaxies with post-starburst
characteristics~---~strong stellar Balmer absorption and weak nebular emission.  Fits to the
restframe UV through near-IR SEDs imply that the galaxies are massive
($M_* = 10^{10.5} - 10^{11.5}~M_{\sun}$; see \S~\ref{section:stellar_masses}), thus, they are likely
to host SMBHs.

\begin{figure}[tbp]
\includegraphics[width=0.48\textwidth]{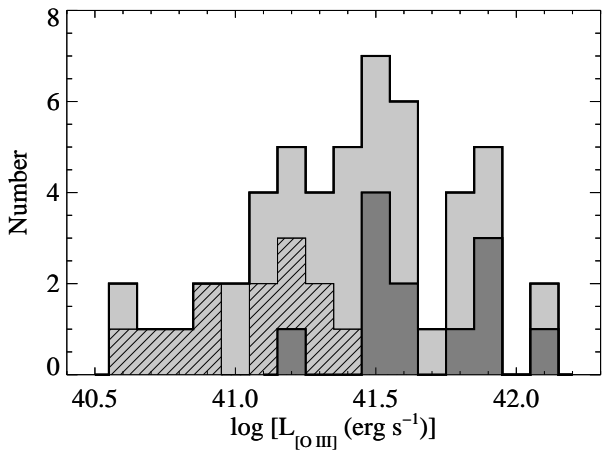}
\caption{[O~III] $\lambda 5007$ luminosities of the 51 galaxies with spectroscopic follow-up
(light grey).  Data that are upper limits are shown with hatched lines overlaid.  The sub-sample of
12 galaxies studied here is shown in dark grey.  These galaxies sample the upper half of the [O~III]
luminosity distribution.}
\label{fig:o3_hist}
\end{figure}

We obtained higher S/N optical spectra of 51 of the galaxies with the MMT Blue Channel spectrograph
and other facilities \citep[See][and \S~\ref{section:data_red_MMT} for details.]{tremonti07}.  The
spectra are dominated by the light of the host galaxy, but in some cases there is evidence
suggestive of AGN activity.  Some galaxies have [O~III]~$\lambda5007$/H$\beta$ emission line ratios
that are higher than expected in massive galaxies that are purely star forming (see \S 4.2), and
three of the galaxies show broad Mg~II~$\lambda\lambda2796,2804$ emission lines.  To explore the
possible SMBH activity in these sources, {\em we selected the 12 galaxies with the strongest AGN
signatures for follow-up with HST and Chandra}.  We included all three of the galaxies displaying
broad Mg~II emission (expected to be Type~I AGN) and nine additional galaxies with strong [O~III]
emission, which could be indicative of an obscured (Type~II) AGN.  At the time of our
{\em Chandra proposal}, these galaxies represented the most [O~III]-luminous galaxies in our sample
with spectroscopic follow-up.  At present, the {\em Chandra} sample represents roughly half of the
galaxies with $L_{[O~III]} > 10^{41.5}$~erg~s$^{-1}$ (Fig.~\ref{fig:o3_hist}).

The galaxy coordinates, redshifts, and identification numbers that we use in the text and figures
are provided in Table~\ref{table:general_info}.  This table also includes stellar mass estimates for
the sample based on fitting the broadband UV, optical, and near-infrared photometry using the
Bayesian spectral energy distribution modeling code {\tt iSEDfit} \citep{moustakas13}, as described
in \citet{diamond-stanic12b}.

\section{Data Reduction and Analysis}
\label{section:data_red}

\subsection{MMT Optical Spectra}
\label{section:data_red_MMT}

\begin{figure*}[tbp]
\centering
\begin{tabular}{lr}
\includegraphics[width=0.48\textwidth]{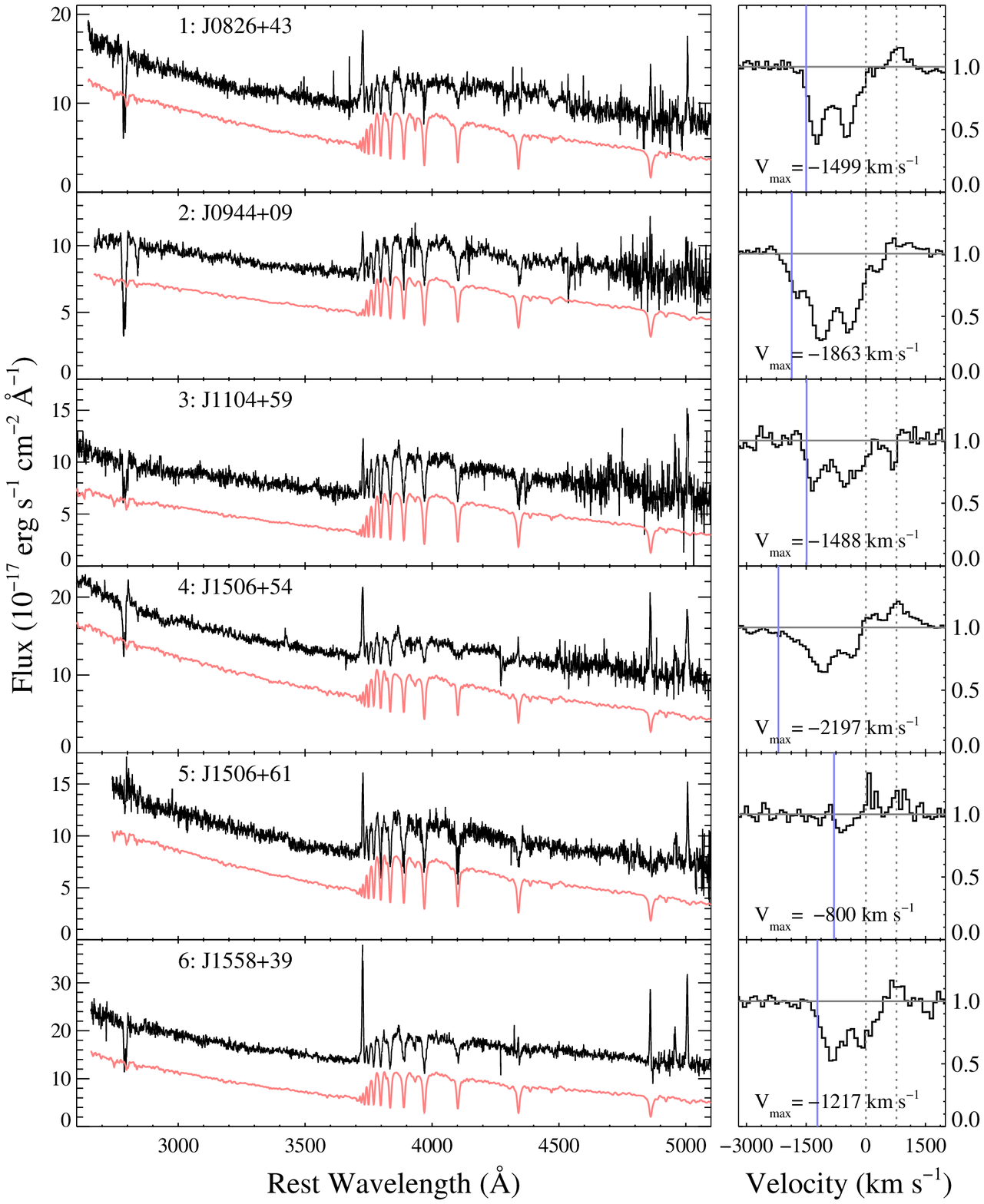} &
\includegraphics[width=0.48\textwidth]{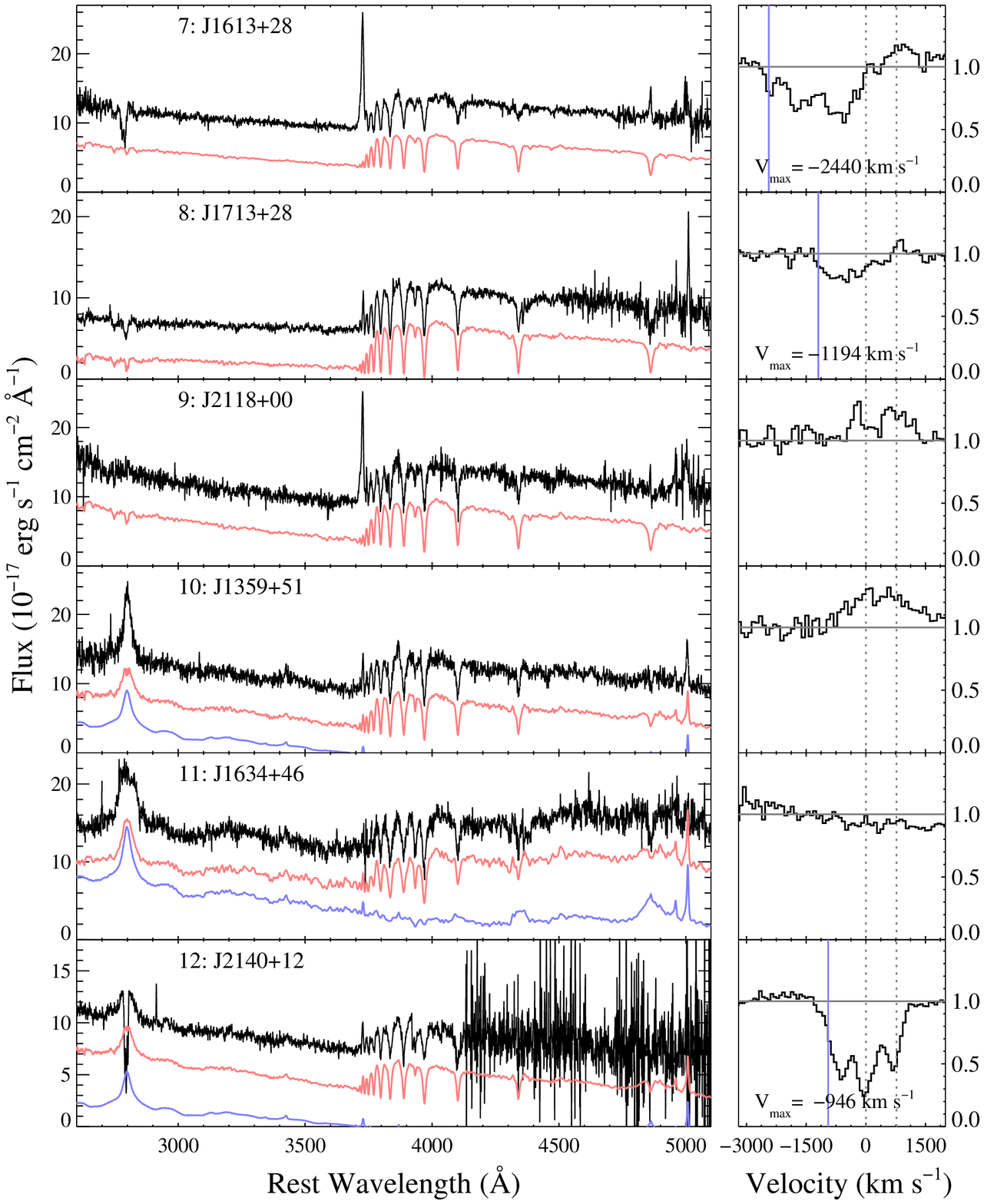} \\
\end{tabular}
\caption{Restframe near-UV and optical spectra of the 12 galaxies.  In the left
panel the black line shows the combined MMT and SDSS spectrum
(joined between 4100 and 5100~\AA) and the red line shows the best fit model
of the continuum, offset in the vertical direction for clarity.  The continuum
model is constructed from a custom grid of stellar population synthesis models
(\S~\ref{section:data_red_MMT}). Since J1359+51, J1634+46, J2140+12 show a broad
Mg~II emission line indicative of a Type~I (unobscured) AGN, a SDSS broad-line
quasar composite spectrum was also included the fit (the blue spectra).
The continuum model is subtracted from each spectrum before measuring the nebular
emission lines of [O~II]~$\lambda3727$, H$\beta$, and [O~III]~$\lambda5007$.  The
region around the Mg~II $\lambda\lambda2796,2804$ lines is enlarged in
the right panel.  In this panel, the continuum is normalized to unity and the x-axis
indicates the velocity of the Mg~II $\lambda2796$ line. The vertical dotted grey lines mark the
restframe wavelength of the Mg~II doublet and the blue line marks the maximum velocity
(see Table~\ref{table:general_info} and \S~\ref{section:data_red_MMT}).  For 9/12 galaxies,
strongly blueshifted interstellar Mg~II is evident.}
\label{fig:MMT_spectra}
\end{figure*}

We obtained high S/N optical spectra of our galaxies with the Blue Channel
Spectrograph on the 6.5-m MMT between December 2004 and July 2007.  We used the
500-line mm grating blazed at 5600, which provided spectral coverage from
approximately 4050~--~7200~\AA\ with a dispersion of 1.19~\AA\ pixel$^{-1}$. For
our $z\sim0.5$ galaxies, this yielded restframe coverage of 2700~--~4800~\AA. 
We observed the galaxies using a 1$^{\prime \prime}$ longslit, which yielded a
FWHM resolution of 3.6~\AA.  Typical exposure times were 45~--~90 minutes.  The
resulting spectra have a S/N of 15~--~30 per pixel.  The spectra were reduced,
extracted, and spectrophotometrically calibrated using the {\sc ispec2d} data
reduction package \citep{moustakas06}.

The main motivation for the \emph{MMT} spectra was to obtain higher S/N
measurements of the Mg~II~$\lambda\lambda2796, 2804$ interstellar medium lines
to look for evidence of gas outflows.  We detected interstellar Mg~II absorption
in 3/4 of the spectroscopic follow-up sample and determined that it is
blueshifted with respect to the starlight indicating gas outflows. 
\citet{tremonti07} reported line centroid velocities ranging from -573 to -2022
km~s$^{-1}$ and highlighted the fact that these outflows are a factor of 2 to 5
times faster than the outflow velocities of typical IR luminous starburst
galaixes \citep[LIRGs and ULIRGs; e.g.,][]{rupke05,martin05b}.

The spectra and the continuum normalized Mg~II lines are shown in
Figure~\ref{fig:MMT_spectra}.  In the 9/12 cases where Mg~II absorption is
observed, the doublet shows complex velocity structure and evidence of multiple
line components. In the present work, we characterize the outflow velocities in
a slightly different manner than in \cite{tremonti07} to avoid the uncertainties
inherent in fitting blended line components.  To compute the average velocity of
all line components, we measure the cumulative equivalent width (EW)
distribution as a function of velocity.  The velocity is defined relative to 
the average wavelength of the doublet ($\lambda_{avg} = 2799.12$) on the
assumption that Mg~II is saturated and thus the 2796 and 2804 lines contribute
roughly equally to the absorption.  The velocity at which the cumulative EW
reaches 50\% is reported as $v_{avg}$ in Table~\ref{table:general_info}.  We
define the maximum velocity, $v_{max}$, as the velocity of the $\lambda2796$
line, at the point where the EW distribution reaches 98\% of the total.  The
value of $v_{max}$ is reported in Table~\ref{table:general_info}, and indicated
by a vertical blue line in Fig.~\ref{fig:MMT_spectra}. In general,  $v_{avg}$
and $v_{max}$ are highly correlated (Pearson r=0.87) with $v_{max}$ being larger
by a factor of $\sim$1.4.  The median values for the sample are $v_{avg} =
-1040$~km~s$^{-1}$ and $v_{max} = -1490$~km~s$^{-1}$.  Determining whether these
fast outflows are driven by starbursts or AGN is the main motivation for the
present work.

The {\em MMT} spectra agree extremely well with the SDSS spectra where there is
overlap.  We therefore join the {\em MMT} and SDSS spectra in order to extend
our spectral coverage redward to include the H$\beta$ and [O~III] $\lambda 5007$
nebular emission lines.  The combined spectra are shown in
Fig.~\ref{fig:MMT_spectra}.

To measure the nebular emission lines, we first model and subtract the stellar
continuum following \cite{tremonti04}.  This is particularly important for the
H$\beta$ line because of the strong underlying stellar Balmer absorption in our
galaxies.  We model the continuum using a linear combination of 10 single-age
stellar population models. At optical wavelengths we use the Charlot and Bruzual
2007 models which are an updated version of the models presented in
\cite{bruzual03}.  At wavelengths less that 3600~\AA\, we use a custom grid of
SSP models built using the UVBLUE theoretical stellar library
\citep{rodriguez05}.  We adopt the \citet{charlot00} dust attenuation curve and
treat dust attenuation as an additional free parameter.

In the three cases where broad Mg~II emission from a Type~I AGN is evident, we
include an additional quasar template in our fitting.  One well-known issue with
the SDSS quasar composite spectrum \citep{vandenberk01} is that the optical
portion of the spectrum was built from low luminosity quasars and therefore it
has some host galaxy contamination.  To avoid this problem, we built our own
quasar composite spectrum using the SDSS DR7 data \citep{abazajian09} and the
quasar catalog of \cite{shen11}.  To best match our Type~I AGN, we selected
non-BAL quasars with Mg~II FWHM = 35~--~85~\AA.  To insure minimal host
contamination we additionally required the quasars to be moderately luminous
($L_{3000} > 10^{45}$~erg~s$^{-1}$).  Our composite spectrum is nearly identical
to the Vanden Berk composite blueward of 4000~\AA, but at redder wavelengths, it
has a steeper (bluer) slope due to the reduced host galaxy contribution.  We
include our quasar composite spectrum as an additional continuum template.  This
enables us to decompose the quasar and starlight in the spectrum and to estimate
the quasar's luminosity (Table~\ref{table:broad-line}).

Our model of the stellar continuum (and QSO continuum, if present) is shown in
red (blue) in Fig.~\ref{fig:MMT_spectra}.  We consider the [O~III] luminosity
and [O~III]/H$\beta$ ratio along with other multi-wavelength diagnostics of AGN
activity in \S~\ref{section:analysis_agn}.

\subsection{HST Optical Observations}
\label{section:data_red_HST}

All 12 galaxies were observed with the Wide Field Camera 3
\citep[WFC3;][]{kimble08} Ultraviolet Imaging Spectrograph (UVIS) channel using
the F814W filter \citep{lupie03} aboard {\em HST}.  The observations were
dithered in a 2-step sequence 1.49" apart.  We used the On-the-Fly Reprocessing
\citep[OTFR;][]{swade01} for the basic reduction and calibrations, but we
reprocessed the images with MultiDrizzle \citep{jedrzejewski05} to resample the
images to a smaller pixel size (0.02$^{\prime \prime}$~pixel$^{-1}$) and drop
size (0.8) so that they have an RMS value in the center of the weight image
$\lesssim 20\%$ the median value, as recommended in the Multidrizzle
handbook\footnote{http://stsdas.stsci.edu/multidrizzle/}.  We experimented with
a grid of values of these two parameters and found that these provided the best
sampling relative to the RMS.

We present 30~kpc~$\times$~30~kpc cutouts of our reduced galaxy images in
Fig.~\ref{fig:cutouts}.  (See the Appendix for additional images.)  All of the
galaxies show a bright, compact central source and are surrounded by irregular
diffuse emission.  In several cases, the galaxies show clear tidal tails
indicative of late-stage major or minor mergers.

In 5/12 galaxies the dominant central source is so compact that the structure of
the {\em HST} point spread function (PSF) is faintly visible, suggesting that
the galaxies are nearly unresolved.   This is remarkable considering that
typical host galaxy half-light radii at $z\sim0.5$ are 0.1~--~0.6 arcsec or
$\sim 10$~--~60 times the {\em HST} PSF FWHM \citep[e.g.,][]{cassatta11}.  To
explore the compactness and the nature of the extended diffuse light further, we
undertake quantitative image analysis.

\begin{figure*}[tbp]
\centering
\includegraphics[width=\textwidth]{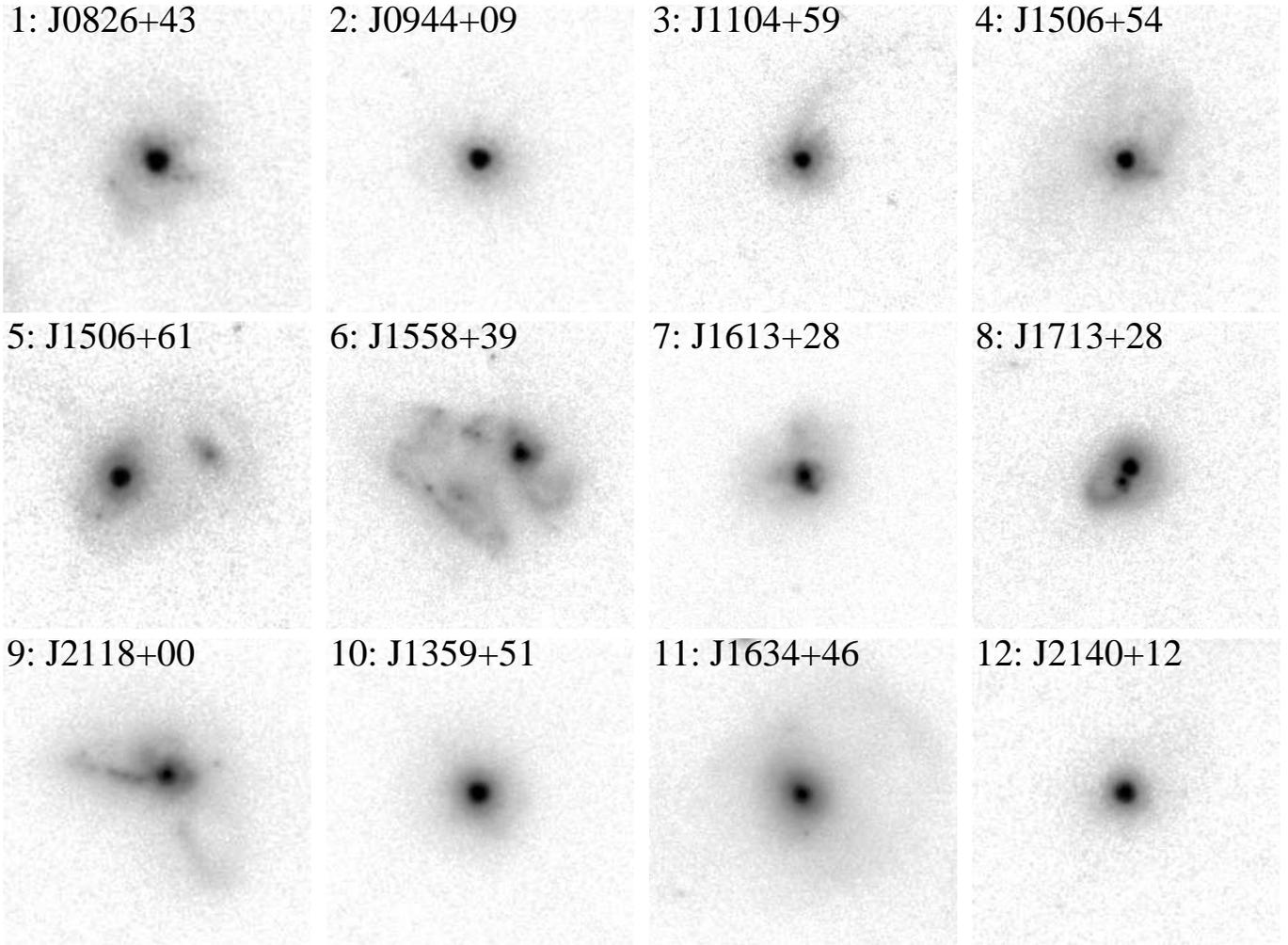}
\caption{30~kpc~$\times$~30~kpc cutouts from the HST/WFC3 F814W (restframe V-band at these
redshifts) observations of our 12 galaxies.  The images are logarithmically scaled using the
best-fit sky level and S\'{e}rsic magnitudes to define the lower and upper scaling limits,
respectively.  North is up and east is left.}
\label{fig:cutouts}
\end{figure*}

\subsubsection{Quantitative Image Analysis with GALFIT}
\label{section:galfit}

To obtain quantitative morphological information about our galaxies,
we employed GALFIT version 3 \citep{peng02,peng10}, a two-dimensional
fitting algorithm.  GALFIT optimizes the fit of various models to an
image (e.g., S\'{e}rsic profile, exponential disk, Gaussian) after
convolving each with a user-supplied PSF.  Because the \emph{HST} PSF
is complex and temporally and spatially variable, care must be taken
in constructing the PSF model when the sources of interest are near
the resolution limit.  The preferred method of generating a PSF is to
observe a nearby bright star at the same position on the CCD as the
science target immediately before or after the science exposure.
Since we did not anticipate that our galaxies would be extraordinarily
compact, we did not take separate PSF images concurrently with our
data.  However, after much experimentation, we arrived at a technique
for generating an empirical PSF from moderately bright stars in our
science images.  Details are discussed in the Appendix.

We fit each galaxy in a 1400~pixel~$\times$~1400~pixel cutout centered
on the brightest core.  This corresponds to 28~arcsec on a side or
151~--~206~kpc on a side for the range of redshifts in our sample.  We
measured radial profiles and visually inspected the images to verify
that these boxes encompass all of the clearly associated diffuse
merger emission (all radial profiles flatten out to the sky level).
To mask the light from the background/foreground galaxies/stars, we
used the SExtractor software \citep[V2.8.6][]{bertin96} to generate
elliptical masks in the vicinity of our main target objects.  After
convolving the images with a Gaussian kernel with FWHM=4 pixels, we
detected objects using DETECT\_THRESH between 1.5 and 3.0 and
requiring DETECT\_MINAREA=17 pixels.  We were conservative in defining
our object masks.  We extended our masks out to 5 times the Kron
radius measured by SExtractor in order to reach the background level.
We inspected the ellipses within our cutouts, removing duplicate or
spurious ones, increasing the ellipse sizes in a few extended haloes,
and added polygons to encompass image artifacts.  We then used the
ellipses and polygons to create image masks for each of our galaxy
cutouts.  The images used in the fitting, radial profiles, and
residual images are provided in the Appendix in
Fig.~\ref{fig:data_fits}.

GALFIT provides the option of fitting the sky simultaneously with
the galaxy model light profile.  However, our galaxies contain diffuse tidal
features that are not well modeled by the simple galaxy light profiles
we assume, and these can skew the sky determination.  Therefore, we opt
to measure the sky independently and freeze it in our fits.
We created radial profiles from our masked cut-out images.  Then we determined
by-eye where the radial profile became effectively flat and took this to be the
sky region.  Since careful inspection showed no evidence of sky gradients, we 
averaged all the unmasked pixels in this radial bin (over a million per galaxy) to 
determine the sky level.  

We fit these galaxies with simple, physically motivated model
combinations.  We began by fitting each distinct, bright galaxy core
with a S\'{e}rsic \citep{sersic63} $+$ sky model (where all sky model
parameters were frozen).  For the two galaxies with two
distinct, clearly associated, bright cores, in the process of major
or minor mergers, J1506+61 and J1713+28, we fit each core with a
S\'{e}rsic profile.  To accommodate the bright,
compact cores of our galaxies, we also tried fitting each core with a
S\'{e}rsic $+$ PSF $+$ sky model (again, all sky model parameters were
frozen).  In the case of the two galaxies with two
distinct, clearly associated, bright cores, J1506+61 and J1713+28, we
only added a PSF to the secondary core if there is a noticeable
improvement to the fit (relatively large change in $\chi^2/\nu$)
and if the best-fit PSF magnitude was non-negligible
(the PSF magnitude had to be $\gtrsim 1\%$ of the S\'{e}rsic
magnitude).  This requirement was only satisfied for J1713+28.

For the galaxies with visible broad-line regions (J1359+51, J1634+46,
and J2140+12), the PSF component can be used to model
unresolved light from the AGN accretion disk.  Our spectroscopic
analysis suggests that the AGN should account for no more than
30~--~35\% of the continuum in the {\em HST} image.  For the remainder
of the galaxies, the PSF component could be used to model extremely
compact star-forming regions (See \S~\ref{section:compact_light} for
further discussion.)  We found that considering any more components
beyond a S\'{e}rsic $+$ PSF model per core was not physically
justifiable and was technically intractable: considerable parameter
degeneracies were encountered, GALFIT became more easily stuck in
local minima, and the fits became very sensitive to the starting
parameters.

To quantitatively assess whether a merger is major or minor, we use
the integrated light ratios from the GALFIT results (the total
integrated magnitudes in Table~\ref{table:HST_info}).  First, J1713+28
appears to be a near-equal mass major merger with two distinct nuclei
with an integrated light ratio of 1.2.  However, the light ratio is skewed by the way
GALFIT defines the two cores (the first is very compact, while the
second is much more extended and includes some of the tidal debris;
see Table~\ref{table:HST_info}).  These nuclei are linked by a
small-scale tidal tail and appear to be separated less than a few kpc.
Second, J1506+61 appears to be an on-going merger
with a much fainter companion core $\sim 10$~kpc away and clearly
associated tidal debris $\sim 70$~kpc away on the opposite side of the
galaxy from the companion.  The integrated light ratio of 1.05 is also
biased similarly to J1713+28 without including a PSF model.  However,
when a PSF is included for the primary core, the bias apparently
disappears and we find an integrated light ratio of 4.9.  Therefore,
this galaxy appears to be a minor merger.

\begin{figure}
\begin{tabular}{lr}
\begin{minipage}{0.238\textwidth}
\includegraphics[width=\textwidth]{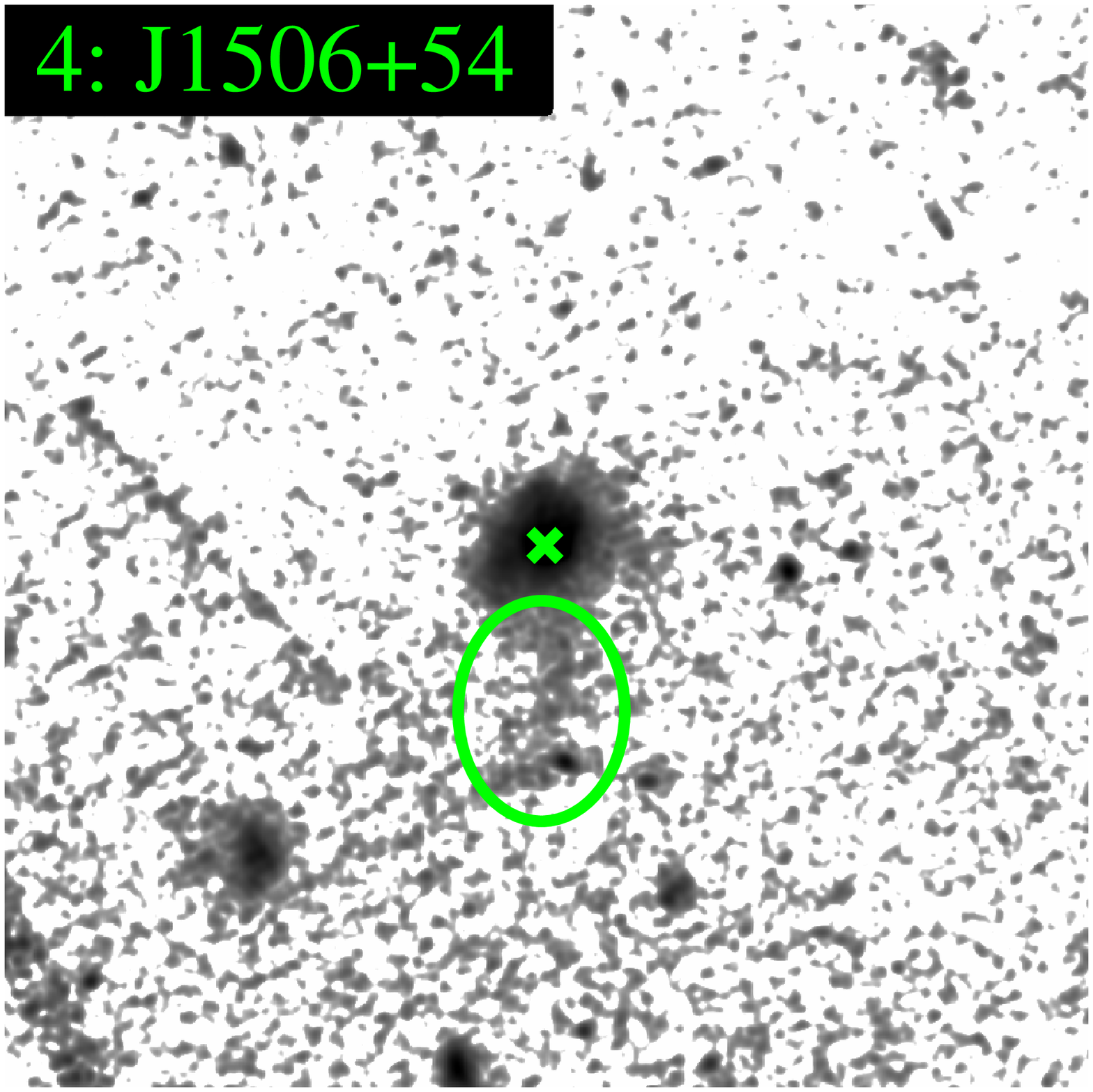}
\end{minipage}
&
\begin{minipage}{0.238\textwidth}
\includegraphics[width=\textwidth]{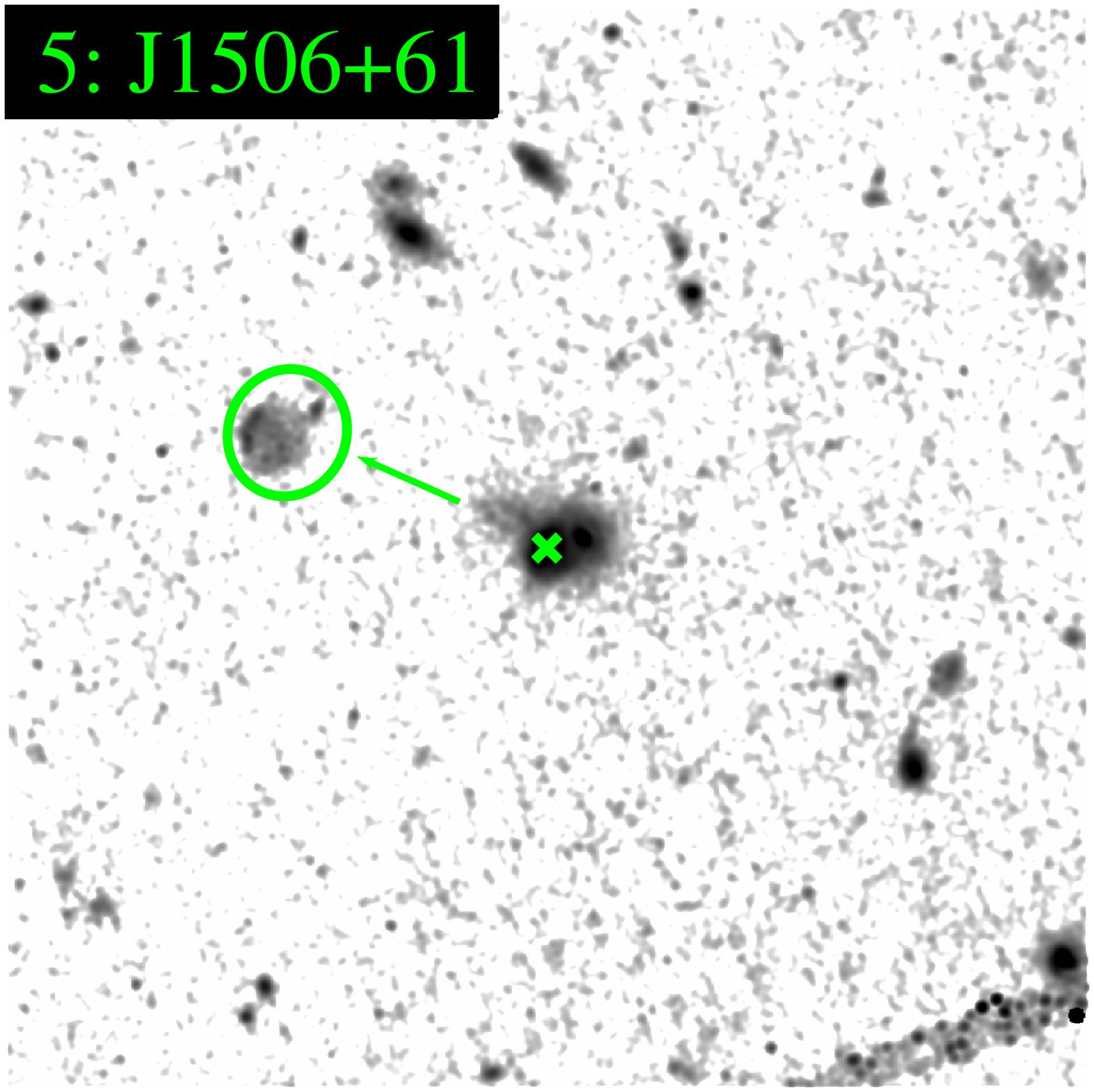}
\end{minipage}
\\
\begin{minipage}{0.238\textwidth}
\includegraphics[width=\textwidth]{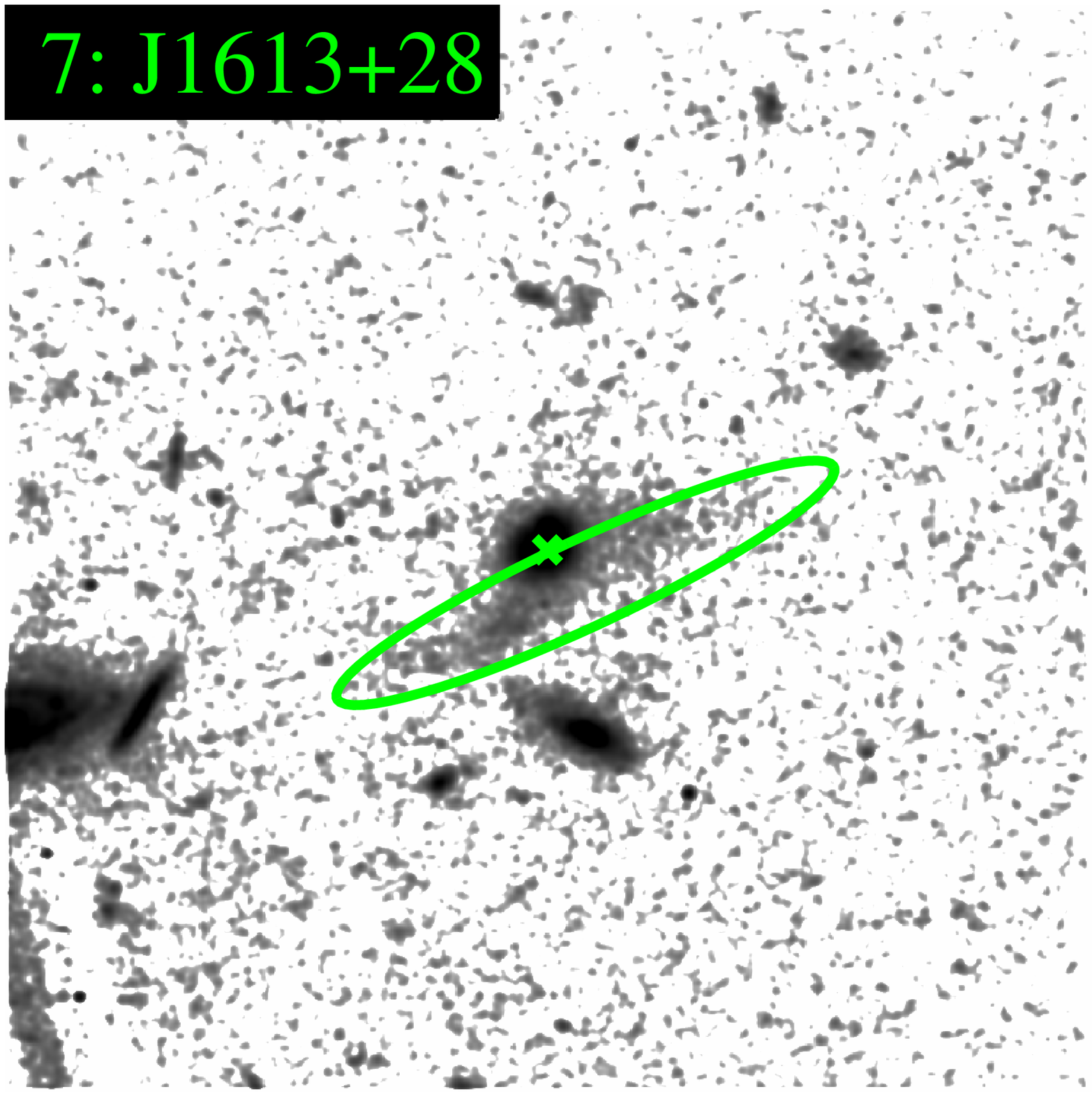}
\end{minipage}
&
\begin{minipage}{0.238\textwidth}
\includegraphics[width=\textwidth]{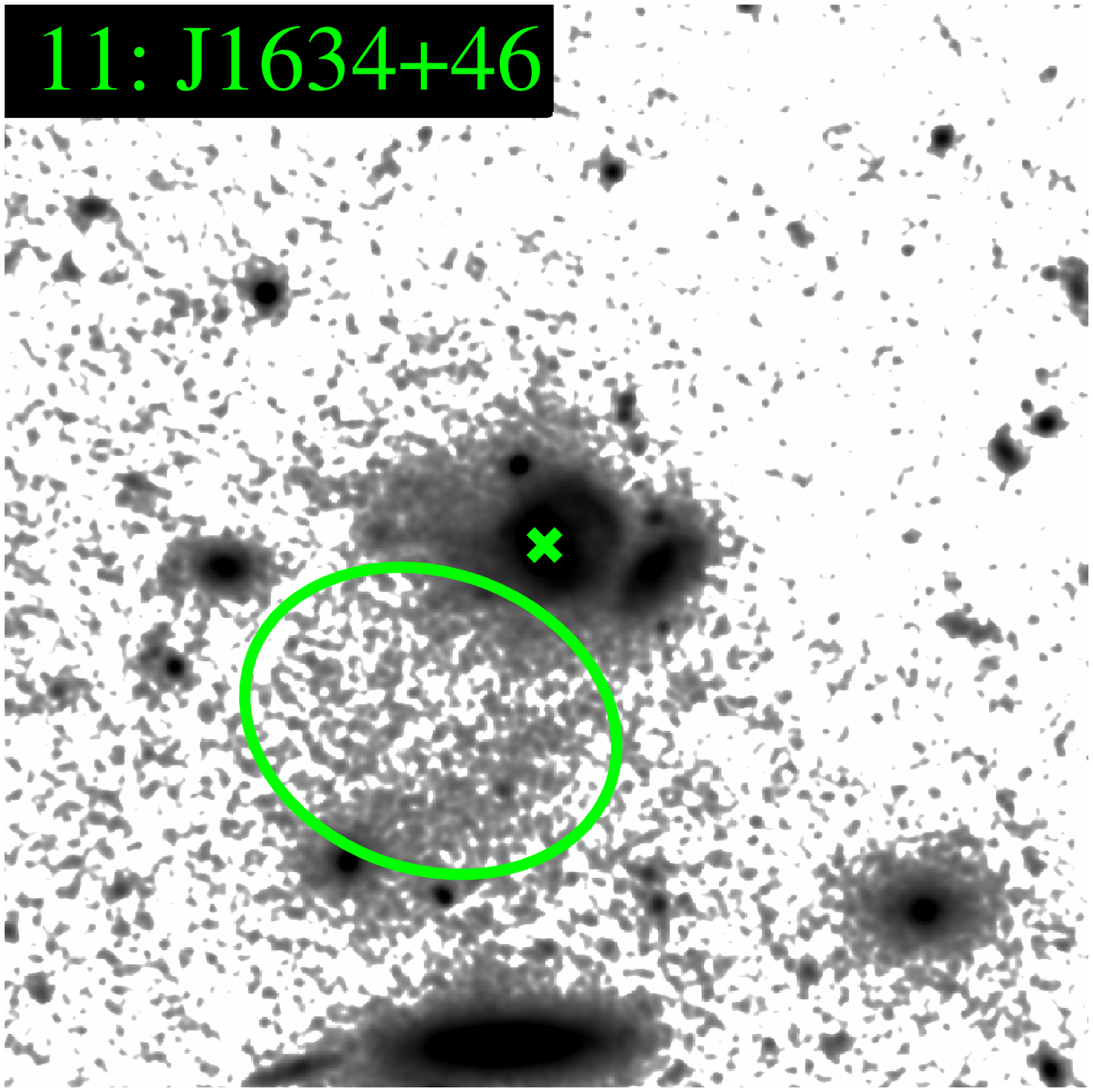}
\end{minipage}
\end{tabular}
\caption{Additional very faint diffuse emission that appears to be associated with a few galaxies,
which is not shown in cutouts elsewhere.  The images have been binned $2 \times 2$, smoothed and
filtered with IDL's {\sc leefilt}, then smoothed again with a 5-pixel wide Gaussian kernel.  The
very straight lines in the corners of the images are artifacts.  The center of each galaxy has been
marked with an ``X" and the extended diffuse emission has been encircled with an ellipse.  Each
stamp is 230~kpc~$\times$~230~kpc.  North is up and east is left.}
\label{fig:smoothed_images}
\end{figure}

We explored fitting one more galaxy in the sample with a second
S\'{e}rsic.
J1634+46 has a second
bright core $\sim 20$~kpc away.  If this core is associated with tidal
debris near the primary core, it would clearly be a minor merger with
an integrated light ratio of 37.  While this core is projected within
the very faint diffuse emission surrounding the primary core (see
Fig.~\ref{fig:smoothed_images}), there is no other clear morpological
link between it and the merger remnant, suggesting that it could be a
chance projection of a faint background galaxy.  Therefore, we have
chosen not to fit this core with a second S\'{e}rsic model.  Not
fitting this core does not affect our GALFIT results for the primary
core.

For these two models, we experimented with either floating or fixing
various model parameters.  In particular, we experimented with
allowing the S\'{e}rsic index, n, to float versus being frozen at
common values of n=1 (exponential) or n=4 (de Vaucouleurs).  However,
the profiles were so peaked with relatively extended wings that the
best-fit S\'{e}rsic index always floated well beyond n=4 up to the
parameter space maximum allowed by GALFIT of n=20.  In addition, when
the S\'{e}rsic index floated, the estimate of the effective radius of
the galaxy became much less constrained, as expected.  The combination
of a compact central source and extended irregular diffuse structure
is inherently difficult to model.  The inner part of the S\'{e}rsic
profile is unresolved, while the faint outer wings, which could in
principle help constrain $n$, are contaminated by the diffuse tidal
structures.  We conclude that we cannot meaningfully measure
both the effective radius and the S\'{e}rsic index simultaneously for
any model combination.  Therefore, we freeze the S\'{e}rsic index to 4
for all of our fits.  Tests of measurements of effective radii at
various S\'{e}rsic radii (n~$\sim 1$~--~5) indicate that freezing the
S\'{e}rsic index to such values does not strongly bias the effective
radius.

We still found that these simplifications were not enough to prevent
GALFIT from sometimes running into unphysical or unusual regions of
parameter space.  Therefore, we made one more model simplification to
reduce the number of free parameters, which we found did help GALFIT
find reasonable minima.  If we fitted a S\'{e}rsic and a PSF to a
single core, we tied the centroids of the models together to reduce
the number of free parameters.  This was consistent with by-eye
inspection of the images that the PSF seems to be centered very close
to the same position as the resolved emission.

One additional issue with GALFIT is that the results can be sensitive
to the choice of starting parameters, suggesting that GALFIT is
finding a local rather than a global minimum.  In order to explore
this issue, we ran GALFIT using a large grid of possible starting
parameters.  We found that our S\'{e}rsic $+$ sky fits were very
robust, returning near identical values of the effective radius and
total magnitude for most values of the starting parameters.  The
S\'{e}rsic $+$ PSF $+$ sky fits were considerably less robust, as the
range of best-fit model parameters had large dispersions.  The best
fit beginning from a single set of starting parameters was
sometimes not the same as or even close to the globally-minimized
$\chi^2$, indicating that GALFIT stopped at a local minimum.  Further
details may be found in the Appendix.

\renewcommand{\thefootnote}{\alph{footnote}}
\begin{deluxetable*}{cccccc|cccccc}
\tablecaption{Quantitative Morphological Fitting Results}
\tablehead{
	\colhead{ID} &
	\colhead{Galaxy} &
	\multicolumn{4}{c|}{S\'{e}rsic Only} &
	\multicolumn{6}{c}{S\'{e}rsic $+$ PSF} \\
	\colhead{} &
	\colhead{} &
	\colhead{S\'{e}rsic Mag} &
	\colhead{R$_e$ (pix)} &
	\colhead{R$_e$ (pc)} &
	\multicolumn{1}{c|}{\% Resid} &
	\colhead{S\'{e}rsic Mag} &
	\colhead{PSF Mag} &
	\colhead{PSF frac} &
	\colhead{R$_e$ (pix)} &
	\colhead{R$_e$ (pc)} &
	\colhead{\% Resid} \\
	\colhead{(1)} &
	\colhead{(2)} &
	\colhead{(3)} &
	\colhead{(4)} &
	\colhead{(5)} &
	\multicolumn{1}{c|}{(6)} &
	\colhead{(7)} &
	\colhead{(8)} &
	\colhead{(9)} &
	\colhead{(0)} &
	\colhead{(11)} &
	\colhead{(12)}
}
\startdata
1  & J0826+43 & 19.23 & 1.60       & 214   & 38\% & 19.50 & 19.64 & 0.47 & 39     & 5200 & 10\% \\
2  & J0944+09 & 19.28 & 1.08       & 133   & 31\% & 19.95 & 19.65 & 0.57 & 23     & 2800 & 14\% \\
3  & J1104+59 & 19.25 & 1.68       & 219   & 20\% & 19.80 & 19.79 & 0.50 & 15     & 2000 &  5\% \\
4  & J1506+54 & 19.06 & 1.08       & 165   & 35\% & 19.25 & 19.38 & 0.47 & 54     & 7300 &  3\% \\
5  & J1506+61 & 19.45 & 1.92       & 217   &  4\% & 19.69 & 19.99 & 0.43 & 28     & 3200 &  5\% \\
   &          & 19.50 & 165$^a$    & 18700 &      & 20.80 &       &      & 48     & 5400 &      \\
6  & J1558+39 & 19.05 & 7.74       & 827   & 48\% & 18.6  & 20.1  & 0.20 & 71     & 7600 &  7\% \\
7  & J1613+28 & 18.69 & 8.55       & 980   & 17\% & 18.72 & 22.27 & 0.04 &  9.6   & 1100 & 16\% \\
8  & J1713+28 & 19.69 & 1.32       & 173   &  8\% & 19.9  & 21.2  & 0.23 &  2.6   &  340 &  1\% \\
   &          & 19.46 & 28.6$^a$   & 3760  &      & 19.5  & 22.2  & 0.08 & 47$^a$ & 6200 &      \\
9  & J2118+00 & 18.73 & 19.3$^b$   & 2240  & 19\% & 18.71 & 21.25 & 0.09 & 31     & 3600 & 10\% \\
10 & J1359+51 & 18.87 & 3.24       & 352   & 17\% & 19.19 & 20.00 & 0.32 &  9.3   & 1000 &  9\% \\
11 & J1634+46 & 18.47 & 12.4       & 1630  & 37\% & 18.40 & 20.09 & 0.17 & 37     & 4900 & 19\% \\
12 & J2140+12 & 19.48 & 1.71       & 251   & 15\% & 20.13 & 20.12 & 0.50 &  8.9   & 1300 &  8\%
\enddata
\tablecomments{
Column 1:  Galaxy identification number used in some figures for brevity.
Column 2:  SDSS short name.
Columns 3 and 7:  Best-fit S\'{e}rsic magnitude for each model.
Columns 4, 5, 10, and 11:  Best-fit S\'{e}rsic effective radius in {\em HST} pixels
(0.02"~pixel$^{-1}$) and parsecs.
Columns 6 and 12:  Percent of light in the residual image.
Column 8:  Best-fit PSF magnitude.  Column 9:  The fraction of light in the PSF, $psf_{frac}$ is
defined as $psf_{flux} / (sersic_{flux} + psf_{flux})$, which is a proxy for how PSF-dominated or
how well the galaxy is resolved.  We freeze the S\'{e}rsic index, n, to 4 in all fits.  We discuss
why we choose not to provide parameter uncertainties in the Appendix.
\\
$^a$ These values of the effective radius for the second core are unusually large because the
best-fit $r_e$ to the primary core is very small, requiring the second S\'{e}rsic to fit to the
larger-scale galactic emission.  In particular for J1506+61, once a PSF is added to the primary
core, the S\'{e}rsic fit to that same core then fits most of the larger-scale galactic emission.
The fit is clearly much better when a PSF is added to the primary core for this galaxy.
\\
$^b$ J2118+00 has the largest single S\'{e}rsic effective radius in the sample because bright tidal
arms are superimposed on the core.}
\label{table:HST_info}
\end{deluxetable*}
\renewcommand{\thefootnote}{\arabic{footnote}}

The results of our GALFIT modeling are listed in
Table~\ref{table:HST_info}.  The values reported are for the model
with the minimum $\chi^2$ drawn from our large grid of models with
different starting parameters.  In this table, we provide the S\'{e}rsic
and PSF magnitudes, effective radii, the percentage of light in the
residual image, and the fraction of light that GALFIT finds for the
PSF:  $psf_{frac}$ = $psf_{flux} / (sersic_{flux} + psf_{flux})$.
These latter two quantities are useful because they highlight for each
galaxy how much of the light can be fit in the simple models
(relatively how disturbed they are), how much of an improvement adding
the PSF makes, and a quantitative measure for how compact, unresolved,
or PSF-like the galaxies are to one another.  We found that when the
PSF flux is approximately greater than or equal to the S\'{e}rsic
flux, the airy ring and diffraction spikes of the PSF become apparent.

The percent residuals provided in Table~\ref{table:HST_info} indicate the
fraction of galaxy light that is not well fit by our simple GALFIT models.
The percent residuals are calculated by summing the pixels in the residual
and original sky-subtracted images and then taking the ratio.  Only pixels
that are unmasked and inside of the sky annulus are included in the sum.
We caution that this method of characterizing the residuals can sometimes
be misleading because negative and positive deviations in the residual
image can cancel out.  In all cases except one, the percent residuals
are smaller when including a PSF model, as expected, because the
additional model provides more freedom for the fit and better matches
what is frequently a large fraction of unresolved light.  In J1506+61, the
percent residuals are smaller without the PSF because the S\'{e}rsic model
over-subtracts the host galaxy and the negative residual cancels out
some of the positive residuals from the tidal features.  However, the
radial profiles of this galaxy (see Appendix) clearly show that the
model is better matched to the data when a PSF model is included.

\setlength{\tabcolsep}{-1pt}
\setlength{\tabcolsep}{6pt}

\subsubsection{The Search for Very Faint, Diffuse, Extended Emission}

To explore the nature of the extended diffuse emission, we employed two
different techniques.  We utilized GALFIT to remove the smooth,
high-surface-brightness features from the images, making the diffuse irregular
emission more visible (see Fig.~\ref{fig:data_fits}).  Our second approach was
to use an image filtering technique called {\sc LEEFILT} in IDL \citep{lee86}
that helped considerably to smooth out image noise to bring out the the faint,
extended, diffuse emission.  Examples of smoothed and filtered images are shown
in Fig.~\ref{fig:smoothed_images}.  While the emission shown in this figure is
as low as $\sim 10-50\%$ above the background, it is extended over hundreds of
pixels, making it highly significant ($\gg 10 \sigma$).  The surface brightness
of this emission (Fig.~\ref{fig:smoothed_images}; $\mu \sim
25$~mag~arcsec$^{-s}$) is approximately an order of magnitude fainter than the
faint emission in the cutouts (Fig.~\ref{fig:cutouts}).  This appears to be the
``fine structure" and tidal debris from recent mergers discussed by
\cite{duc13}.  Note that two images (J0826+43 and J2140+12) have very faint
artifacts from internal camera reflections that we have masked in our analysis. 
Diffuse structure is evident in all the images with a wide range of morphologies
(see also Fig.~\ref{fig:data_fits} in the Appendix).

\subsection{Chandra X-ray Observations}
\label{section:data_red_Chandra}

\begin{deluxetable*}{cccccccc}
\tablecaption{Chandra X-ray Observation Data}
\tablehead{
	\colhead{ID} &
	\colhead{Galaxy} &
	\colhead{\em Chandra} &
	\colhead{Exposure} &
	\colhead{Src} &
	\colhead{Bkg} &
	\colhead{PNS} &
	\colhead{Log(L$_{\rm X}$)} \\
	\colhead{} &
	\colhead{} &
	\colhead{ObsID} &
	\colhead{Time (s)} &
	\colhead{Cnts} &
	\colhead{Cnts} &
	\colhead{} &
	\colhead{(erg~s$^{-1}$)} \\
	\colhead{(1)} &
	\colhead{(2)} &
	\colhead{(3)} &
	\colhead{(4)} &
	\colhead{(5)} &
	\colhead{(6)} &
	\colhead{(7)} &
	\colhead{(8)}
}
\startdata
1  & J0826+43 &  11698  &  9902   &  0   &  0.046  &  $1.0$                 &  $<42.9$               \\
2  & J0944+09 &  11702  &  6398   &  0   &  0.029  &  $1.0$                 &  $<42.9$               \\
3  & J1104+59 &  11696  &  8265   &  0   &  0.030  &  $1.0$                 &  $<42.9$               \\
4  & J1506+54 &  11699  &  10780  &  4   &  0.086  &  $2.4 \times 10^{-6}$  &  $41.9_{-1.0}^{+1.0}$  \\
5  & J1506+61 &  11705  &  13263  &  1   &  0.047  &  $0.046$               &  $<42.6$               \\
6  & J1558+39 &  11706  &  6139   &  1   &  0.022  &  $0.022$               &  $<42.8$               \\
7  & J1613+28 &  11700  &  4982   &  4   &  0.043  &  $1.6 \times 10^{-7}$  &  $42.2_{-0.6}^{+0.8}$  \\
8  & J1713+28 &  11703  &  5832   &  1   &  0.024  &  $0.023$               &  $<43.2$               \\
9  & J2118+00 &  11707  &  5655   &  4   &  0.055  &  $4.0 \times 10^{-7}$  &  $42.5_{-0.6}^{+0.4}$  \\
10 & J1359+51 &  11697  &  8215   &  97  &  0.082  &  $0.0$                 &  $43.6_{-0.3}^{+0.1}$  \\
11 & J1634+46 &  11704  &  4982   &  67  &  0.045  &  $0.0$                 &  $44.0_{-0.1}^{+0.3}$  \\
12 & J2140+12 &  11701  &  11365  &  0   &  0.051  &  $1.0$                 &  $<43.1$               \\
M  & M        &  M      &  46653  &  15  &  0.277  &  $3.1 \times 10^{-21}$ &  $42.2_{-0.4}^{+0.3}$  
\enddata
\tablecomments{Column 1:  Galaxy identification number used in some figures for brevity.  ``M" is
the merged X-ray spectrum of all sources with 1~--~4 counts.
Column 2:  SDSS short name.
Column 3:  {\em Chandra} observation identification number.
Column 4:  Exposure time of each {\em Chandra} observation.
Column 5:  Counts in {\em Chandra} source extraction region enclosing 95\% of the PSF in the range
0.5~--~8.0~keV.
Column 6:  Expected background counts in the source region from a nearby annular extraction.
Column 7:  PROB\_NO\_SOURCE:  AE's probability that there is no source in the range 0.5~--~8.0~keV
as defined in section 5.10.3 of the AE users manual.
Column 8:  2.0~--~10.0~keV K-corrected X-ray luminosity.}
\label{table:chandra_info}
\end{deluxetable*}

The 12 targets in Table~\ref{table:general_info} were observed on the S3 chip of
the Advanced CCD Imaging Spectrometer \citep[ACIS;][]{garmire03} aboard the {\em
Chandra X-ray Observatory} for a total of 95.78~ksec, $\sim 5$~--~13~ksec each. 
Data were taken in timed exposure mode with the standard frame time at the
default location on the S3 chip and telemetered to the ground in very faint
mode.  Data reduction and point source extractions were completed using CIAO
version 4.2 \citep{fruscione06}, and ACIS Extract version
2010-02-26\footnote{http://www2.astro.psu.edu/xray/docs/TARA/ \\
ae\_users\_guide/ae\_users\_guide.html} \citep[AE;][]{broos10}, respectively.

To maximize our {\em Chandra} observing efficiency, exposure times were
estimated from both the SMBH mass (using the stellar mass; to detect a source
accreting at $>1\%$ of Eddington), and from $L_X$ / $L_{[O~III]}$ scaling
relations for Type~I AGN \citep{heckman05} with an additional obscuring screen
of $10^{22\text{~--~}23}$~cm$^{-2}$.  The goal was to detect each source with
$\sim 10$~--~150 counts (0.5~--~8.0~keV).  However, only two galaxies are
clearly bright, Type~I AGN:  J1359+51 and J1634+46, which are detected with 97
and 67 counts, respectively.  The remaining ten galaxies were either not
detected or barely detected by {\em Chandra} (J2140+12 shows faint AGN
broad-line Mg~II emission, but is among those not detected by {\em Chandra}). 
Three of these ten galaxies had only four counts each (0.5~--~8.0~keV).  Using
these sources and the one-count sources (0.5~--~8.0~keV; J1506+54, J1506+61,
J1558+39, J1613+28, J1713+28, J2118+00), we created a merged spectrum that we
will use to ascertain the nature of the faint X-ray emission.

We use AE's PROB\_NO\_SOURCE (section 5.10.3 of the AE users manual) to assess
if these galaxies with four counts each are significant detections (the single
count sources are consistent with the background).  This statistic gives the
probability that the observed counts in the source extraction region are
background counts.  Given the low background levels of the {\em Chandra}
observations and, especially that we know the location of the source a priori,
the probabilities of $\lesssim 10^{-6}$ are highly significant.  Furthermore,
the merged data supports this conclusion.  PROB\_NO\_SOURCE is extremely small
and a K-S test \citep{kolmogorov41} between the merged source and background
spectra (p=$3.5 \times 10^{-7}$ for 0.3~--~9.886~keV) strongly suggests that the
merged spectrum is different from the background.

For the two brightest sources, the source position was adjusted based on the
mean position of the extracted counts to more accurately calculate the point
source photometry.  For the remaining sources where possible, we aligned our
exposures using other sources within a few arcminutes of our target that are
detected in both the {\em HST} and {\em Chandra} images.  This adjustment for
the faint or undetected sources resulted in a shift in the coordinates of the
extraction region $\lesssim 1^{\prime \prime}$, which is consistent with the
combined expected astrometric accuracy of {\em Chandra} ($\lesssim 0.8 ^{\prime
\prime}$)\footnote{http://cxc.harvard.edu/cal/ASPECT/celmon/} and {\em HST}
\citep[$\sim 0.4$~--~$0.8 ^{\prime \prime}$;][]{morrison01}.

We used Sherpa version 4.5 \citep{freeman01} to jointly fit the unbinned source
and background spectra at 0.5~--~8.0~keV of each source and of our merged source
and background spectra.  We used the C-statistic for the fitting, which is
similar to the \cite{cash79} statistic but with an approximate goodness-of-fit
measure.  We used this statistical method for the fitting to avoid losing the
little spectral information that we have for most of our spectra
\citep[e.g.,][]{nousek89}.  Since degeneracies in the fit parameters will
frequently arise for low-count sources, we implemented a simple fitting scheme. 
We fit each source spectrum with an absorbed power-law model ({\small \sc
xsphabs $\times$ xspowerlaw}).  The column density, $N_H$, was fixed at the
Galactic foreground value \citep{dickey90} for each source and the photon index,
$\Gamma$ was fixed at 1.7, consistent with typical values found for AGN
\citep[e.g.,][]{page05} and X-ray binaries \citep[XRBs; e.g.,][]{sell11}.  Each
background spectrum was simultaneously fit with a power-law with $\Gamma = 1.4$,
consistent with the hard X-ray background \citep[e.g.,][]{tozzi06}.  Note,
however, that, because the background is very low, approximately half of the
flux in the background region is
instrumental\footnote{http://cxc.harvard.edu/proposer/POG/html/ \\
chap6.html\#tth\_fIg6.21}.

If we allow $\Gamma$ to float for the two Type~I AGN and the merged source
spectra, the best-fit value is consistent with $\Gamma = 1.7$ within the
uncertainties calculated by Sherpa's {\sc
conf}\footnote{http://cxc.harvard.edu/sherpa/ahelp/conf.html}.  We also consider
possible attentuation in these sources.  Including intrinsic, redshifted
absorption for these three sources does not change the luminosity more than 10\%
as such a model strongly prefers low instrinsic $N_H$; we find $3\sigma$ upper
limits on the intrinsic $N_H$ for J1359+51, J1634+46, and the merged spectrum,
respectively (in units of $10^{21}$~cm$^{-2}$):  1.6, 5.6, and 4.9.  For these
three sources, $N_H$ is 2~--~3 orders of magnitude smaller than expected for a
highly obscured or Compton-thick AGN \citep[e.g.,][]{vignali10}.  This is also
supported by hardness ratio analysis of the merged spectrum.  Using the BEHR
code \citep{park06}, we calculate a hardness ratio, (H-S)/(H+S), of
$-0.58_{-0.13}^{+0.30}$ (H=2.0-8.0~keV, S=0.5~--~2.0~keV).   However, this value
is not unexpected for XRBs, which can be a wide range of X-ray colors
\citep[e.g.,][we return to this in \S~\ref{section:xrays}]{trouille08}.  These
results are consistent with the finding that these are relatively soft sources
exhibiting very little intrinsic absorption.

For comparison to other studies, the 2.0~--~10.0~keV X-ray luminosities listed
in Table~\ref{table:chandra_info} were calculated from each of the model fits
and the uncertainties were calculated as follows.  The luminosity is evaluated
at each point in a two-dimensional grid 200 points on a side, each dimension
corresponding to the power-law photon index and power-law normalization.  The
absorption was frozen at the galactic foreground value and not allowed to vary
since we can only constrain $N_H$ in the Type~I AGN and merged sources; the
Type~I AGN and merged source uncertainties were calculated in the same way for
consistency.  The photon index was allowed to vary $1.7 \pm 0.4$ to encompass
the typical observed photon indices for AGN \citep[e.g.,][]{xue11}.  Then the
minimum and maximum luminosity was selected within the confidence interval for
the appropriate change in statistic value \citep[e.g.,][]{avni76}.

\subsection{JVLA Radio Observations}
\label{section:data_red_VLA}

Radio observations were obtained for 10 of the 12 galaxies with the JVLA in
C-configuration during fall 2010.  Observations were made using the L-band
continuum mode with a spectral range of 1536-1664 MHz made up of 256 channels. 
We began each set of observations by looking at a bright flux calibrator
followed by a phase calibrator.  We then spent 80~--~90 minutes of total
integration time on each target galaxy, alternating between the target and the
phase calibrator every 20 minutes.

The observations were reduced using standard calibration techniques in CASA
\citep{petry12}\footnote{http://casa.nrao.edu/}.  Unfortunately, the radio
frequency interference in our chosen wavelength regime required us to flag
60-80\% of the bandpass, which prevented us from achieving our target noise
limit of 15~$\mu$Jy~beam$^{-1}$.  Consequently, we achieved similar detection
limits and uncertainties to the FIRST survey \citep{becker95}.  We only detected
one galaxy (J1634+46, one of the Type I AGN) at the same flux level as FIRST. 
This galaxy is clearly a radio-loud AGN (see Table~\ref{table:other_info}).

\begin{deluxetable*}{cccccccc}
\tablecaption{Supplemental Multiwavength Data}
\tablehead{
	\colhead{ID} &
	\colhead{Galaxy} &
	\colhead{Log(L$_{\rm R}$)} &
	\colhead{L$_{\rm [O~III]}$} &
	\colhead{L$_{\rm [Ne~V]}$} &
	\colhead{$3.6 - 4.5 \mu$m} &
	\colhead{L$_{12 \mu m}$} &
	\colhead{SFR$_{\rm IR}$} \\
	\colhead{} &
	\colhead{} &
	\colhead{(W~Hz$^{-1}$)} &
	\colhead{($10^{40}$~erg~s$^{-1}$)} &
	\colhead{($10^{40}$~erg~s$^{-1}$)} &
	\colhead{Color} &
	\colhead{($10^{44}$~erg~s$^{-1}$)} &
	\colhead{(M$_\odot$~yr$^{-1}$)} \\
	\colhead{(1)} &
	\colhead{(2)} &
	\colhead{(3)} &
	\colhead{(4)} &
	\colhead{(5)} &
	\colhead{(6)} &
	\colhead{(7)} &
	\colhead{(8)}
}
\startdata
1  & J0826+43 & $<23.39$         & $72.9\pm6.6$  & $<6.6$       & $0.18\pm0.05$ & $4.90\pm0.49$  & 380    \\
2  & J0944+09 & $<23.63$         & $16.4\pm3.3$  & $<2.0$       & $0.35\pm0.05$ & $3.81\pm0.31$  & 220    \\
3  & J1104+59 & $<23.73$         & $73.3\pm7.9$  & $5.2\pm1.2$  & ---           & ---            & 70     \\
4  & J1506+54 & $<23.68$         & $133.5\pm6.0$ & $16.3\pm1.6$ & $0.35\pm0.05$ & $11.77\pm1.21$ & 250    \\
5  & J1506+61 & $<23.41$         & $32.2\pm1.9$  & $<2.1$       & $0.28\pm0.05$ & $0.59\pm0.05$  & 210    \\
6  & J1558+39 & $<23.27$         & $66.9\pm2.2$  & $<2.3$       & $0.57\pm0.05$ & $2.69\pm0.26$  & 610    \\
7  & J1613+28 & $<23.40$         & $31.8\pm3.4$  & $<2.8$       & $0.61\pm0.05$ & $7.83\pm0.98$  & 230    \\
8  & J1713+28 & $<23.80$         & $88.1\pm5.7$  & $4.2\pm0.9$  & ---           & ---            & 500    \\
9  & J2118+00 & $<23.39$         & $43.4\pm3.5$  & $<5.2$       & $0.57\pm0.05$ & $7.20\pm0.76$  & 130    \\
10 & J1359+51 & $<23.25$         & $32.2\pm2.1$  & $6.5\pm1.0$  & $0.58\pm0.05$ & $0.71\pm0.05$  & $<350$ \\
11 & J1634+46 & $25.14 \pm 0.02$ & $37.8\pm8.3$  & $13.1\pm2.6$ & $0.40\pm0.05$ & $1.42\pm0.09$  & $<400$ \\
12 & J2140+12 & $<23.94$         & $7.1\pm15.2$  & $<8.3$       & $0.49\pm0.05$ & $7.63\pm0.39$  & $<500$
\enddata
\tablecomments{
Column 1:  Galaxy identification number used in some figures for brevity.
Column 2:  SDSS short name.
Column 3:  K-corrected radio luminosity.  
Column 4:  [O~III] ($\lambda5007$) luminosity.
Column 5:  [Ne~V] ($\lambda3426$) luminosity.
Column 6:  The IRAC $3.6 - 4.5 \mu$m observed color (Vega magnitudes).
Column 7:  K-corrected $12 \mu$m luminosity.
Column 8:  IR-based star formation rate using the \citet{kennicutt98} calibration converted to a
           \citet{chabrier03} IMF.  The uncertainties are approximately a factor of 2~--~3.  The
           SFRs for the last three (broad-line AGN) are listed as upper limits because we have
           not accounted for an AGN contribution to the IR.
}
\label{table:other_info}
\end{deluxetable*}

\subsection{WISE Infrared Observations}
\label{section:WISE}

We derive the galaxies' restframe 12~$\mu$m luminosity from
observed-frame 12~$\mu$m and 22~$\mu$m photometry from the WISE bands
W3 and W4.  We do not report results for J1104+59, which is
contaminated by a nearby star, and J1713+28, which is not
detected.  The median S/N of the remaining sources is 7 and 3 for
bands W3 and W4, respectively.  We fit the data with a range of
star-forming galaxy templates \citep[for more details,
see][]{diamond-stanic12b}. We consider templates from \citet{chary01},
\citet{dale05}, and \citet{rieke09} and use the average of our results
and use the standard deviation of the three model values to estimate
our errors.  We also estimate the SFR from the IR using the
\citet{kennicutt98} calibration converted to a \citet{chabrier03} IMF.
Both the IR luminosities and SFRs are listed in
Table~\ref{table:other_info}.

\subsection{Stellar Masses}
\label{section:stellar_masses}

We derive stellar masses from the galaxies' 0.1~--~5~$\mu$m SEDs.
We use far- and near-UV photometry from GALEX and $ugriz$ optical
photometry from the SDSS Data Release~8 \citep{aihara11}.  We obtained
3.6~$\mu$m and 4.5~$\mu$m images with the Infrared Array Camera
\citep[IRAC;][]{Fazio04} as part of Spitzer GO program 60145.
We used the post-basic calibrated data to perform aperture photometry
on all sources and point-source photometry on sources in crowded fields.
The SED analysis is preformed using \texttt{iSEDfit}, a code which
implements a simplified Bayesian framework to derive galaxy physical
properties \citet{moustakas13}.  A \citet{chabrier03} IMF is assumed.
Stellar masses are reported in Table~\ref{table:general_info}.
A more detailed description of the SED
modeling of the broadband spectra of these galaxies is beyond the
scope of this paper and will be discussed in a future publication.

\section{Assessment of Supermassive Black Hole Activity}
\label{section:analysis_agn}

\subsection{Broad Line AGN}
\label{section:BL_AGN}

\begin{deluxetable*}{cccccccccc}
\tablecaption{Properties of Type~I AGN}
\tablehead{
    \colhead{ID} &
    \colhead{Galaxy} &
    \colhead{$f_{AGN,3000}$} &
    \colhead{$f_{AGN,F814W}$} &
    \colhead{$\log L_{3000}$} &
    \colhead{Mg~II} &
    \colhead{Mg~II} &
    \colhead{$\log (M_{BH}/M_{\odot})$} &
    \colhead{$L_{bol}^{3000}/L_{Edd}$} &
    \colhead{$L_{bol}^{X}/L_{Edd}$} \\
    \colhead{} &
    \colhead{} &
    \colhead{} &
    \colhead{} &
    \colhead{(erg~s$^{-1}$)} &
    \colhead{FWHM (\AA)} &
    \colhead{EW (\AA)} &
    \colhead{} &
    \colhead{} &
    \colhead{} \\
    \colhead{(1)} &
	\colhead{(2)} &
	\colhead{(3)} &
	\colhead{(4)} &
	\colhead{(5)} &
	\colhead{(6)} &
	\colhead{(7)} &
	\colhead{(8)} &
	\colhead{(9)} &
	\colhead{(10)}
}
\startdata
10 & J1359+51 & $0.62\pm0.02$ & $0.31\pm0.05$ & $44.16\pm0.01$ & $39.6\pm1.0$ & $28.1_{-2.9}^{+3.5}$ & $8.10\pm0.02$ & $0.068^{+0.068}_{-0.039}$ & $0.042^{+0.041}_{-0.025}$  \\
11 & J1634+46 & $0.74\pm0.09$ & $0.30\pm0.05$ & $44.66\pm0.05$ & $82.2\pm1.6$ & $43.0_{-3.9}^{+4.7}$ & $9.04\pm0.04$ & $0.017^{+0.017}_{-0.010}$ & $0.014^{+0.014}_{-0.009}$  \\
12 & J2140+12 & $0.58\pm0.03$ & $0.33\pm0.05$ & $44.64\pm0.02$ & $54.8\pm0.7$ & $18.4_{-1.1}^{+1.2}$ & $8.65\pm0.02$ & $0.037^{+0.037}_{-0.021}$ & $<0.018$
\enddata
\tablecomments{
Column 1:  Galaxy identification number used in some figures for brevity.
Column 2: SDSS short name.
Column 3: Fraction of the continuum light at 3000~\AA\ contributed by the AGN.
Column 4: Fraction of the continuum light through the F814W HST filter contributed by the AGN.
The values in columns 3 and 4 were estimated from the SED modeling.
Column 5: AGN continuum luminosity at 3000~\AA\ ($\lambda L_{\lambda}$).
Column 6: Restframe FWHM of the broad Mg~II line (fit using a single Gaussian).
Column 7: Restframe EW of the broad Mg~II line.
Column 8: SMBH mass derived using Eqn.~12 of \citet{trakhtenbrot12}.
Since the listed random error is much smaller than the approximate systematic
error (0.32~dex; \citealt{trakhtenbrot12}), the systematic error is used to estimate the
uncertainties in $L_{bol}^{3000}/L_{Edd}$ and $L_{bol}^{X}/L_{Edd}$.  This systematic error
dominates the error budget in these two quantities.
Column 9: Ratio of the bolometric to Eddington luminosity.  $L_{bol}^{3000}$ is inferred form the
optical continuum following \citet{trakhtenbrot12}.
Column 10: Same as Column 8 except that $L_{bol}^{X}$ is the bolometric luminosity estimated from
$L_X$ and $L_{3000}$ using \cite{brightman13}.}
\label{table:broad-line}
\end{deluxetable*}

The rest-frame optical spectrum of our targets is dominated by the
light of the host galaxy (Fig.~\ref{fig:MMT_spectra}).  However, three of the 12
galaxies (J1359+51, J1634+46, and J2140+12) display broad Mg~II emission indicative
of a Type~I AGN.  For these sources, we obtain virial SMBH mass
estimates from the Mg~II linewidth and the AGN continuum luminosity
at 3000~\AA\ following \citet{trakhtenbrot12}.  We have been careful
to decompose the continuum of the host galaxy and that of the AGN using
stellar population synthesis modeling.
Results are reported in Table~\ref{table:broad-line}.

The SMBH masses that we infer using the viral technique are
large, $M_{SMBH}=10^{8\text{~--~}9}~M_{\odot}$.  The masses are in reasonable agreement with
the SMBH -- stellar mass correlation \citep{marconi03,mcconnell13,schramm13}.
We detect J1359+51 and J1634+46 in X-rays, while J2140+12 is undetected (see
\S~\ref{section:data_red_Chandra} for details.)

We use the optical continuum light and X-rays separately to estimate the AGN
bolometric luminosities.  We use the luminosity-dependent bolometric
corrections of \citet{trakhtenbrot12} for the optical and \cite{brightman13} for
the X-rays.  We calculate Eddington luminosities using
$L_{Edd} = (1.26 \times 10^{38}) (M_{SMBH}/M_\odot)$.
We find ratios of $\sim 1$~--~7\% of Eddington
for these galaxies, which are typical for samples of broad-line AGN, but lower than
typical SDSS quasars \citep{kelly10,steinhardt10,kelly13}.

One surprising result is that J2140+12 is not X-ray-detected, yielding a
$L_{bol}^{X}/L_{Edd} \ 3\sigma$ upper limit for this broad-line AGN $\sim 2 \times$ lower
than $L_{bol}^{3000}/L_{Edd}$.  This could be explained by either variability
between the time of the optical and X-ray observations, which is causally possible on these
timescales \citep[e.g.,][]{ulrich97,mchardy13}, or by considerable differential attenuation
by the large amounts of cold gas that could still be present shortly after a galaxy merger.
There could be X-ray absorption from dust-free gas inside the dust sublimation radius
\citep{maiolino07}.  In addition, since the X-rays are likely concentrated much more centrally
than the Mg~II $\lambda\lambda2796,2804$ broad-line emission and the UV continuum
(e.g., in the case of a simple, disk blackbody model), a couple of high-density,
AU-scale gas clouds could more easily obscure the X-ray emitting
region as compared to the UV emitting BLR.  This could also explain the strong
absorption dips seen in the broad Mg~II line (see Fig.~\ref{fig:MMT_spectra}).

\subsection{Narrow-line AGN or Star-forming Galaxies?}
\label{section:non_BLAGN}

\begin{figure*}[htp]
  \centering
  \begin{tabular}{cc}
    \includegraphics[width=0.48\textwidth]{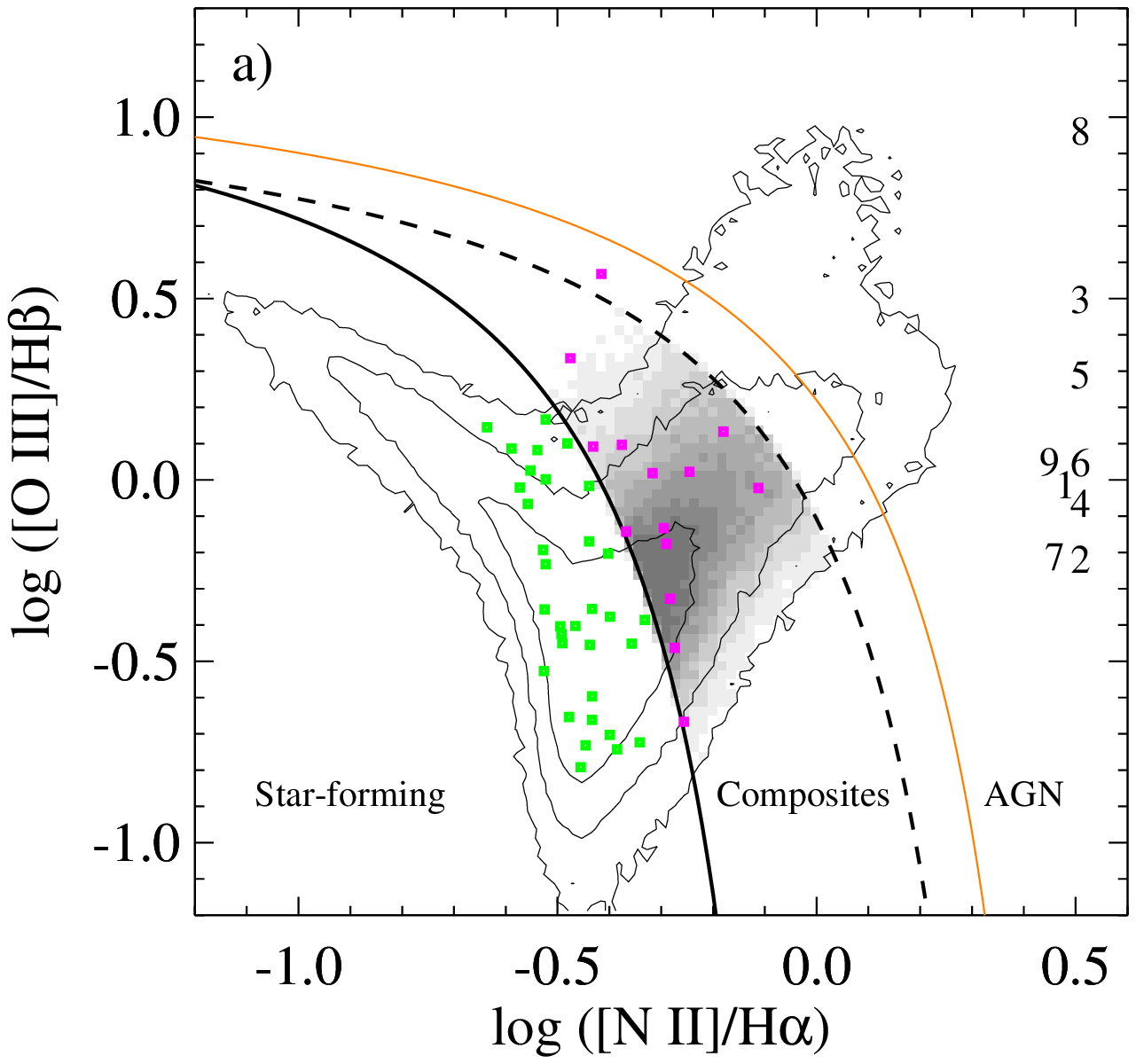}&
    \includegraphics[width=0.48\textwidth]{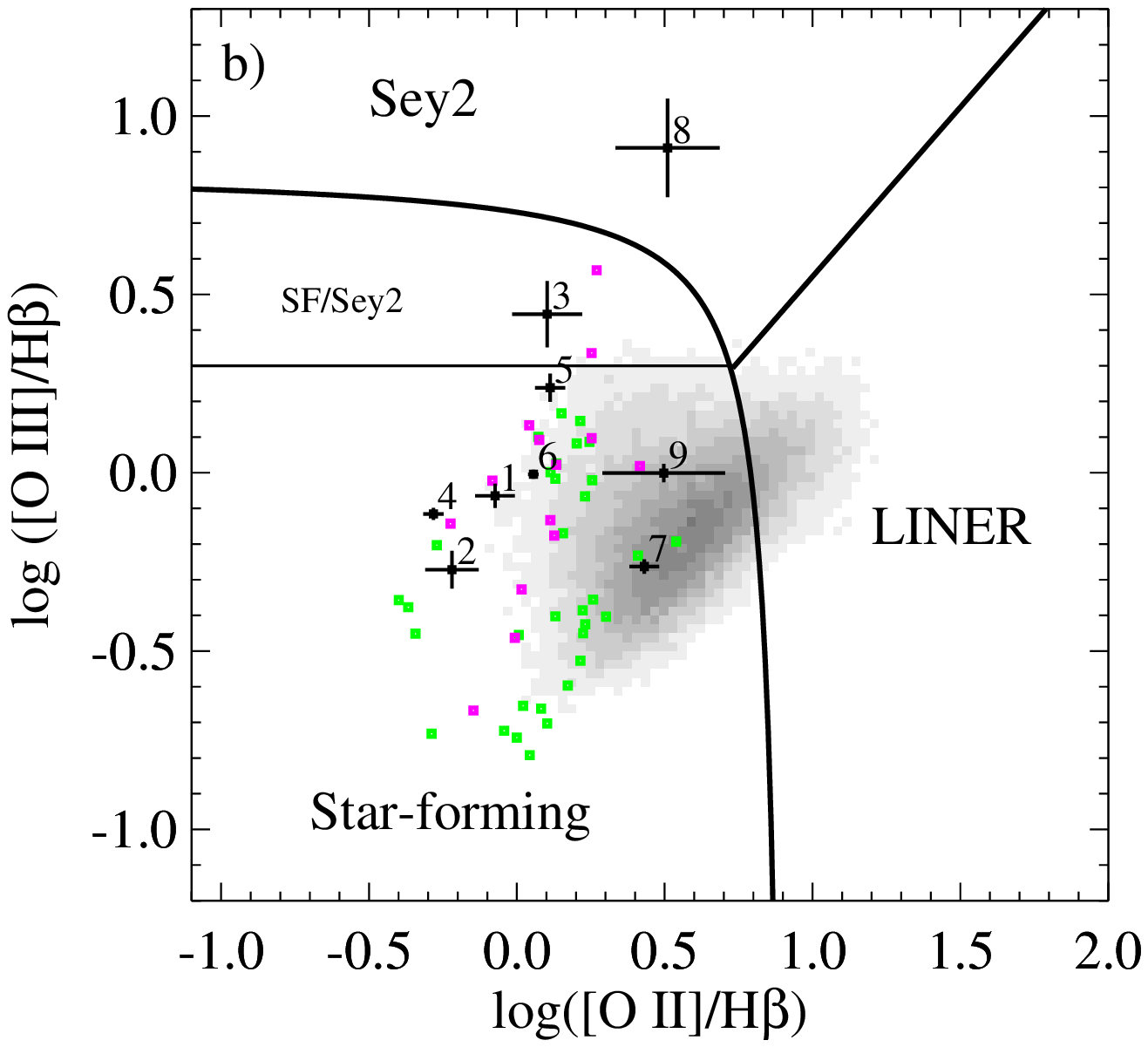}\\
    \includegraphics[width=0.48\textwidth]{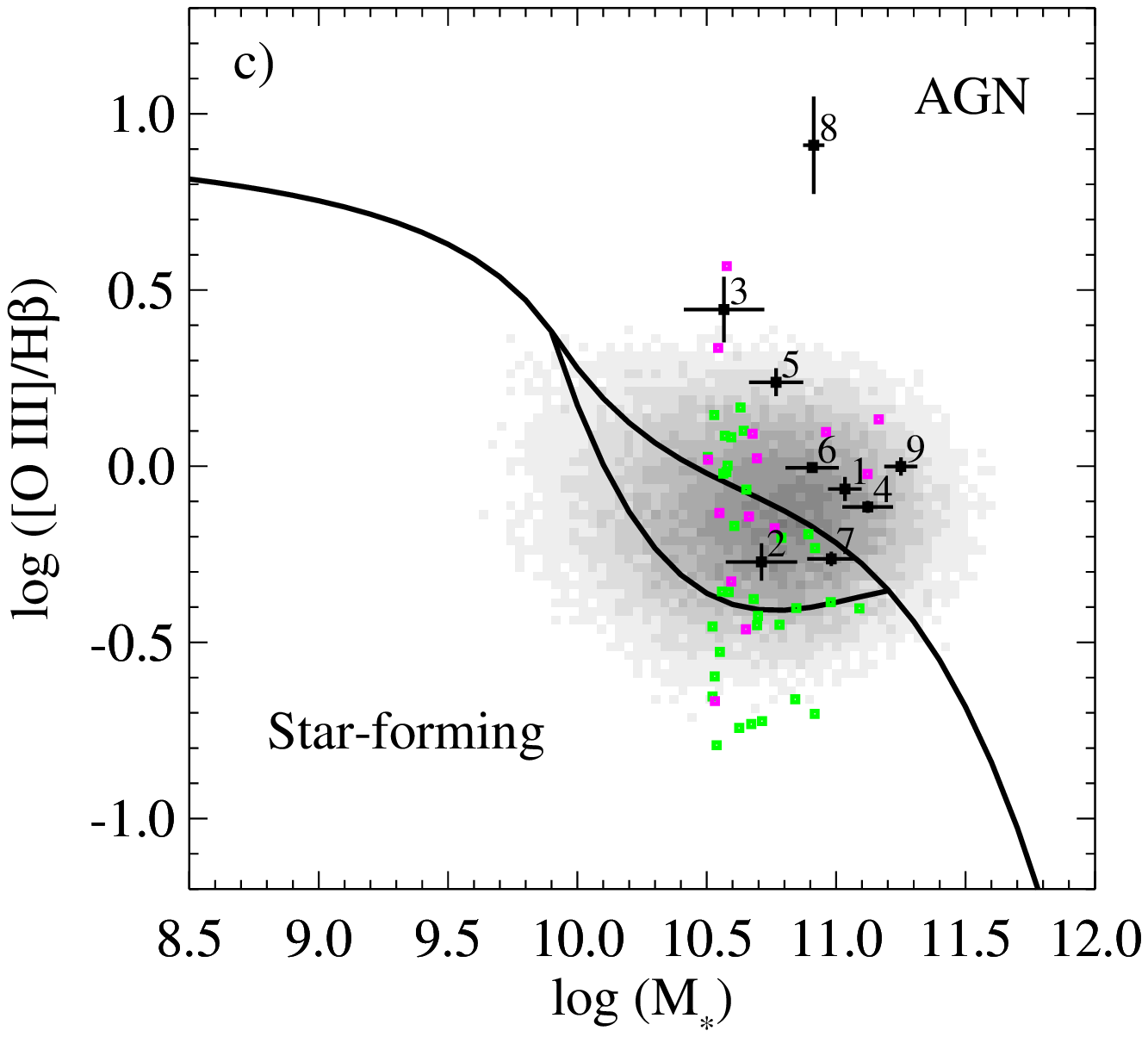}&
    \includegraphics[width=0.48\textwidth]{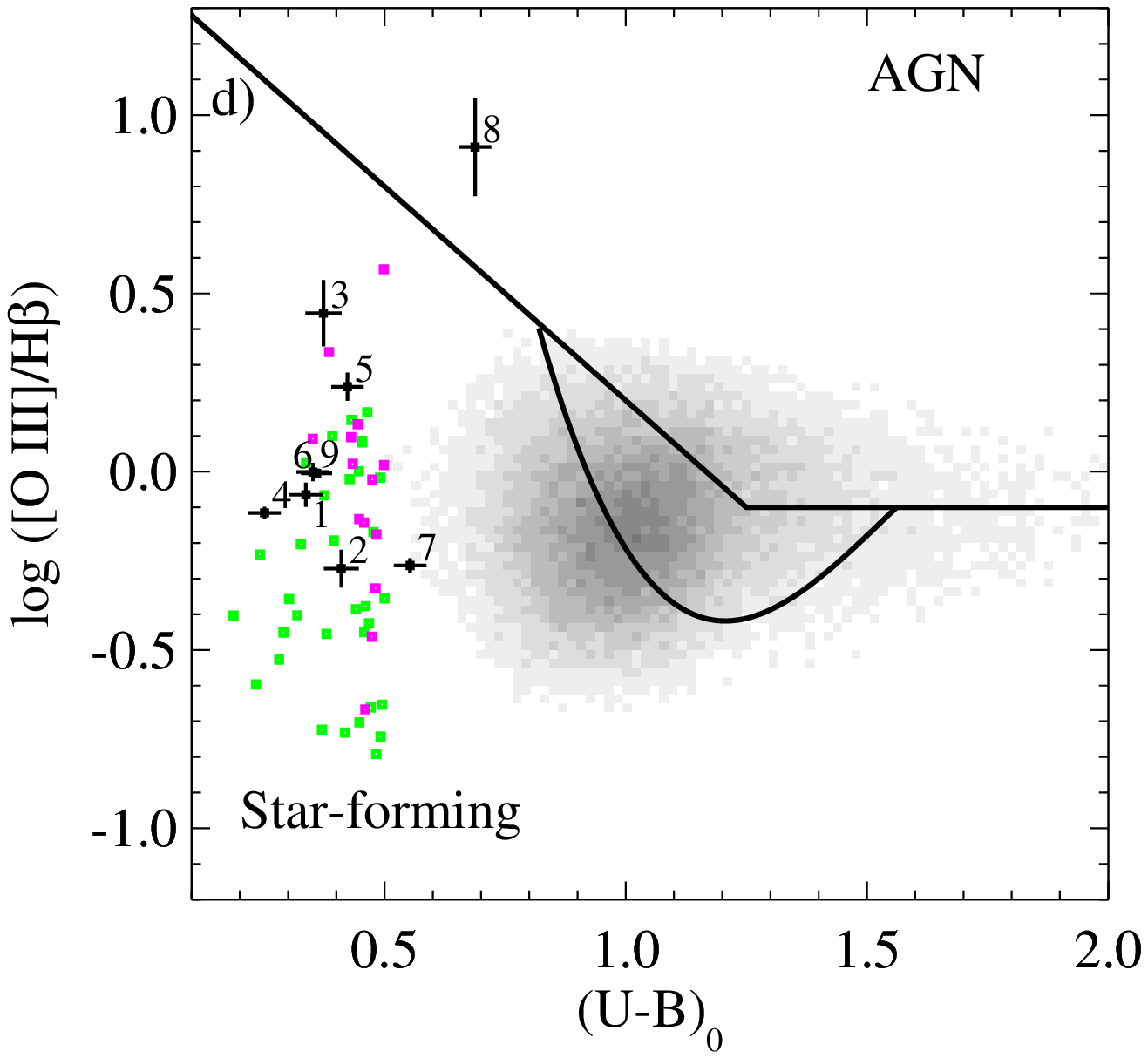}\\
  \end{tabular}
  \caption{Diagnostic diagrams designed to separate
  star-forming galaxies, AGN (Seyfert 2s/LINERs), and
  composite systems. The nine galaxies with narrow emission
  lines are labeled with their galaxy ID number (Table 1).  Panel a) shows
  the BPT diagram \citep{baldwin81} of $\sim338,000$ galaxies at
  $z=0.02-0.25$ from the SDSS-I MAIN sample (contours). The black solid
  and dashed lines show the divisions between star-forming galaxies
  and AGN proposed by \citet{kauffmann03} and \citet{kewley01}, respectively.
  The solid orange line shows the theoretical upper limit star-forming
  abundance sequence for $z=3$ from \cite{kewley13a}.
  Our $z>0.4$ galaxies lack [N~II]/H$\alpha$ measurements; their
  [O~III]/H$\beta$ values are indicated at the right of the BPT diagram
  by their ID number.
  Panel b) shows the `Blue' diagnostic of \citet{lamareille10}; panel c)
  the mass--excitation diagnositc (MeX) of \citet{juneau11}; and panel d)
  the color--excitation diagnostic (CeX) of \citet{yan11}.  In all panels,
  BPT-composites are shown in greyscale.  We highlight
  47 SDSS-I MAIN galaxies that have masses and colors similar to
  our galaxies as green (BPT star-forming) and magenta (BPT
  composite and AGN) points.  J1713+28 (ID~8) is likely an AGN; the other
  eight galaxies are not consistently classified, nor are the
  galaxies in the comparison sample (green and magenta points). See
  \S~\ref{section:BPT} for details.}
\label{fig:bpt}
\end{figure*}

Nine of the twelve galaxies in the sample do not show evidence of
broad Mg~II or H$\beta$ emission lines. Here we use multi-wavelength
diagnostics to assess the relative contribution of star formation
and obscured AGN activity to their bolometric luminosities.

\subsubsection{Optical Narrow-line Diagonsitcs}
\label{section:BPT}

The classic line ratio diagnostic diagram of [N~II]~($\lambda6584$) / H$\alpha$
vs. [O~III]~($\lambda5007$) / H$\beta$, commonly called the `BPT' diagram, has been
widely used as a diagnostic to separate star-forming galaxies and AGN
\citep{baldwin81,veilleux87} and various sub-classes of
AGN \citep{kewley06}.  In Fig.~\ref{fig:bpt}a, we show the BPT
diagram for a sample of $\sim338,000$ low redshift galaxies drawn from
the SDSS-I MAIN sample \citep{strauss02}.  These galaxies have
redshifts between $z=0.02-0.25$ and stellar masses between $M_*=
10^{8.5-12}~M_{\sun}$.  The solid line is the empirical dividing line
proposed by \citet{kauffmann03} to separate pure star-forming galaxies
from AGN.  The dashed line is the theoretical maximum starburst limit
proposed by \citet{kewley01}.  Galaxies falling between the solid and
dashed lines are classified as `composites' because most are believed
to be ionized by a mixture of star formation and AGN activity.  We
do not have measurements of [N~II] and H$\alpha$ for our $z>0.4$
galaxies.  Their [O~III]/H$\beta$ fluxes are indicated at the
righthand edge of the BPT diagram by the galaxy ID number.
$L_{[O~III]}$ is listed in Table~\ref{table:other_info}.

For comparison with our $z\sim0.6$ galaxies, we select a
sub-sample of SDSS-I MAIN galaxies that are comparably massive ($M_* >
10^{10.5}~M_{\sun}$) and blue ($(U-B)_0 < 0.5$).  In
Fig.~\ref{fig:bpt}, the 33 star-forming galaxies that meet this
criterion are shown in green and the 14 composites and AGN are shown
as magenta points. Note that such galaxies are very rare at low
redshift, comprising only 0.01\% of the parent sample. (The
photometry and spectroscopy of the comparison galaxies has been
hand-checked to remove contaminants due to photometry errors from
bright stars, etc.).

Several alternate diagnostic diagrams have been devised for use
at $z>0.4$, where [N~II] and H$\alpha$ have redshifted out of the
observed-frame optical \citep{lamareille10, trouille11, yan11,
juneau11}. We show three of these diagnostics in
Fig.~\ref{fig:bpt}b-d.  These pseudo-BPT diagrams typically have difficulty
identifying composite sources that contain both significant star
formation and AGN activity. To illustrate this, we have overplotted
the SDSS-I composites in greyscale. All panels also include the
mass- and color-matched comparison sample (green and magenta points).

Fig.~\ref{fig:bpt}b shows the `Blue' AGN diagnostic
\citep{lamareille10}, which substitutes [O~II]/H$\beta$ for
[N~II]/H$\alpha$.  To mitigate extinction effects on the widely
separated [O~II] and H$\beta$ lines, this diagram uses the ratio of
line EWs rather than line fluxes (this assumes that the attenuation of
the continuum and emission lines is the same). In Fig.~\ref{fig:bpt}c
we show the mass-excitation diagram (MeX) of \citet{juneau11}, which
substitutes stellar mass for the [N~II]/H$\alpha$ ratio.  In this
diagram, composites dominate the wedge-shaped region between the
star-forming galaxies and AGN.  Fig.~\ref{fig:bpt}d shows the
color-excitation diagram (CeX) of \citet{yan11} which utilizes the
rest-frame $U-B$ color in AB magnitudes in place of [N~II]/H$\alpha$. 
The region between the star-forming galaxies and AGN was identified by
\citet{juneau11} as being dominated by composite galaxies.

Based on these diagrams together, we classify J1713+28 (ID~8) as the
most-likely galaxy to host a Type~II AGN, followed by J1104+59 (ID~3).
To place the most likely galaxy (J1713+28 or ID~8) in context, we estimate
the bolometric Eddington fraction using the measured [O~III] luminosity in
Table~\ref{table:other_info}.  Unfortunately, our bolometric estimate
will be very uncertain because we do not have enough information
to correct [O~III] for extinction and we do not know how much the [O~III]
emission from an AGN is contaminated by star formation.
The bolometric correction for [O~III] typically
ranges from 600 \citep{heckman04} to 3500 \citep{kauffmann09}, depending on the
extinction correction.  From these conversions, we estimate
$L_{bol,[O~III]} \approx 0.5$~--~$3 \times 10^{45}$~erg~s$^{-1}$.  Then, using the
SMBH -- bulge mass relation of \citet{mcconnell13} to estimate the SMBH mass
and hence $L_{Edd}$, we find a rough estimate for the bolometric Eddington fraction:
$L_{bol}/L_{Edd} \approx 0.02$~--~0.13. 
This estimate is comparable to the bolometric Eddington fractions found for the
Type~I AGN in our sample ($\sim 1$~--~7\%; see Table~\ref{table:broad-line}) and
to J1506+54 (ID~4; see \S~\ref{section:j1506}).  The remaining seven narrow-line
galaxies, along with the mass and color-matched, low-$z$ comparison sample, are
classified inconsistently or ambiguiously by the Blue, MeX, and CeX diagrams.
For the low-$z$ comparison sample, the MeX diagram does the best job of reproducing
the BPT classifications, but still classifies over half of the BPT star-forming
galaxies as AGN or composites.

This analysis indicates that the pseudo-BPT diagrams (Fig.~5b~--~5d) may not be
reliable for certain types of galaxies.  Future near-IR spectroscopy of the sample
will enable us to measure their [N~II]/H$\alpha$ ratios so that we can place them
on the BPT diagram.  However, recent studies indicate even BPT classification must
be treated with some caution.  In fact, studies of rest-frame optical emission lines
of star-forming galaxies at $z\sim 1$~--~2 \citep[e.g.,][]{shapley05,erb06,liu08}
have shown that high-$z$ galaxies have elevated [O~III]/H$\beta$ ratios relative
to local star-forming galaxies.  The small fraction of local galaxies with similar
line ratios tend to have larger electron densities, star-formation rates, and SFR
surface densities \citep[e.g.,][]{kewley01,liu08,brinchmann08}, which suggests H~II
region physical conditions influence a galaxy's position on the BPT, sometimes
leading to mis-classification.  This is particularly germane to our sample because
we expect elevated electron densities and interstellar pressures in compact
starbursts \citep{liu08,verdolini13,rich13}.

A new set of theoretical models \citep{dopita05,dopita06b,dopita06c}, which are
compared to high redshift samples, show how these types of galaxies move on the
BPT diagram \citep{kewley13a,kewley13b}.  Compact starburst galaxies at higher
redshift, which have larger pressures and densities of ionizing photons,
fall in the same position on the diagram as galaxies labeled as ``Composites"
at low redshift.  In particular, some of the \cite{groves08} models predict
$\log($[O~III]/H$\beta$) ratios for some compact starbursts up to 0.9,
statistically consistent with all of the [O~III]/H$\beta$ ratios in our sample.
Thus, it is not possible to draw firm conclusions at the present time about the
nature of the excitation in any of the non-broad-line galaxies from these
diagnostics.

\subsubsection{SDSS J1506+54:  A Case Study of a Galaxy with [Ne V]}
\label{section:j1506}

[Ne~V] ($\lambda3426$) is typically an order of magnitude weaker than
[O III] \citep{ferland86}, but it has a much higher ionization energy making it a
valuable tracer of AGN activity \citep{schmidt98}\footnote{Photons with energies
above 97~eV are required to create [Ne~V], wheres stars typically do not produce
photons beyond 54~eV \citep{abel08}.}.  Indeed, it is not too surprising that we
strongly detect [Ne~V] in the two X-ray-detected broad-line AGN (J1359+51 and
J1634+46); the third broad-line AGN, J2140+12, may be
strongly variable or likely suffers from considerable attentuation
(see \S~\ref{section:BL_AGN} and Table~\ref{table:other_info}).

We detect [Ne~V] in three of the nine narrow-line galaxies.  The line is very
strongly detected ($\sim 10 \sigma$) in one of these galaxies (J1506+54) and is only
weakly detected ($\lesssim 4 \sigma$) in two other cases (J1104+59 and J1713+28).
The significant detection of [Ne~V] in the two latter galaxies is consistent with the finding
in \S~\ref{section:BPT} that these galaxies are most likely to host obscured AGN activity.
However, assessing the [Ne~V] contribution in these latter two cases could be
controversial and extremely challenging:  1) the significance and luminosity of the
[Ne~V] line is sensitive to the fit of the galaxy continuum model, which has
an additional uncertainty not encapsulated above that is very difficult to quantify;
2) the measured strength of a line near the sensitivity limit can be
overestimated \citep{rola94}; and 3) without an AGN, very young stellar populations
containing Wolf-Rayet and other O-stars (less than a few Myr) can produce some
detectable [Ne~V] because much higher fractions of high-energy photons are produced
\citep{abel08}.  Given these challenges applied to our unusual galaxies,
we only analyze the very strong detection of [Ne~V] in J1506+54 in detail.

First, we consider the case where the narrow-line and X-ray emission is produced by
an AGN.  \citet{gilli10} suggest that the ratio of X-ray to [Ne~V] luminosity is a good
diagnostic of AGN nuclear obscuration.  Following \citet{gilli10} we do not correct
narrow-line emssion for extinction.  They found $L_X/L_{[Ne~V]} \sim 400$ for Type~I
(unobscured) AGN and $L_X/L_{[Ne~V]} < 15$ for a Compton-thick AGN.  We measure
$L_X/L_{[Ne~V]} = 4.9$, which implies a Compton-thick AGN ($N_H > 10^{24}$~cm$^{-2}$).
Then, we assume $L_{bol}/L_{[Ne~V]} = (L_X/L_{[Ne~V]})_{Type~I} \times (L_{bol}/L_X) = 400
\times 20 = 8000$ for the bolometric correction.  For [Ne~V], we find
$L_{bol} \approx 1.3 \times 10^{45}~\textrm{erg~s}^{-1}$ and $L_{bol}/L_{Edd} \approx 0.05$,
using the SMBH -- bulge mass relation of \citet{mcconnell13} to estimate the SMBH mass
and hence $L_{Edd}$.  Similarly for [O~III], we calculate
$L_{bol,[O~III]} \approx 0.8$~--~$4.8 \times 10^{45}~\textrm{erg~s}^{-1}$ and
$L_{bol}/L_{Edd} \approx 0.05$~--~0.27 in the same manner as in the previous section
for J1713+28.
This bolometric Eddington fraction is about a factor of two larger than found for J1713+28
(see \S~\ref{section:BPT}), but the [O~III] in both galaxies could be contaminated by
star formation.

In fact, further inspection of J1506+54 reveals that its situation is even more unusual.
Even for our sample of galaxies, this galaxy appears to have an unusually young ($\sim 3$~Myr)
and extreme ($\Sigma_{\textnormal{\scriptsize{SFR}}}\approx3000$~M$_{\odot}$~yr$^{-1}$~kpc$^{-2}$)
stellar population.  We, therefore, investigate the possibility that this galaxy has a
starburst capable of exciting the [Ne~V] line we observe.  We compare this galaxy to blue
compact dwarf galaxies, which can exhibit considerable [Ne~V] emission without any other
expected or conclusive signs of AGN activity \citep{izotov12}.  For example, \cite{izotov04}
found that $L_{[Ne~V]} \approx 7 \times 10^{38}$~erg~s$^{-1}$ for Tol~1214$-$27.  To produce
this emission, they calculated that $L_{>0.14~\text{keV}} \sim 10^{39-40}$~erg~s$^{-1}$ is
required or about a factor of 10 higher than $L_{[Ne~V]}$.  For J1506+54, we find
$L_{[Ne~V]} = 1.5 \times 10^{40}$~erg~s$^{-1}$ and $L_X = 10^{41.9 \pm 1.0}$~erg~s$^{-1}$.
The X-ray luminosity likely suffers from Eddington bias \citep{eddington13}.  This arises
because we sample counts from the Poisson distribution, which is especially asymmetric for
low numbers of counts.  A correction for this possible upward bias in this luminosity would
only bring the X-ray luminosity more in line with the expected value based on the [Ne~V]
prediction.  Therefore, it is plausible that all of the [Ne~V]
emission is produced by the very young, ultra-compact starburst.  This analysis emphasizes
the overall conclusion in \S~\ref{section:BPT} that standard nebular diagnostics using
high excitation lines as a tracer of AGN activity are frequently not useful diagnostics
for these extreme galaxies.

While there is ambiguity about the origin of the [Ne~V] in J1506+54, we
can make the assumption that it traces obscured AGN activity and ask
whether the AGN would be bolometrically dominant.  Since both the AGN
and starburst are heavily obscured, we consider their relative
contributions to the 12~$\mu$m luminosity.  We adopt
$L_{12\mu m,AGN}$ = $L_{bol,AGN}/9$ from \citet{richards06} and estimate
$L_{12\mu m,AGN} = 1.4 \times 10^{44}$~erg~s$^{-1}$ for J1506+54.
Compared to the observed value (Table~\ref{table:other_info}), this implies
that roughly 11\% of the galaxy's mid-IR luminosity is powered by
the AGN and $\sim$89\% is powered by star formation.  This finding
is consistent with the fact that starburst templates provide better
fits to the IR SED than AGN templates \citep{diamond-stanic12b}.

\subsubsection{X-ray Diagnostics}
\label{section:xrays}

X-ray observations are among the most efficient and unbiased ways of
detecting and characterizing AGN \citep[e.g.,][]{mushotzky04}.  Here
we discuss the \emph{Chandra} observations of the nine galaxies in our sample
without a Mg~II broad line.  Our observations were designed to detect
$10^{8.2-9.2}~M_{\odot}$ SMBHs radiating at $\sim 1\%$ of their
Eddington limit with an obscuring column as high as
$N_H=10^{23}$~cm$^{-2}$.  Based on the measured [O~III] luminosities,
we surmised that the galaxies were radiating at even higher rates,
although as discussed in \S~\ref{section:BPT} some of the [O~III] could
come from star formation.

\begin{figure}
\centering
\includegraphics[scale=0.32, angle=90]{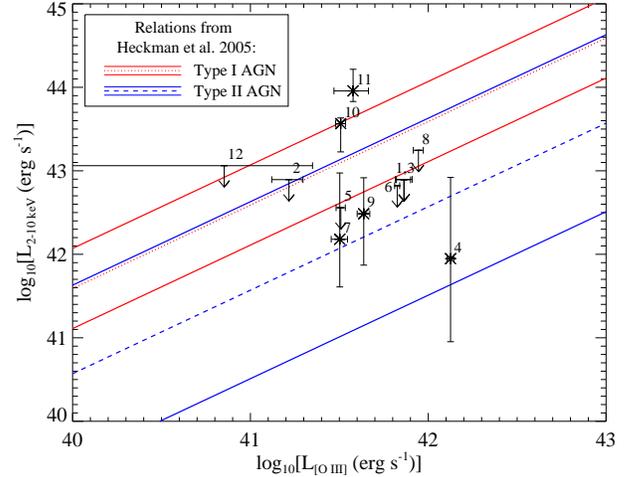}
\caption{The X-ray luminosities and upper limits for our galaxies compared to the Type~I and Type~II
AGN relations from \cite{heckman05}.  Note that the X-rays from all sources except J1359+51 and
J1634+46 (IDs~10 and 11) are consistent with the level of X-rays from XRBs in these galaxies
(see \S~\ref{section:xrays}).  Also, the [O~III] emission from most of the sources is consistent
with that from extreme starbursts (see \S~\ref{section:BPT}).}
\label{fig:Heckman_comparison}
\end{figure}

As described in \S~\ref{section:data_red_Chandra}, we detect only
three of the nine narrow-line objects (J1506+54, J1613+28, and J2118+00) with
{\em Chandra} with four counts each.  We compare the [O~III] and
X-ray luminosities and upper limits in Fig.~\ref{fig:Heckman_comparison}.
The three detected narrow-line objects have large error bars, but are
consistent with the relationship found for local Type~II AGN by \cite{heckman05}.

However, in galaxies lacking powerful AGN, XRBs are
responsible for the bulk of the 2~--~10~keV emission
\citep[e.g.,][]{kong03,li11}.  In star-forming
galaxies, the emission is dominated by high-mass XRBs, that are
associated with young ($<100~$Myr) stellar populations.  As a
consequence, X-ray emission has been shown to scale with the SFR:
$L_X \approx 3.5 \pm 0.4 \times 10^{39}$~erg~s$^{-1}$ per
M$_\sun$~yr$^{-1}$ \citep{grimm03,mineo12c}.  Star formation in a
merger event also produces measurable amounts of X-ray bright hot gas
\citep{cox06a}, which has been shown to correlate with the star
formation rate: $L_X \approx 8 \times 10^{38}$~erg~s$^{-1}$ per
M$_\sun$~yr$^{-1}$ \citep{owen09,mineo12b}.

For the galaxies with 4-count X-ray detections (J1506+54, J1613+28, J2118+00),
the IR-based SFRs from restframe WISE data are 250, 230, and 130 M$_\sun$~yr$^{-1}$
(see Table~\ref{table:other_info} and \S~\ref{section:WISE} for details).
The predicted X-ray luminosities from star formation for these three galaxies are
$\log(L_X(\textrm{erg~s}^{-1}))$ = 41.9, 41.9, and 41.6, respectively.  Given
that these numbers agree with the observations within the error bars and that
the spectra are relatively soft (see \S~\ref{section:data_red_Chandra}),
we conclude that the X-rays from all of the galaxies except the two Type~I AGN
(J1359+51 and J1634+46) are likely from intense starbursts, not heavily obscured SMBHs.

\subsubsection{Infrared Diagnostics}
\label{section:infrared}

To explore whether the narrow-line sources are Compton-thick AGN, we
consider their IR luminosity.  We derive restframe 12~$\mu$m
luminosities from fits to the broadband SED
of the galaxies, which are constrained by WISE W3 and W4 in this
spectral range.  We also calculate the IRAC $3.6 - 4.5 \mu$m (Vega magnitudes) colors
for our sources.

\begin{figure}
\centering
\includegraphics[scale=0.32, angle=90]{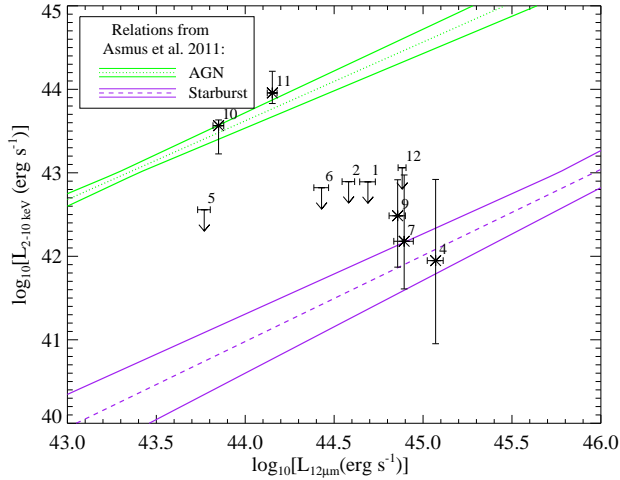}
\caption{We compare our galaxies to the AGN and starburst relations from \cite{asmus11}.
The best-fit and one sigma intervals for the relations are shown.  The X-ray luminosities are not
absorption-corrected (see \S~\ref{section:xrays}).}
\label{fig:Asmus_comparison}
\end{figure}

In Fig.~\ref{fig:Asmus_comparison}, we plot the 2~--~10~keV luminosity
versus the 12~$\mu$m luminosity of all the galaxies in our sample
(narrow- and broad-line), with mid-IR photometry.  Both luminosities
are K-corrected.  We do not correct the X-ray luminosities for absorption
because we do not have enough information to do so, and fits to the X-ray
spectra for the two Type~I AGN and the hardness ratio of the merged
spectrum of the faint X-ray sources suggest only minor absorption.
However, we do consider photon indices that encompass both obscured and
unobscured AGN when we calculate the X-ray uncertainties
(see \S~\ref{section:data_red_Chandra}).  We overplot the relationship
found by \citet{asmus11} for star-forming galaxies (purple) and
absorption-corrected AGN (green).

As expected, the two X-ray-detected Type~I AGN (J1359+51 and J1634+46) lie
near the AGN line.  J2140+12 appears to be a considerably attenuated broad-line
AGN based on differential attentuation of the optical and X-ray bolometic estimates and
its position in Fig.~\ref{fig:Asmus_comparison}, consistent with the suggestion in
\S~\ref{section:BL_AGN}.  The three X-ray detected narrow-line sources are statistically
consistent with the relation for starburst galaxies.

We also compare the available 3.6~--~4.5$\mu$m color
(in Vega magnitudes; see Table~\ref{table:other_info}) for each source to the common color
cuts for mid-IR-selected AGN \citep[see][]{mendez13}.  We find 3.6~--~4.5$\mu$m colors of
0.18~--~0.61.  The observed $5.8 - 8.0 \mu$m color is unavailable in our {\em Spitzer} warm
mission observations, but this does not significantly affect our comparison.  The mid-IR
colors of 4/7 of the narrow-line galaxies with IRAC data with are relatively red
(3.6~--~4.5$\mu$m$\sim0.6$), consistent with the colors of AGN defined by \citet{stern05}.
However, they are in a region of color space heavily contaminated by star-forming galaxies
\citep{donley12}.  None of the galaxies make new, higher-fidelity AGN cut of
3.6~--~4.5$\mu\textrm{m} > 0.8$ adopted by \cite{stern12}.  Therefore, there is no clear
evidence for obscured AGN in this sample based on the available near IR colors.  Overall,
this IR analysis suggests that, even if Compton-thick AGN are present, they must not be a
major contributor to the mid-IR luminosity.  This is consistent with the finding for
J1506+54 (\S~\ref{section:j1506}).

\section{Discussion}\label{section:discussion}

\subsection{Evidence for Mergers}
\label{section:mergers}

The $z=0.4-0.75$ galaxies in our sample are unresolved in SDSS imaging.
Our \emph{HST/WFC3} restframe V-band ($\sim550$~nm) observations
enable us to study their morphologies. Tidal tails and other debris indicative of a
recent major or minor merger are evident in two-thirds of the sample
(J0826+43, J1104+59, J1506+54, J1558+39, J1613+28, J1713+28, J2118+00, and J1634+46;
see Figs.~\ref{fig:cutouts} and \ref{fig:smoothed_images} and
Fig.~\ref{fig:data_fits} in the Appendix).
Our single-orbit images are fairly shallow, probing down to
surface brightness levels of $\mu \sim 25$~mag~arcsec$^{-2}$.
Since tidal features are commonly at least one magnitude fainter than this
\citep{duc13}, we cannot rule out their presence in the remaining three galaxies.
Ten of the 12 galaxies have a single bright core, consistent
with a late-stage merger where nuclear coalescence has already occurred, while
two of the galaxies have another clearly associated, distinct, bright core within
$\sim 20$~kpc.

\subsection{Extremely Compact Light Profiles:  Evidence of Unobscured AGN or Compact Starbursts?}
\label{section:compact_light}

One of the most significant and unexpected results of this work is the
compact nature of the galaxy light profiles.  Most of the objects in our
sample have half-light radii less than a few hundred parsecs based on our
S\'{e}rsic-only fits.  The median value is $r_e = 251$~pc (using the brightest
cores in the case of double nuclei).  For comparison, a typical early-type
galaxy at $z\sim0.5$ has $r_e \sim 2000$~pc \citep[e.g.,][]{huertas-company13}.
We do not have sufficient data to correct our sizes for dust attenuation, but
we estimate that such a correction would only make the $r_e$ values smaller,
depending on the magnitude of $A_V$ \citep{arribas12}.

For many of the galaxies, our S\'{e}rsic $+$ PSF fits suggest that a
large fraction of the nuclear light ($\sim 20$~--~60\%) is unresolved.
First, we consider if the source of this unresolved light for the three galaxies
where we detect broad Mg~II and H$\beta$ emission lines
(J1359+51, J1634+46, and J2140+12)
is consistent with the amount of light expected for the broad-line AGN.
We estimate the quasar contribution to the light in the
WFC3/F814W filter by fitting galaxy and quasar templates to the
UV-optical spectra (\S~\ref{section:data_red_MMT}).  Based on the spectra, we infer 
quasar light fractions of 31\%, 30\%, and 33\% in the F814W filter.
The corresponding PSF light fractions measured from the {\em HST} data
are 32\%, 17\%, and 50\%.  The agreement
is reasonable considering measurement uncertainties and the
possibility of AGN variability in the $\sim5$~years separating the
acquisition of the spectra and images \citep[e.g.,][]{ulrich97,mchardy13}.

\begin{figure}[tbp]
\includegraphics[width=0.48\textwidth]{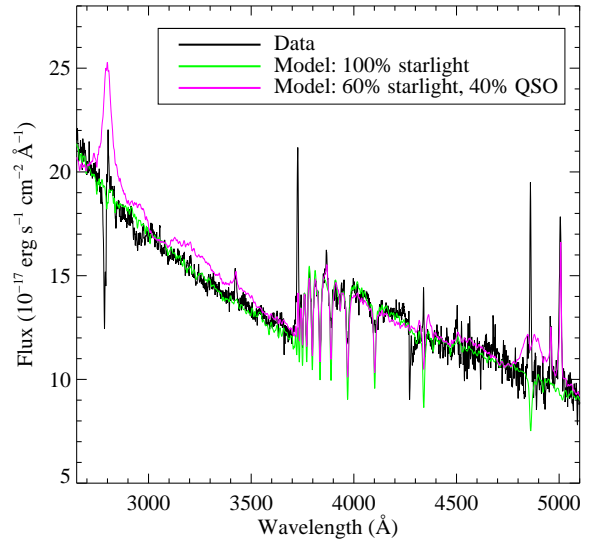}
\caption{The black line shows the spectrum of J1506+54, a galaxy with
an unresolved core that accounts for 47\% of the light in the
restframe $g$-band \emph{HST} image. The green line shows a stellar
population model fit
to the continuum.  The magenta line shows a fit where a QSO template has
been included. The QSO template accounts for 40\% of the light in
the $g$-band.  The broad Mg~II and H$\beta$ emission lines are clearly visible
in the template, but not in the data, ruling out the idea
that the point source in the \emph{HST} image is due to the presence
of an unobscured AGN with a typical spectrum.}
\label{fig:qso_spec_test}
\end{figure}

For the remaining nine galaxies, inspection of our UV-optical spectra show no evidence
of a typical unobscured AGN (Fig.~\ref{fig:MMT_spectra}).  In five of these galaxies
(J0826+43, J0944+09, J1104+59, J1506+54, J1506+61), the PSF light
fraction is substantial (40~--~60\%), and thus we would expect broad
Mg~II and H$\beta$ emission lines to be visible if the PSF light were
due to an unobscured AGN with a normal UV-optical spectrum.
Fig.~\ref{fig:qso_spec_test} illustrates this point for J1506+54.

However, the spectra do show an unexpected very blue continuum with weak nebular
emission lines.  We first investigate if the unresolved light could be consistent
with weak emission-line quasars \citep[WLQs; e.g.,][]{diamond-stanic09,plotkin10}.
To compare our galaxies to WLQs, we calculate $\Delta \alpha_{ox}$, a diagnostic
commonly used to compare WLQs to other similar populations of galaxies (e.g., BL Lacs).
This quantity is defined as the X-ray brightness relative to typical radio-quiet quasars
\citep[for more details,][]{wu12}.  We compare our
$\Delta \alpha_{ox}$ values to the 11 WLQs from \cite{wu12}\footnote{We only use these
WLQs because they are at lower redshifts where the weak line is Mg~II (as opposed to Lya
at $z>2.2$) and because these WLQs are X-ray weak, more similar to our objects.} using
the expectation from equation 3 of \cite{just07}.  In the astronomy survival statistical
package (ASURV)\footnote{http://astrostatistics.psu.edu/statcodes/sc\_censor.html},
we use various two-sample statistical tests \citep{feigelson85,lavalley92}:
Gehan's generalized Wilcoxon test (permuation and hypergeometric variances),
logrank test, Peto and Peto generalized Wilcoxon test, and Peto and Prentice
generalized Wilcoxon test.  The probabilities that the samples are drawn from
the same parent distribution are 0.2~--~0.5\%.  Overall, this analysis
leads us to conclude that it is unlikely that these spectra are consistent with WLQs.

The remaining likely explanations for the unresolved light in the HST images
of the non-broad-line AGN are ultra-compact starbursts.  The
presence of bolometrically weak, obscured AGN would not have a significant
contribution to the optical continuum.  Together,
our emission-line, X-ray, and infrared diagnostics (see \S~\ref{section:analysis_agn})
support this conclusion.  In this case, there are a few possible explanations for the
unusually blue continuum.  One of the most plausible is
very strong differential dust attenuation (i.e., the ionizing O-stars
are more heavily attenuated than the B-stars).  There is some
evidence for this in two galaxies where it is possible to compare
the nebular extinction using the H$\gamma$ and H$\beta$ emission lines
to the extinction to the galaxy's broadband SED continuum.  Other
possibilities include leakage of ionizing continuum
radiation or very abruptly truncated starbursts.

Therefore, we conclude that each of the PSF components in the galaxies
that we have not classified as broad-line AGN are very compact stellar
populations produced in central starbursts.  When characterizing each
galaxy light distribution, we do not wish to remove the central
starburst, thus we quote the half-light radius measured from the
S\'{e}rsic fit rather than the S\'{e}rsic+PSF fit.  While our sample
is relatively small (12 galaxies), \emph{HST} follow-up of an
additional 17 galaxies has demonstrated that compactness is a
near-ubiquitous property of our galaxies \citep{diamond-stanic12b}.

\subsection{Comparison with Theory}

Numerical simulations suggest that compact central starbursts
(0.01~--~1 kpc) are produced in dissipational (gas-rich) major mergers
\citep{mihos94,cox06b,hopkins08c,hopkins09a}.  The remnants of such mergers
are predicted to have two-component light profiles: an outer profile
established by violent relaxation of stars present in the progenitors
before final coalescence, and an inner stellar population formed from
gas driven to the nuclear regions by strong tidal torques.  A large
fraction of local ellipticals appear to be consistent with this
picture: their light profiles show central `cusps' or light that is in
excess of an inward extrapolation of an $r^{1/4}$ law fit to the outer
regions \citep[][and references therein]{kormendy09}.  Cusps with
$r \lesssim 100$~pc have also been identified in NIR imaging of local
ULIRGs and recent merger remnants \citep{rothberg04,haan13}.  We
hypothesize that the unresolved light in the galaxies not classified as
broad-line AGN represents the central cusp predicted by simulations ---
i.e., stars formed from gas that sank to the inner regions of the
potential well.  At $z=0.4-0.75$ the FWHM of the WFC3/F814 PSF is
400~--~540~pc, and thus it is unsurprising that we do not resolve these
features.

\citet{hopkins08c} found that the mass fraction of the central
starburst correlates with the initial gas fraction of the progenitor
disks. Given our single-band \emph{HST} imaging, we have not attempted
to compute central starburst mass fractions.  However, the light fraction
represented by the PSF frac in Table~\ref{table:HST_info} can provide a
rough estimate of the mass fraction, given that much of the light in the
outer dissipationless component comes from stars formed in the interaction
but prior to nuclear coalescence.  Indeed, 100~Myr after the final merger,
\citet{hopkins08c} show that radial B-band mass-to-light variations are,
at most, a factor of $\sim2$~--~3, with the inner regions sometimes having
higher $M_*/L_B$ due to dust attenuation.

Five of the nine narrow-line galaxies have PSF fractions in the range
of 40~--~60\%, while the remaing four have PSF fractions less than 20\%.
Notably, galaxies in this later group show indications of having not
reached complete nuclear coalescence including double nuclei
(J1713+28), elliptical inner light profiles with large GALFIT
residuals that may be consistent with dual nuclei slightly below the
resolution limit (J1613+28 and J2118+00), and very bright inner tidal
features (J1558+39, J1713+28, and J2118+00).  If we assume that
the 40~--~60\% PSF fractions of the more relaxed sources imply comparable
central starburst mass fractions, the Hopkins simulations suggest that
the progenitor disks must have had gas fractions in the range of
40~--~80\%.  These gas fractions are at the upper end of the
distribution for massive disks at $z\sim0.5$ \citep{combes13},
consistent with the fact that our galaxies are very rare objects.

\subsection{Supermassive Black Hole Activity}
\label{section:SMBH_activity}

It is widely believed that all massive galaxies with bulge-like cores
contain SMBHs at their centers \citep[e.g.,][]{richstone98}.  All of
our galaxies are massive ($\log (M_*/M_{\odot})$ = 10.5~--~11.5) and have
centrally concentrated light distributions, thus they are likely to
host SMBHs.  Recent theoretical work indicates that powerful AGN may
be able to drive massive outflows from galaxies that quench star
formation \citep[e.g.,][]{king11,faucher-giguere12}.  Our goal is to
determine what role, if any, SBMHs play in powering the galaxy-scale
fast outflows that we observe in Mg~II absorption in our galaxies.
(The 12 galaxies in our sample were selected from a larger sample to be
most-likely to host AGN activity; see \S~\ref{section:sample} for
details).  The first stage in this analysis is to determine the
activity level of the SMBHs.  We summarize our findings below.

Three of our 12 galaxies host broad-line AGN (see
\S~\ref{section:BL_AGN}).  Virial SMBH mass estimates suggest
that they have masses of $\log (M_{SMBH}/M_{\odot})$ = 8~--~9 as expected.
Two are X-ray-detected and the third is undetected suggesting that it
may be partially obscured.  Estimates of the bolometric luminosity
based on X-rays and the optical continuum suggest that the sources are
radiating at $\sim 1$~--~7\% of Eddington.  Notably, in spite of the fairly
high luminosity of the AGN, they provide only $\sim 30$\% of the optical
continuum.  This is due to the fact that the massive host galaxy has
very recently experienced a strong starburst.

The remaining nine galaxies exhibit narrow-line emission, but lack broad lines.
For these galaxies, we consider the obscured AGN scenario, which is quite plausible
for the narrow-line AGN given that 1) we know these are gas-rich,
highly dissipative mergers where cold gas seems to be efficiently
funneled to the cores of the galaxies that have produced strong cusps
and 2) broad-line AGN have been detected in three cases.  A large
accretion rate can produce a Compton-thick torus, where, in the
unified AGN model \citep{urry95}, the inclination of the torus can
strongly affect our ability to see the SMBH accretion disk.
There is some evidence for large numbers of heavily-obscured or
Compton-thick nuclei in gas-rich galaxies
\citep{daddi07,vignali10,treister10b,fiore12}
that only very hard X-rays can pierce \citep{koss11}.  However,
we only find evidence for heavily obscured AGN in a small fraction of
our sample, and our analysis presents a consistent picture that none
of them are bolometrically dominant as compared to the
compact central starbursts.  We review these findings below.

We explored the narrow line emission for the nine galaxies lacking broad 
lines to look for evidence of obscured AGN (see \S~\ref{section:non_BLAGN}).
We employed a variety of diagnostic diagrams designed to be similar to
the classic BPT diagram (Fig.~\ref{fig:bpt}), but found that they produced 
inconsistent classifications both for our sample and a color and 
mass-matched comparison sample.  Based on the strength of the 
high-excitation emission lines, [O~III] and [Ne~V], we consider the 
following three galaxies as candidate Type~II AGN: J1104+59 (ID~3), 
J1506+54 (ID~4), and J1713+28 (ID~8).   Both J1104+59 and 
J1713+28 have high [O~III]/H$\beta$ ratios relative to our purely 
star forming comparison sample (Fig.~\ref{fig:bpt}).  However, the line ratios could 
still be consistent with star formation given the unusual physical 
conditions in the galaxies \citep{kewley13a,kewley13b}.  J1506+54 has a 
much lower [O~III]/H$\beta$ ratio but the highest [Ne~V] luminosity of 
the sample (Table~\ref{table:broad-line}).  [Ne~V] has a very high ionization potential 
(97~eV) and is generally considered a good AGN indicator, but it is 
possible that the [Ne~V] in J1506+54 is produced by the galaxy's hot 
young stars (\S~\ref{section:j1506}). If we use the [Ne~V] luminosity to estimate the 
AGN bolometric luminosity we find $L_{bol}/L_{Edd}= 0.05$.  Most 
importantly, we estimate that the AGN would produce only $\sim10$\% of 
the galaxy's mid-IR luminosity, with the remainder powered by the young 
starburst.  Thus, while obscured AGN may be present in our sample, we 
conclude that they are not significant contributors to the galaxies' 
total luminosities.

Another possibility is that the combined duty cycle and Eddington
ratios of these galaxies produce a population where less than half of
the SMBHs have sufficiently high accretion rates to observe.  This
could arise because SMBHs experience state transitions similar to
stellar mass BHs in X-ray binaries, but on much longer
timescales given the large size scales involved
\citep[e.g,][]{merloni03,markoff08}.
In uniformly selected samples of galaxies, it is not surprising to
find a low active fraction of SMBHs represented by a duty cycle of
$\sim 10$~--~$20\%$ \citep{fu10,diamond-stanic12b}.  In fact, only
$\sim 1$\% of the galaxy population has $L/L_{Edd}>0.01$ \citep{aird12}.
In additon, complete, distance-limited samples of galaxies probing SMBHs
at X-ray wavelengths find that most galaxies have Eddington ratios
$\lesssim 10^{-5}$ \citep{miller12}.  Furthermore, only a fraction of
SMBHs are radio-loud \citep[e.g.,][]{chen13}.  This likely all
arises from strong (factors of thousands to millions) AGN variability
over long ($\gtrsim 10^4$~yr) timescales that can only be constrained
indirectly through observations of large populations of galaxies
\citep{hickox13}.  There is considerable evidence for this
\citep[e.g., AGN light echoes, the ``{\em Fermi} bubbles"; e.g.,][]{keel12,su12}.

The fact that our galaxies appear to be late-stage mergers might suggest
enhanced AGN activity.  However, much is still unknown about how gas
accretes onto the AGN from the galaxy. The theoretical work of \citet{cen12} suggests
that a SMBH does not enter a rapid accretion phase until $\sim
100$~Myr after the starburst peak when the AGN can capture material
from the slow winds of post-asymptotic giant branch stars.  Most of
our galaxies have on-going star formation, and thus it may
be that the SMBHs are not accreting at a high rate yet.

\subsection{Is AGN feedback responsible for driving the galaxy-wide outflows and shutting off star formation in these galaxies?}

A number of theoretical models have tried to link the SMBH to its
larger-scale surroundings.  \cite{wagner12} suggests that feedback
from AGN can be efficient in a clumpy, spherical medium if the AGN
power is sufficiently close to L$_{Edd}$.  However, a key issue that
is still not resolved is how AGN can inject considerable momentum to
couple to a large fraction of the gas to drive it out of the galaxy
\citep{debuhr12}.  \cite{wagner13} suggests that this can be
accomplished through ram pressure with dense clouds embedded in a
tenuous, hot, hydrostatic medium.  Alternatively, the potential for feedback may
depend very strongly on the ability of outflowing energy from either
starbursts or AGN to couple to dust \citep{novak12}.  AGN feedback
is frequently modeled on small scales \citep[e.g.,][]{liu13}, which is
difficult to causally connect to galaxy wide outflows.

Recent observational work has had a difficult
time disentangling the effects of stars from the SMBH to resolve
whether AGN or starburst feedback plays a more critical role in
shutting off star formation and driving galactic-scale, fast outflows
\citep{harrison12}.  There is considerable observational evidence for
dominant AGN feedback on small scales, where it can be better
separated from starburst feedback: NGC~1266 \citep{alatalo13}, Mrk~231
\citep{feruglio10}.  However, whether the SMBH is primarily
responsible for driving the galaxy-wide outflows in star-forming galaxies has not been
fully resolved.  In fact, when the SMBH injects energy into the
surrounding gas, it is not clear if it has a net negative or positive
effect on these galaxies.  In some situations, the SMBH can
actually help to trigger a starburst \citep[e.g.,][]{zubovas13}.

In a companion paper, \cite{diamond-stanic12b} analyzed 29 galaxies
(of which our 12 galaxies are a subsample) drawn from our larger galaxy sample.
Their analysis combined the available broadband SEDs
compared to a few starburst and quasar models, the compactness of the
galaxies deduced from \emph{HST} observations, highly blueshifted
absorption lines indicative of massive outflows, and model comparisons
to the estimated star formation rate surface density.  They concluded
that AGN feedback is not \emph{required} by arguing that compact
starbursts are capable of driving massive, galaxy-wide outflows at
$\sim 1000$~km~s$^{-1}$.  We revisit this issue here in a more
detailed presentation of our multiwavelength data including critical,
new information on the AGN content for our subsample.

We detect high-velocity outflows based
on Mg~II absorption line measurements in 9/12 galaxies in our sample.
Of the three galaxies hosting broad-line AGN, only J2140+12 shows
Mg~II absorption.  We hypothesize that outflows may be present in the
other two broad-line AGN but the magnesium in the outflow may exist in
a higher ionization stage due to exposure to the AGN's hard
photoionizing continuum \citep{hennawi13}.  Notably, J2140+12
is not detected in the X-rays suggesting partial obscuration of the
nuclear source, which may explain why Mg~II is present in absorption
(see \S~\ref{section:BL_AGN}).  The outflow velocity measured for
J2140+12 is the second slowest in our sample.

Of the nine narrow-line galaxies, eight have outflows based on highly
blueshifted Mg~II absorption.  We have found evidence for Type~II AGN
in only three galaxies (J1104+59, J1506+54, and J1713+28),
but the results are somewhat ambiguous (see \S~\ref{section:SMBH_activity}
for a discussion); these galaxies have a range of outflow velocities
consistent with the rest of the narrow-line galaxies.  As noted previously,
we cannot conclusively rule out the presence of a Compton-thick AGN in any
of these narrow-line galaxies.

\begin{figure*}
\includegraphics[scale=0.33, angle=90]{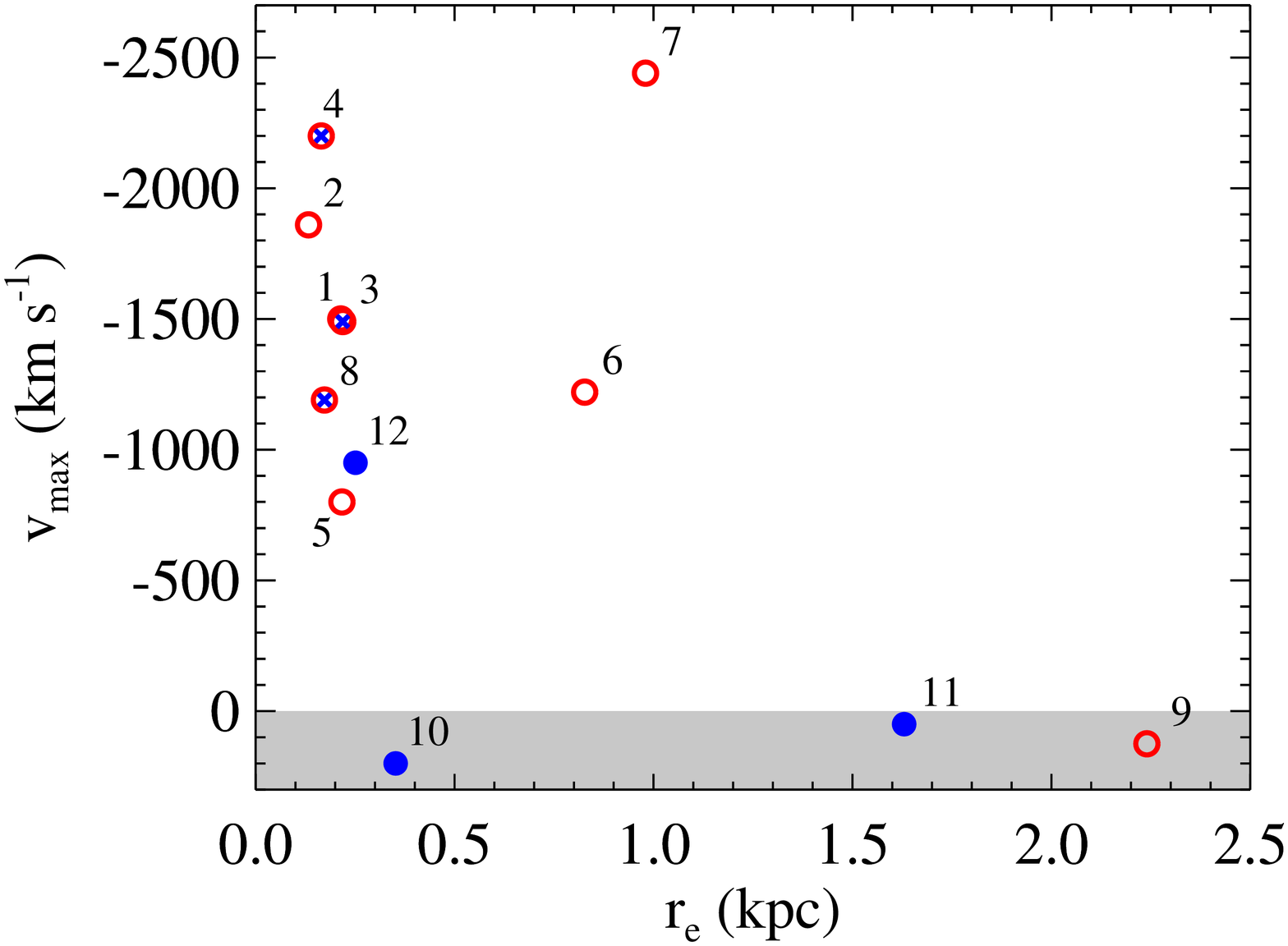}
\includegraphics[scale=0.33, angle=90]{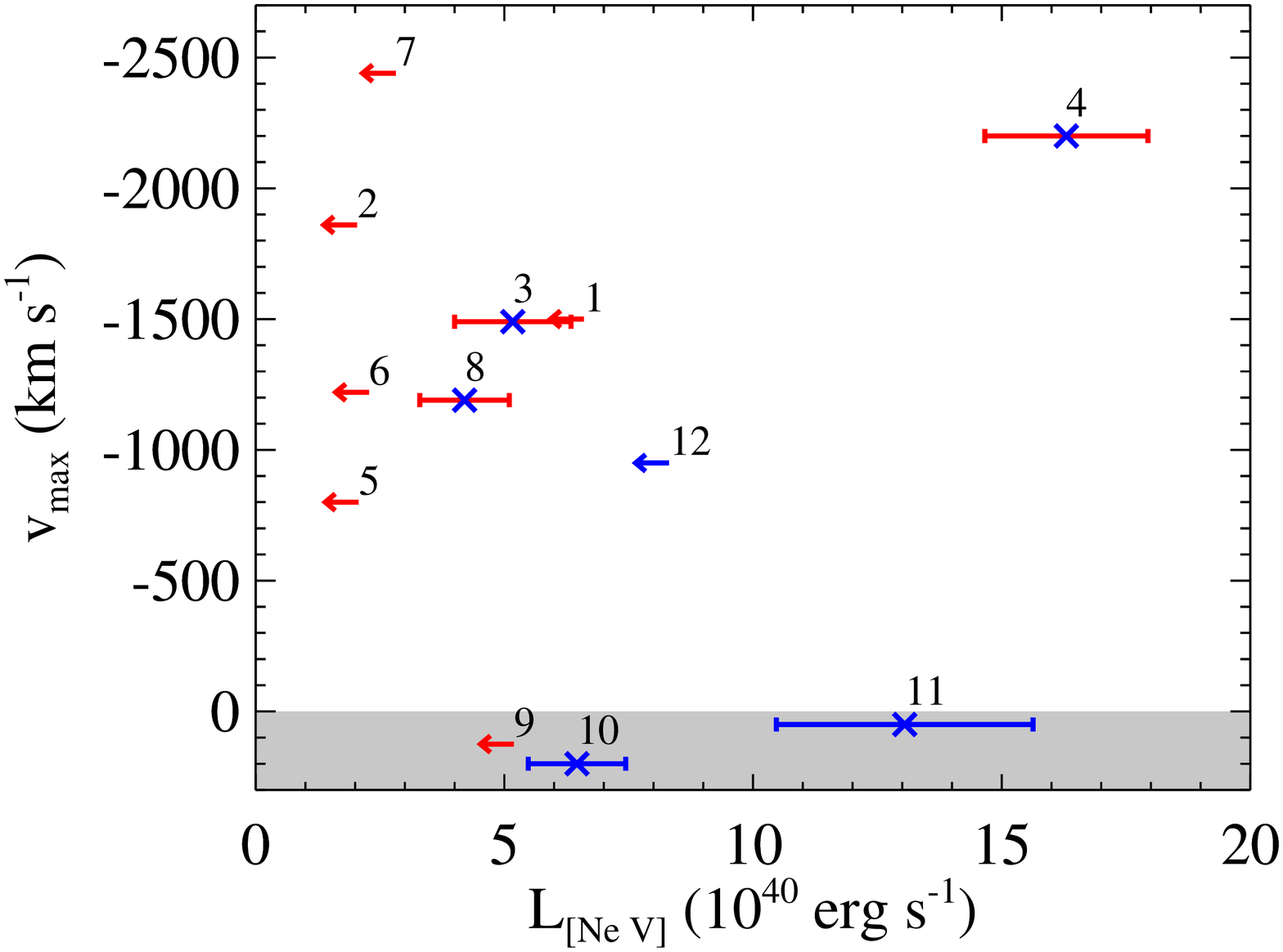}
\caption{These plots bring together our spectroscopic, AGN, and morphological
analysis to examine the correlation or lack thereof between outflow velocities,
size, and AGN content.  The colors in both plots highlight the different
source classifications:  blue for AGN and red for starbursts.  Blue-only
points (filled circles on the left) have broad-line AGN; red-only points
(open circles on the left) are consistent with pure star formation; points with
a combination of blue and red (with ``X's'' on the left) indicate emission
lines consistent with extreme star formation and Type~II AGN (ambiguous).
In both parts of the figure, we plot the maximum outflow velocity (the trends
are the same when using the average velocity, as the two velocities are
highly correlated --- see \S~\ref{section:data_red_MMT}).  In 3/12 cases where
no outflow is detected, we assigned slightly different positive velocities
for clarity only.  On the left, we plot the outflow velocity as a function of
the best-fit single S\'{e}rsic model effective radius (a measure of compactness;
where there are two cores, we use the radius of the brighter one).  On the right,
we plot the outflow velocity as a function of the [Ne~V] luminosity
(high-excitation [Ne~V] is generally a good tracer of AGN activity but see
\S~\ref{section:j1506} for caveats).  If [Ne~V] is tracing AGN activity and the
AGN drive the outflows, we would expect to see a correlation, but we do not.
These plots support our conclusion that the galaxy-wide outflows appear to be
driven by extreme starbursts, not AGN.}
\label{fig:summary}
\end{figure*}

Figure~\ref{fig:summary} best summarizes our findings.  We plot the maximum outflow
velocity as a function of the galaxy half-light radius and [Ne~V]
luminosity to explore whether the outflow velocity is more closely
related to galaxy morphology or nuclear activity.  We find that all of
the galaxies with the fastest outflows  ($< -1000$~km~s$^{-1}$) are
compact starbursts ($r_e < 1$~kpc).  The median half-light radius of
these galaxies is $r_e = 251$~pc.  Three of the seven galaxies with
$v_{max} < -1000$~km~s$^{-1}$ have some evidence for obscured AGN
activity, but the highest velocity outflow ($v_{max} = -2440$~km~s$^{-1}$)
is found in a galaxy with no evidence for AGN activity (J1613+28).
Only one of the three unequivocal (broad-line) AGN has
a detected outflow (J2140+12); this galaxy has one of the slowest outflows
in the sample ($v_{avg} = -490$~km~s$^{-1}$, $v_{max} = -950$~km~s$^{-1}$).
We also find that [Ne~V], which is a common tracer of AGN activity given the high
excitation energies required, shows no correlation with outflow velocity, as would
be expected if AGN were driving the outflows.  If we plot the [Ne V] bolometric
Eddington fraction, we find similar results.  Furthermore, the possible obscured
AGN in the galaxy with the highest [Ne~V] luminosity (J1506+54) accounts for only
$\sim11$\% of the galaxy's mid-IR luminosity, suggesting that obscured star formation
dominates the galaxy's bolometric luminosity.  Therefore, we conclude that the
presence of a powerful outflow seems to be more closely linked to the star formation
properties of the host galaxies rather than the AGN.  This is reminiscent of
Fig.~4 of \cite{diamond-stanic12b}, who showed that the outflow is linked
with the SFR surface density.

Despite the apparent lack of correlation between the high outflow
velocities and AGN activity in this sample, it remains possible that
the outflows were driven in the recent past by AGN activity that has
since switched off and is no longer visible. A strong correlation has
recently been observed between nuclear star formation and AGN activity
\citep[e.g.,][]{diamond-stanic12a,esquej13}, suggesting that our sources,
which exhibit strong nuclear starbursts, may be more likely to host a
powerful AGN.  If the AGN experiences rapid variability over a large
dynamic range of several orders of magnitude
\citep[see][for a discussion]{hickox13}, the possibility exists that a wind
could have been launched by an AGN that has rapidly decreased in luminosity,
while the outflow persists over longer timescales \citep{zubovas14}. This
scenario cannot be conclusively ruled out from our observations;
however, the ubiquity of rapid outflows among our sample would suggest
that essentially all our galaxies have hosted an AGN in the recent
past, which imply a remarkably strong connection between nuclear star
formation and AGN activity.  Given that the highly compact starbursts
themselves may be capable of producing the high-velocity winds 
\citep[e.g.,][]{heckman11}, a more straightforward explanation is that AGN
driving is simply not required, and this is the interpretation that we
favor here.

The additional lack of bright radio emission in these galaxies
suggests that radio jets do not play a major role, at least shortly after
the merger, but that the powerful, compact starburst initially drives
out the gas in a galaxy-wide outflow.  While
considerable non-luminous, mechanical energy is contained in SMBH jets and
disk winds, which could be larger during low-luminosity states
\citep[e.g.,][]{heinz07c,kording08}, the relative efficiency of SMBH mechanical
energy is typically modeled to be approximately two orders of magnitude below
that of radiation pressure feedback for PSBGs \citep[e.g.,][]{ciotti10}.
Furthermore, \cite{karouzos13} find that radio jets can suppress
star-formation in their host galaxies but appear not to totally quench it.

These findings together generally support the major merger evolutionary scenario
first developed by \cite{sanders88}, where a gas-rich merger produces a
shrouded, dusty starburst.  Then after some time, the SMBH is uncovered
as the remnant ages to become an early-type galaxy.  This scenario is also
consistent with more recent models where the AGN is efficiently fueled
\emph{after} the starburst has ended \citep[e.g.,][]{cen12}.  Later, the
SMBH launches powerful radio jets from the center of an elliptical remnant
into a hot, tenuous ISM, which can act to maintain the state of the gas as
a low-excitation radio AGN \citep[e.g.,][]{smolcic09}.

Our results agree with the primary result of \cite{diamond-stanic12b} that the
$v\propto r^{-1/2}$ scaling for winds driven by either supernovae or radiation pressure
from massive stars can drive outflows up to the extreme velocities we observe.
Furthermore, we find that the presence of an AGN does not imply a
higher outflow velocity.  This is consistent with the results of
\citet{coil11} who explored outflows in samples of X-ray selected AGN
and classic post-starburst (K$+$A) galaxies at intermediate redshifts and
found $v \sim -200$~km~s$^{-1}$, irrespective of nuclear activity.  Given the low
inferred contributions of the AGN to the galaxies' bolometric luminosities and the
lack of conclusive AGN activity in the galaxies with the fastest outflows, we conclude that
AGN feedback does not appear to be the dominant mode of feedback for these galaxies.
Altogether, this discussion suggests that the galaxy-wide outflows in this sample of
galaxies are primarily driven by extreme starbursts.

\section{Summary and Conclusions}\label{section:conclusions}

We have analyzed \emph{HST} and \emph{Chandra} images, UV-optical
spectra, UV-MIR photometric data, and JVLA radio data on a
a sub-sample of massive galaxies at $z=0.4$~--~0.75 in the midst of
star formation quenching.  These galaxies were selected from a larger parent sample
as the most likely to host AGN.  Our primary goal is to understand the activity
of their SMBHs and the morphology of their host galaxies to gain insight into
whether the SMBHs drove the galaxy-wide outflows and played the primary role in shutting
down their recent star formation.  A summary of our findings is presented below.

\begin{itemize}

\item{Restframe V-band \emph{HST} imaging reveals tidal tails or
disturbed morphologies indicative of a recent major or minor merger
in 9/12 galaxies.  Given the shallow depth of our images, we cannot
rule out the presence of such features in the remaining galaxies.
We conclude that the recent starburst in all of our galaxies was likely
triggered by a merger.}

\item{All of the galaxies have very compact light profiles.  J1558+39, J1613+28,
	J1713+28, and J2118+00 appear to be in the midst of nuclear coalescence
	and have not relaxed yet.  Excluding these galaxies and the three
    Type~I AGN, we measure effective radii of 0.1~--~0.2~kpc using a single
    S\'{e}rsic fit.  These objects are better fit by a combination of a
    S\'{e}rsic profile and a nuclear point source which contains 40~--~60\%
    of the total light.  We argue that the unresolved light is not due
    to an AGN because we see no evidence of broad Mg~II or H$\beta$
    emission lines in the optical spectrum and the probability that these galaxies
    are consistent with weak-lined AGN is small based on their $\alpha_{ox}$ ratios.
    We conclude that the nuclear light is likely due to a compact ($<0.5$~kpc)
    central starburst triggered by the dissipative collapse of
    very gas rich progenitor disks, as suggested by theoretical
    models \citep[c.f.,][]{hopkins09a}.}

\item{Three of the galaxies (J1359+51, J1634+46, and J2140+12) are broad-line AGN.
	We use the width of the broad Mg~II emission lines to derive virial masses of
    $M_{SMBH} \sim 10^8$~--~$10^9$~M$_\odot$.  Two of the AGN are X-ray-detected
    (J1359+51 and J1634+46) and one is radio-loud (J1634+46).  We calculate
    that they are radiating at $\sim 1$~--~7\% of their Eddington luminosities.}

\item{Based on high-excitation emission-line diagnostics, only 3/9 narrow-line
    galaxies (J1104+59, J1506+54, and J1713+28) exhibit signs of
    obscured AGN.  The bolometric Eddington fractions are similar to those found for
    the broad-line AGN, but even more uncertain because the emission is also
    consistent with the presence of an ultra-compact starburst; this leads to
    somewhat ambiguous results.  In one case, J1506+54, we find that only 11\% of the
    mid-IR luminosity of is due to the AGN with the remainder coming from star formation.}

\item{The other six galaxies in the sample are completely consistent with compact
    starbursts.  We have shown that analysis of BPT-type diagrams does not
    provide a clear indication of AGN activity for these galaxies
    primarily because the [O~III]/H$\beta$ ratio does not discriminate
    extreme starbursts well.  In addition, although faint levels of X-ray emission
    are observed in three of these galaxies, the derived X-ray luminosities and
    stacked hardness ratio are entirely consistent with emission from
    XRBs given the large recent SFRs of these galaxies.  While we
    cannot conclusively rule out Compton-thick AGN in these sources,
    we suggest that star formation is likely to be the dominant
    contributor to the galaxies' bolometric luminosities.}

\item{Nine of our galaxies show evidence of ultra-fast
    ($v_{max} \gtrsim 1000$~km~s$^{-1}$) Mg~II outflows.  Only one of the three
    unequivocally identified (broad-line) AGN
    has a detected outflow, and it is the slowest or second slowest in the sample,
    depending on how the outflow velocity is measured.  The light from
    5/9 of the galaxies with the ultra-fast outflows is completely consistent with
    compact starbursts (and in 3/9 cases, we are unable to conclusively differentiate
    between starburst and AGN activity).  We conclude that outflow properties are
    not linked to ongoing AGN activity.}

\end{itemize}

These results support the primary conclusion of \cite{diamond-stanic12b},
who argued that our ultra-compact galaxies have the physical conditions necessary to
launch the high-velocity outflows we observe by highlighting the
$v\propto r^{-1/2}$ scaling for winds driven by either supernovae or radiation
pressure from massive stars.  To this argument we add the fact that the presence
of a luminous AGN does not appear to have any positive correlation with Mg~II
absorption strength or outflow velocity.  This study and \cite{diamond-stanic12b}
cannot and did not conclusively rule out that AGN play at least a minor role
at some point in the evolution of these galaxies.  However, we do not find any
evidence directly in support of AGN feedback in this sample of galaxies, and AGN
feedback is unnecessary to explain the observations.  Overall, we conclude that
these galaxies are massive merger remnants with high-velocity outflows primarily
driven by powerful, unusually compact starbursts.

\acknowledgments

Support for this work was provided by the National Aeronautics and Space
Administration (NASA) through {\em Chandra} Award Number GO0-11135A issued by the
{\em Chandra X-ray Observatory} Center, which is operated by the Smithsonian
Astrophysical Observatory (SAO) for and on behalf of the NASA under contract
NAS8-03060.  Support was also provided
by {\em HST} Programs HST-G0-12019 and HST-GO-12272, which
were funded by NASA through a grant from the Space Telescope Science Institute (STScI),
which is operated by the Association of Universities for Research in Astronomy (AURA),
Incorporated, under NASA contract NAS5-26555.  This work is based in part on
observations made with the {\em Spitzer Space Telescope} (Spitzer-GO-60145), which
is operated by the Jet Propulsion Laboratory (JPL), California Institute of Technology
under a contract with NASA. Support for this work was provided by NASA through an
award issued by JPL/Caltech.  AMD acknowledges support from The Grainger Foundation
and from the Southern California Center for Galaxy Evolution, a multi-campus research
program funded by the University of California Office of Research.  ALC acknowledges
support from NSF CAREER award AST-1055081.  CT and GR acknowledges the support of the
Alexander von Humboldt Foundation.

\appendix
\renewcommand{\thefigure}{A\arabic{figure}}

\subsubsection{Creation of the PSFs for GALFIT}
\label{psfs}

From the Multidrizzled images, we created high signal-to-noise (S/N),
representative PSFs for each galaxy, critical for performing two-dimensional
galaxy fitting of our ultra-compact galaxies (Fig.~\ref{fig:data_fits}).  We
used a strategy simliar to that of \cite{canalizo07} to create our PSFs and come
to similar conclusions regarding PSF selection.  Since we did not have separate
stellar PSF images, we hand-selected single, isolated stars by eye that were not
too faint ($>1000$ peak counts) or saturated ($\lesssim 60000$ peak counts).  We
verified that each star was indeed an unsaturated point source by plotting its
radial profile and calculating its FWHM.  We identified 4~--~36 PSF stars per
image, which depended on the stellar density of the observed field.

For each star, we extracted $2^{\prime \prime} \times 2^{\prime \prime}$ stamps
approximately centered on each star, subsampled each stamp by a factor of 10 in
each dimension, and centroided each star in its stamp.  Using these stars, we
calculated a single stacked PSF for each galaxy,  weighting the stars by their
integrated counts.  We sampled the resulting PSF back up by a factor of 5 so
that it was $2 \times$ oversampled.  Then, we smoothed the outermost sections of
each stacked PSF using a median-filtered box that increases in radius starting
$0.2^{\prime \prime}$ away from the center of the stacked
PSF\footnote{halo\_smooth.pro, http://132.248.1.102/$\sim$morisset/idl/pro/ \\
starfinder/}.  Finally, we background-subtracted and normalized the PSF so that
it could be used for the galaxy fitting.

\begin{figure*}[btp]
\centering
\begin{tabular}{cccc}
\includegraphics[scale=0.4]{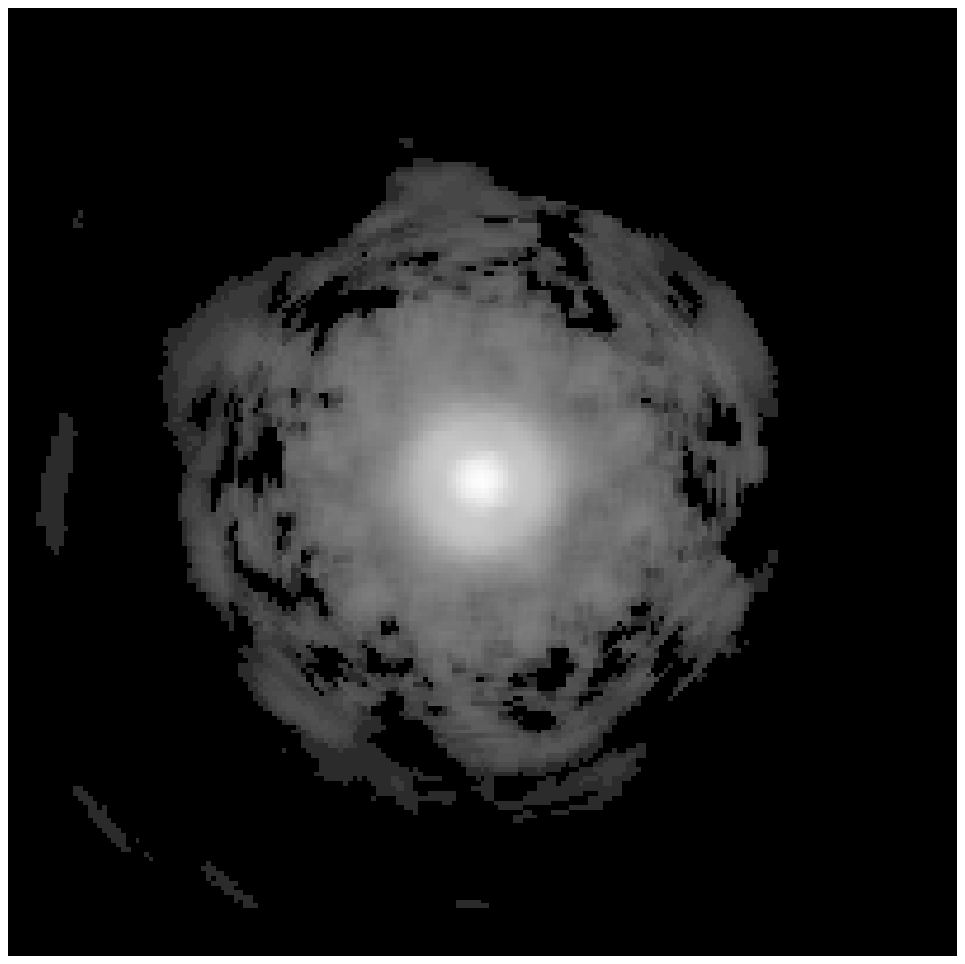} &
\includegraphics[scale=0.4]{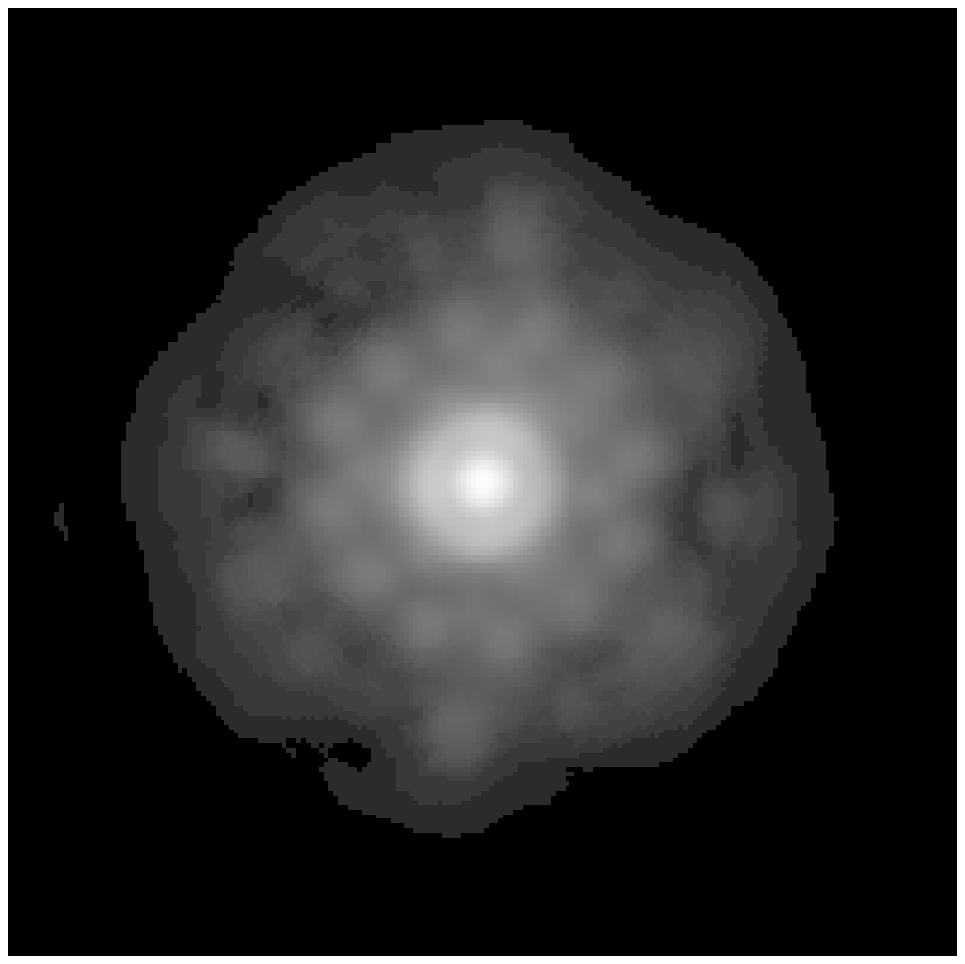} &
\includegraphics[scale=0.4]{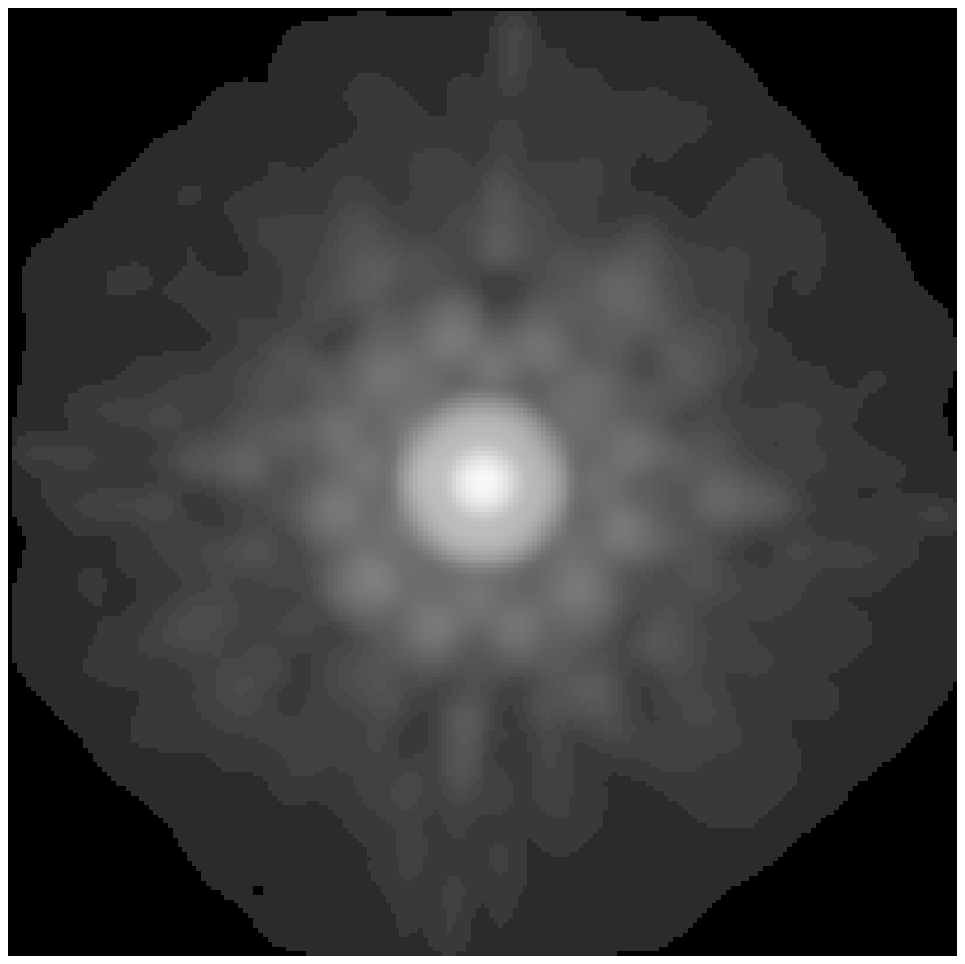} &
\includegraphics[scale=0.4]{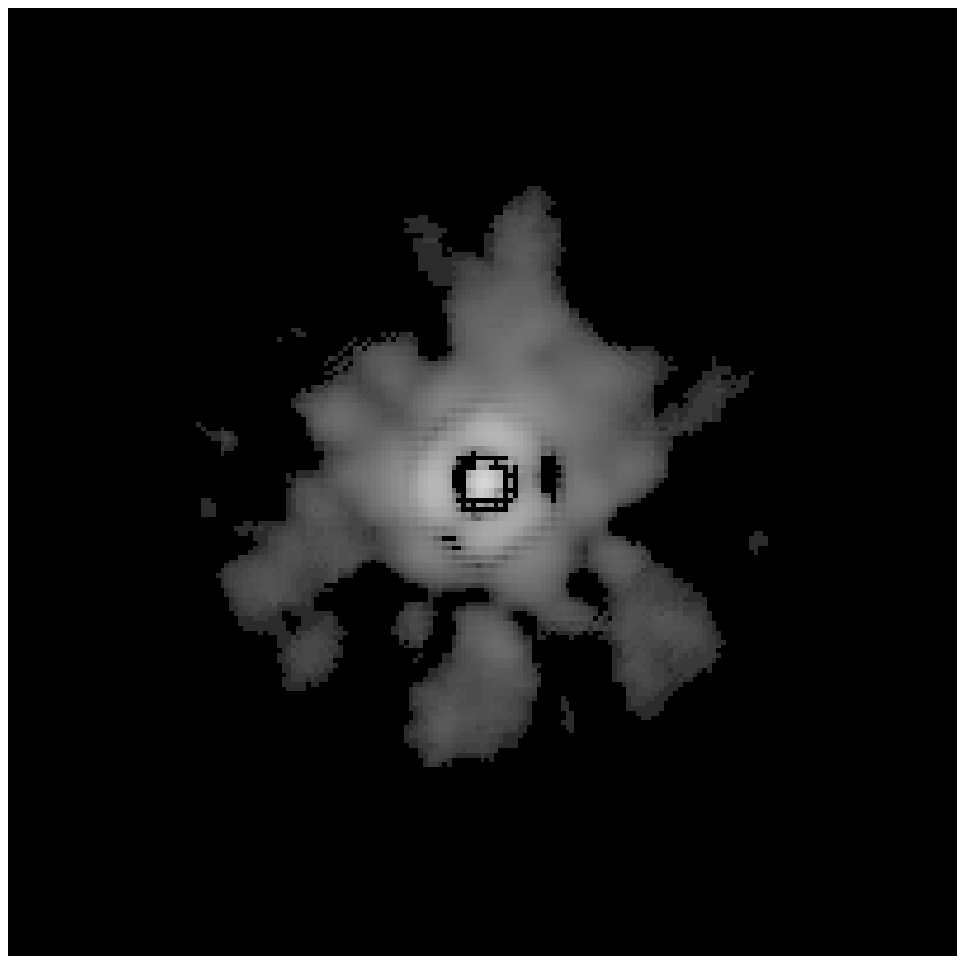} \\
\includegraphics[scale=0.14]{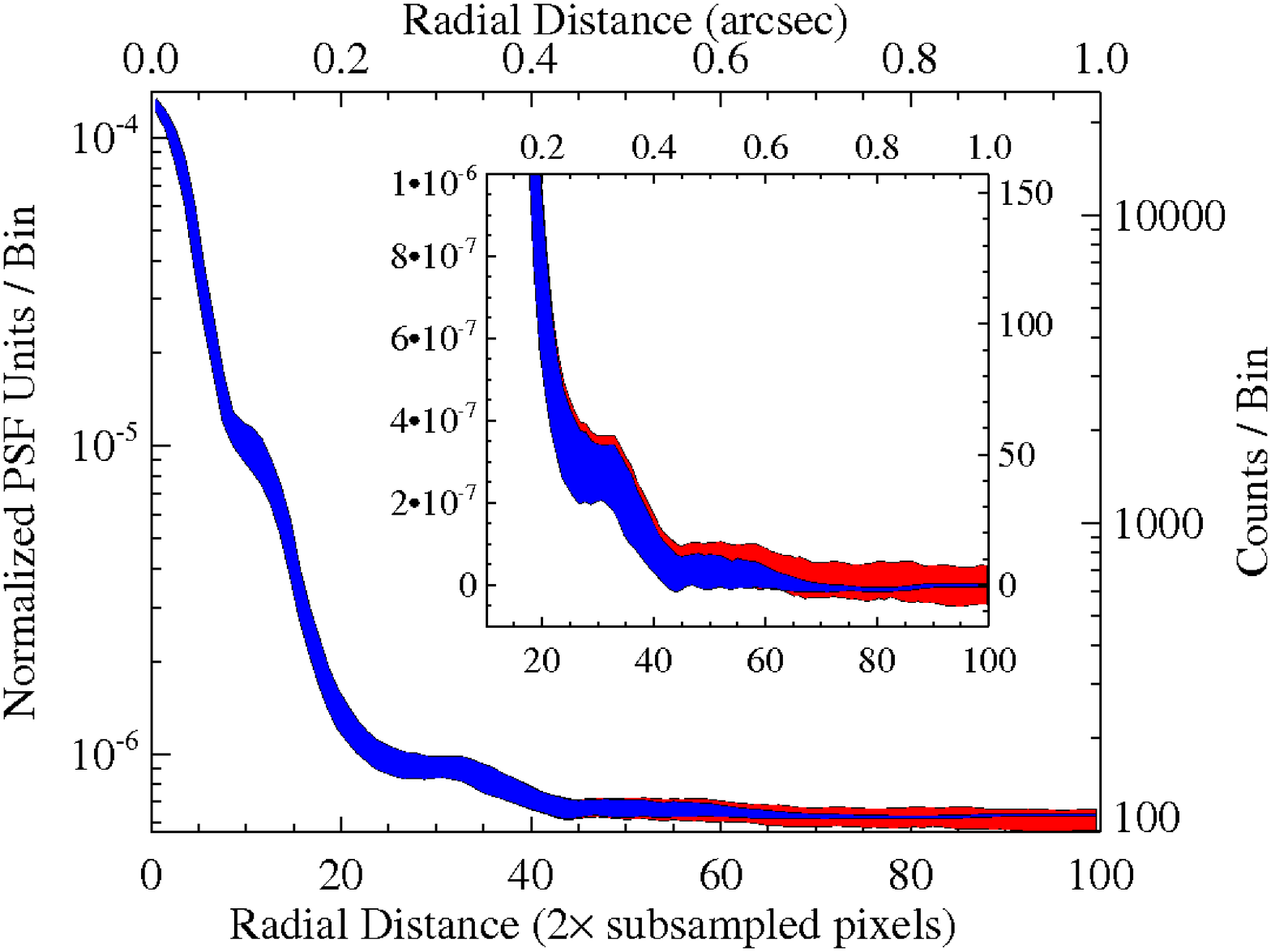} &
\includegraphics[scale=0.14]{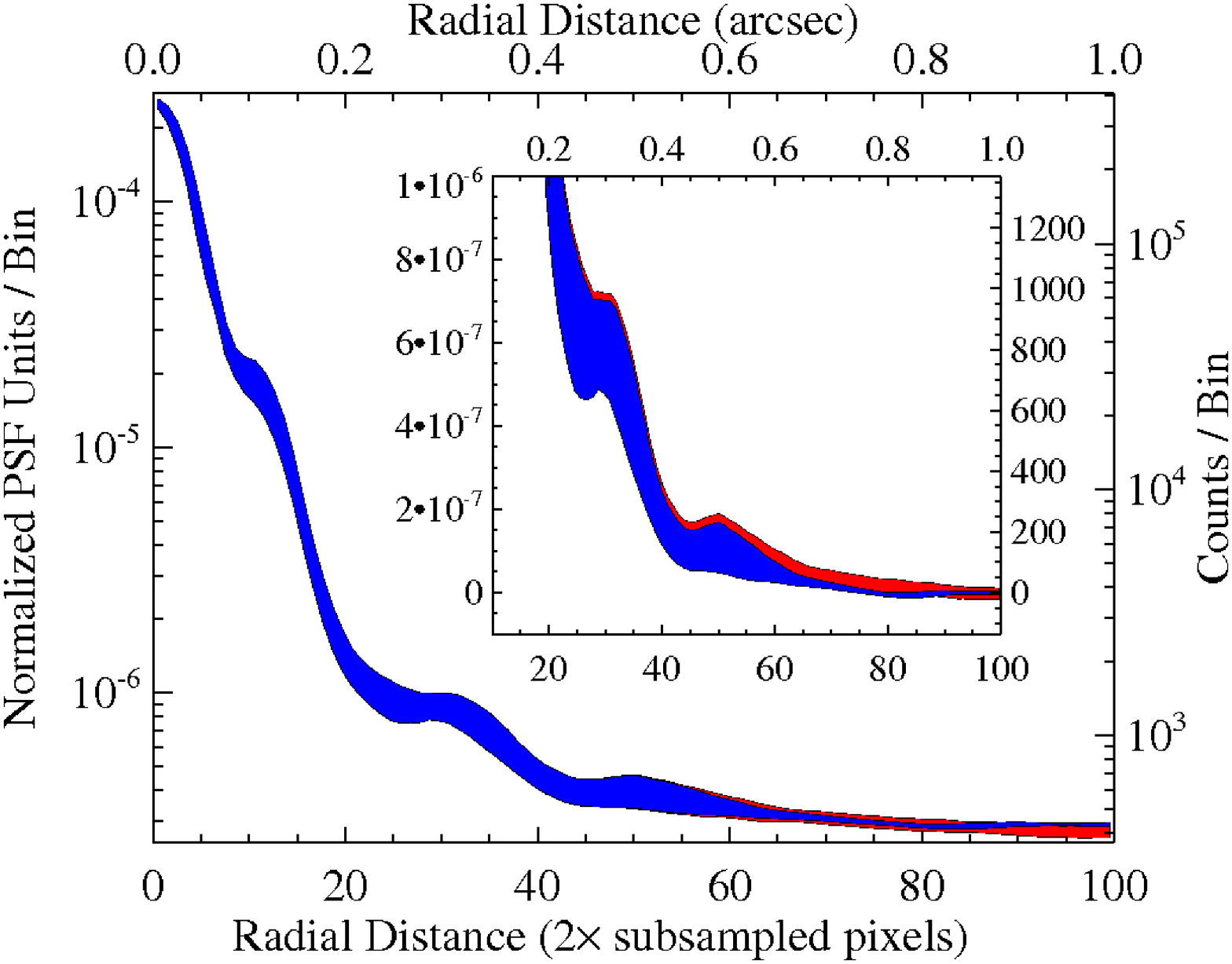} &
\includegraphics[scale=0.14]{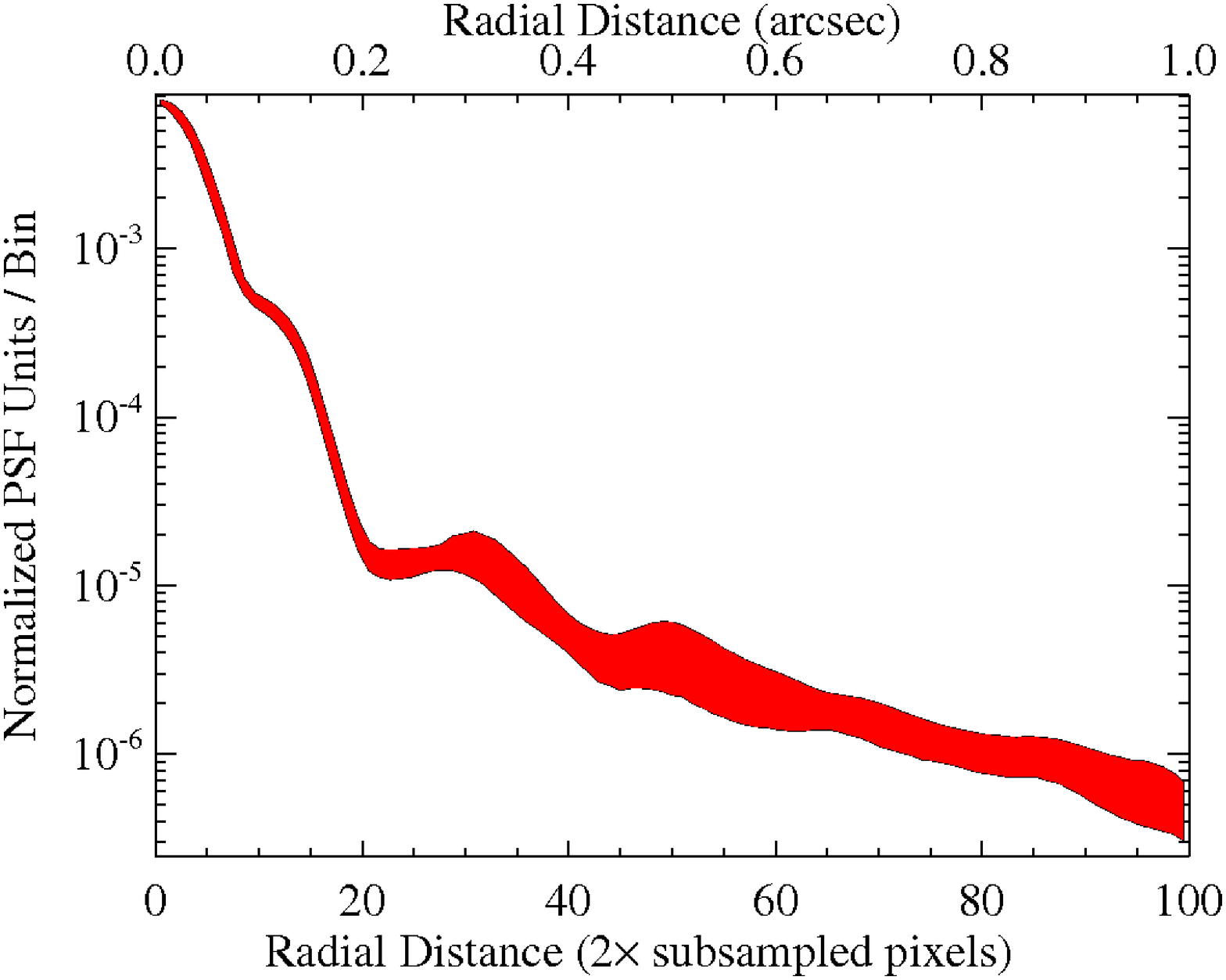} &
\includegraphics[scale=0.14]{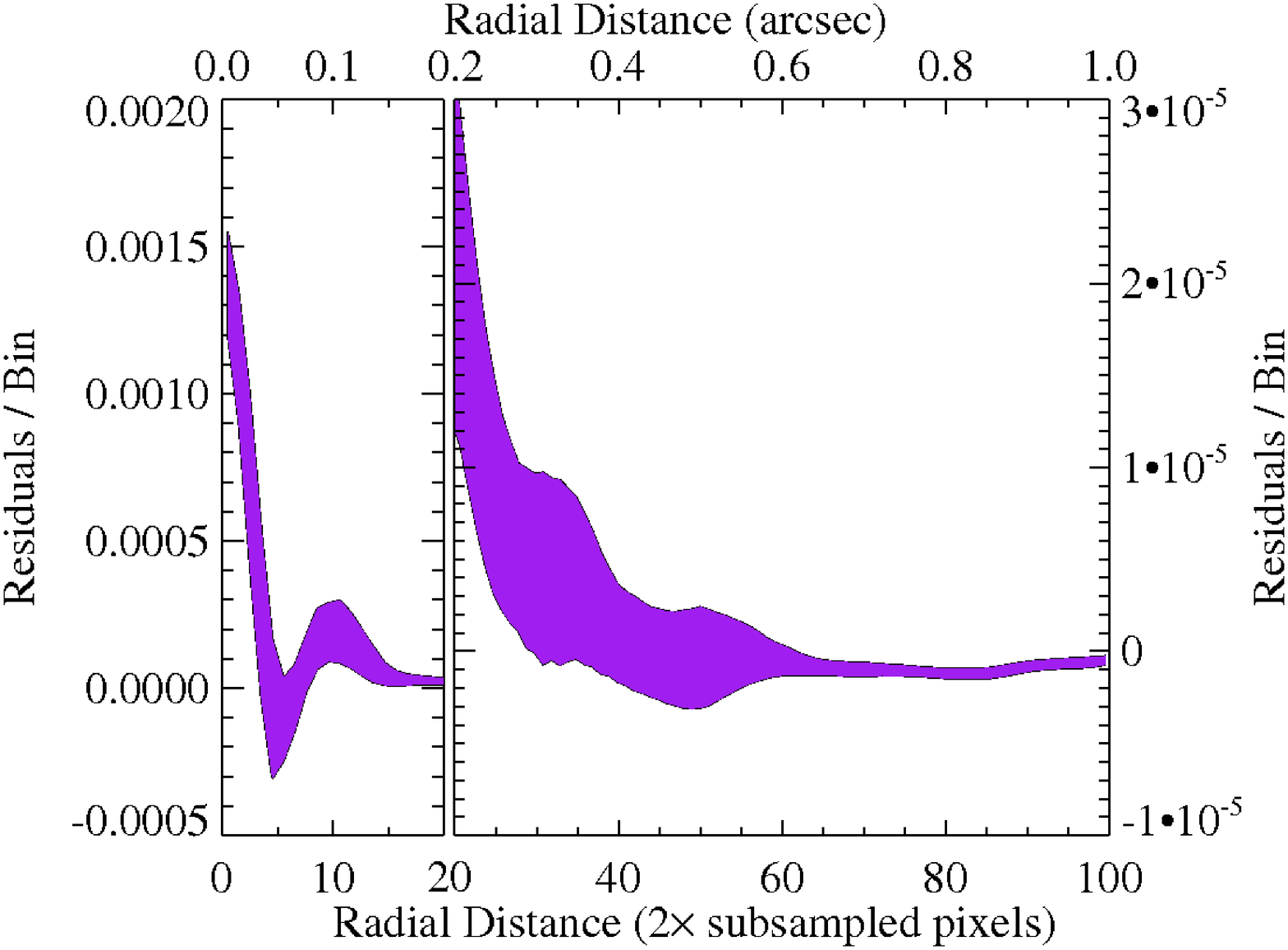}
\end{tabular}
\caption{First row (images from left to right):  J0826+43, lowest S/N stacked PSF; J2118+00,
highest S/N stacked PSF; TinyTim (``tiny2" stage) PSF convolved with a Gaussian to try to match
J2118+00's PSF in GALFIT; residuals between previous two PSFs.  The images have been considerably
stretched to show all parts of the PSF out to the faint wings.  The scales in all images have been
matched to each other.  All images are $2^{\prime \prime} \times 2^{\prime \prime}$.  Second row:
The radial profiles corresponding to each respective image.  For the first two plots from the left,
the larger plot contains the stacked PSF profile that has not been background-subtracted and the
inset contains the background-subtracted profile, focusing on the faint end of the PSF at large
radii.  The inset axes have the same units as their larger counterparts, except that each right
y-axis is the background-subtracted counts.  The third plot from the left is the radial profile of
the model PSF, which was not smoothed.  The fourth plot from the left is the radial profile of the
GALFIT residuals for when we convolved the TinyTim (``tiny2") PSF with a Gaussian to try to match
J2118+00's PSF.  The width of all profiles represent the standard deviation of the pixel values at
each radial bin.  The red and blue colors are the unsmoothed and smoothed PSFs, respectively.}
\label{fig:psfs}
\end{figure*}

Some PSFs are better determined than others because of the presence or lack of
bright stars in each image and the positions of those bright stars relative to
the galaxy.  This results in differences in S/N between the PSFs, which does not
appear to significantly bias our galaxy fits.  We show our highest and lowest
S/N PSFs in Fig.~\ref{fig:psfs}.  The J2118+00 PSF has the most
background-subtracted counts ($\gtrsim 11$~million) while the J0826+43 PSF is
comprised of the fewest background-subtracted counts ($\sim 700,000$).  While it
is tempting to substitute the PSF from J2118+00 for the PSF from J0826+43, we
chose not to do this for a number of reasons:  the focus clearly changes
significantly from image to image, the orientation of the PSF clearly changes
through the rotation of the slightly asymmetric airy pattern, etc.  We conclude
that the uncertainty introduced from using a PSF from a different image (e.g.,
having to rotate it) is probably at least as large as the uncertainty from using
the lower S/N PSF.

We also attempted to make use of the {\em HST} PSF modeling tool, TinyTim
\citep{hook08}, as it provides infinite S/N.  Because the final step in creating
the TinyTim PSF (the ``tiny3" stage) currently does not resample and distort the
PSF in the same way as Multidrizzle, we fit this PSF in GALFIT with our highest
S/N stacked image PSF (from J2118+00) using a Gaussian (to broaden the PSF to
match the image better).  Note that the TinyTim PSF was first rotated to the
nearest degree to match the orientation of the stacked PSF image so that
$\chi^2$ was minimized.  This matched the roll orientation of the telescope to
within a few degrees, as expected.

The output model (convolved) PSF and the residuals from the fit are shown in
Fig.~\ref{fig:psfs}.  The focus offset for the creation of the model PSF with
TinyTim was not well-determined as the model and measured focus differed
considerably\footnote{http://www.stsci.edu/hst/observatory/focus/FocusModel}. 
However, tests indicated that $\chi^2$ returned by GALFIT changed negligibly for
various reasonable focus offsets (within $\pm 0.4$~microns).  We conclude that
determining the correct rotation is important for creating a PSF with TinyTim,
but that the differences in focus offsets for reasonable focus values is
negligible.

As is evident in Fig.~\ref{fig:psfs}, the TinyTim PSF does not match the image
PSF very well.  The largest source of uncertainty in the TinyTim PSF model as
noted by the {\em HST}
team\footnote{http://www.stsci.edu/hst/observatory/focus/TinyTim} appears to be
the aberration coefficients used to generate the PSF, which are not well-modeled
for WFC3.  Because of these many complications and the fact that the {\em HST}
team recommends the use of empirical PSFs, we chose not to use the convolved
TinyTim PSF in our analysis.  For each galaxy, we used the stacked stellar PSF
from the respective image only.

\subsection{2-D Image Fitting Parameter Uncertainties}
\label{appendix:galfit_errors}

The statistical uncertainties provided by GALFIT are not very meaningful because
the rigorous meaning of $\chi^2$ is violated to a large extent during
fitting\footnote{http://users.obs.carnegiescience.edu/peng/work/galfit/ \\
CHI2.html}.  The errors reported by GALFIT underestimate the true error. 
However, it is important to quantify degeneracies in the model fits so that we
can quantitatively ascertain how meaningful the fits are, especially when we add
a PSF to the model.

\begin{figure*}[tbp]
\centering
\begin{tabular}{lr}
\includegraphics[width=0.48\textwidth]{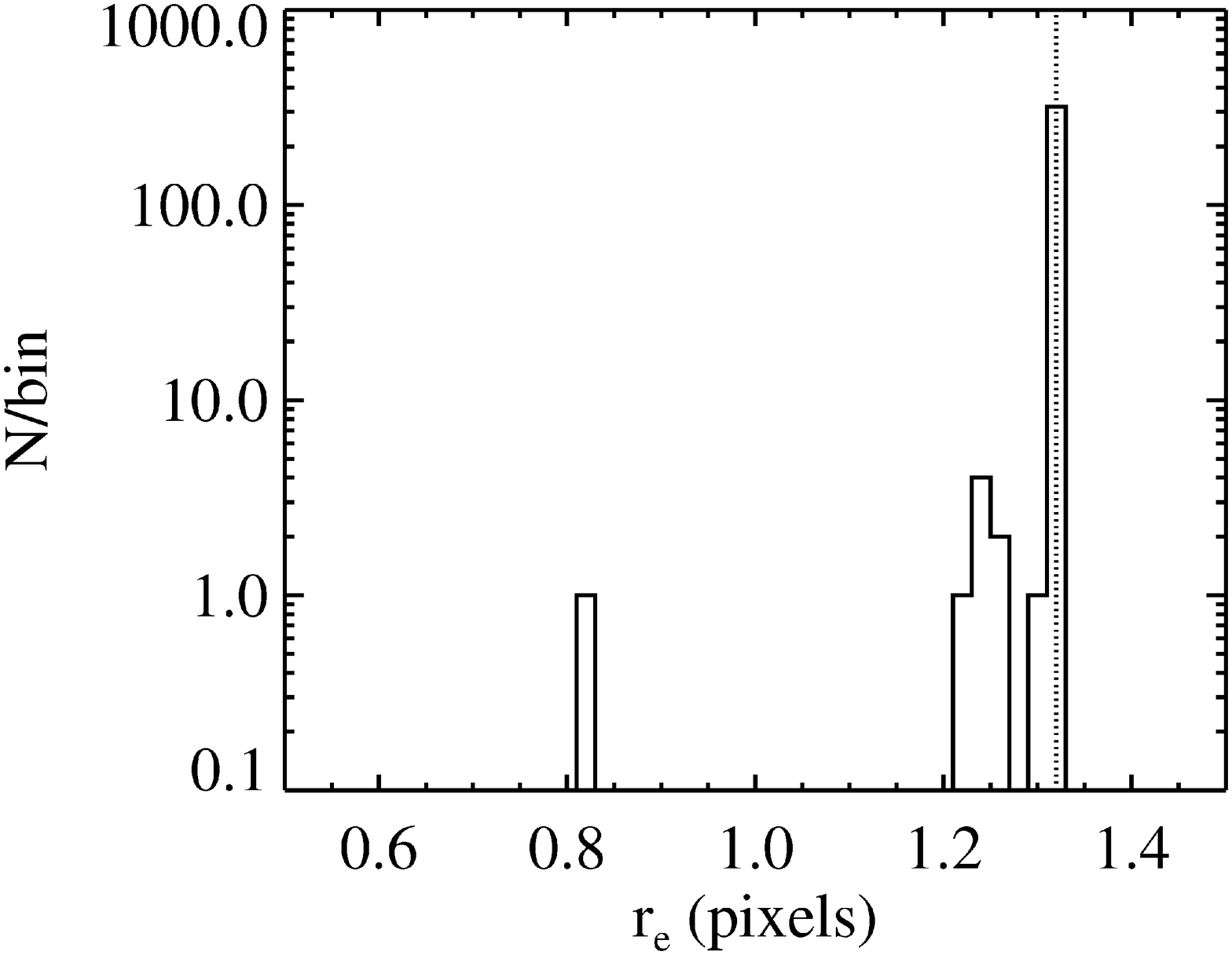} &
\includegraphics[width=0.48\textwidth]{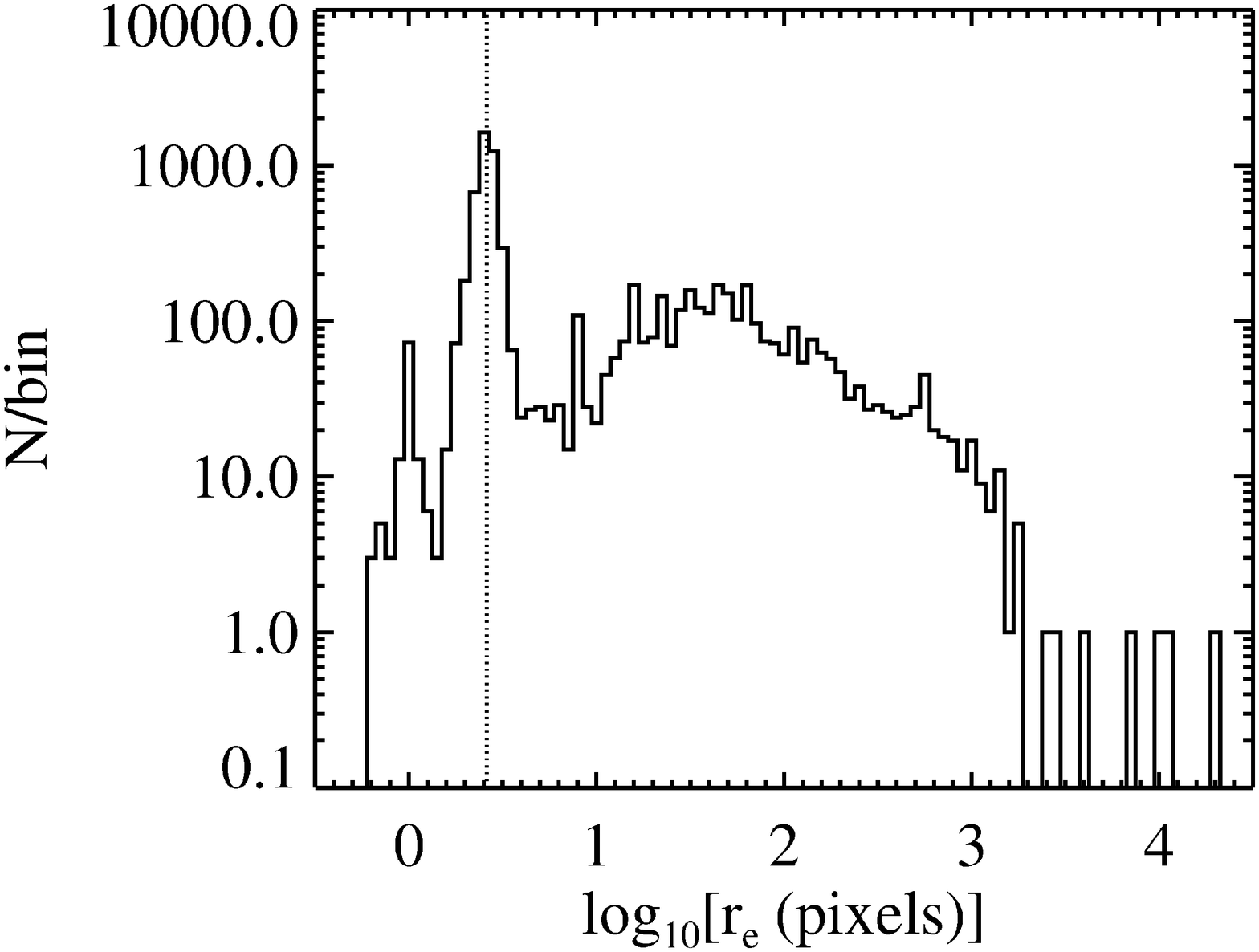} \\
\includegraphics[width=0.48\textwidth]{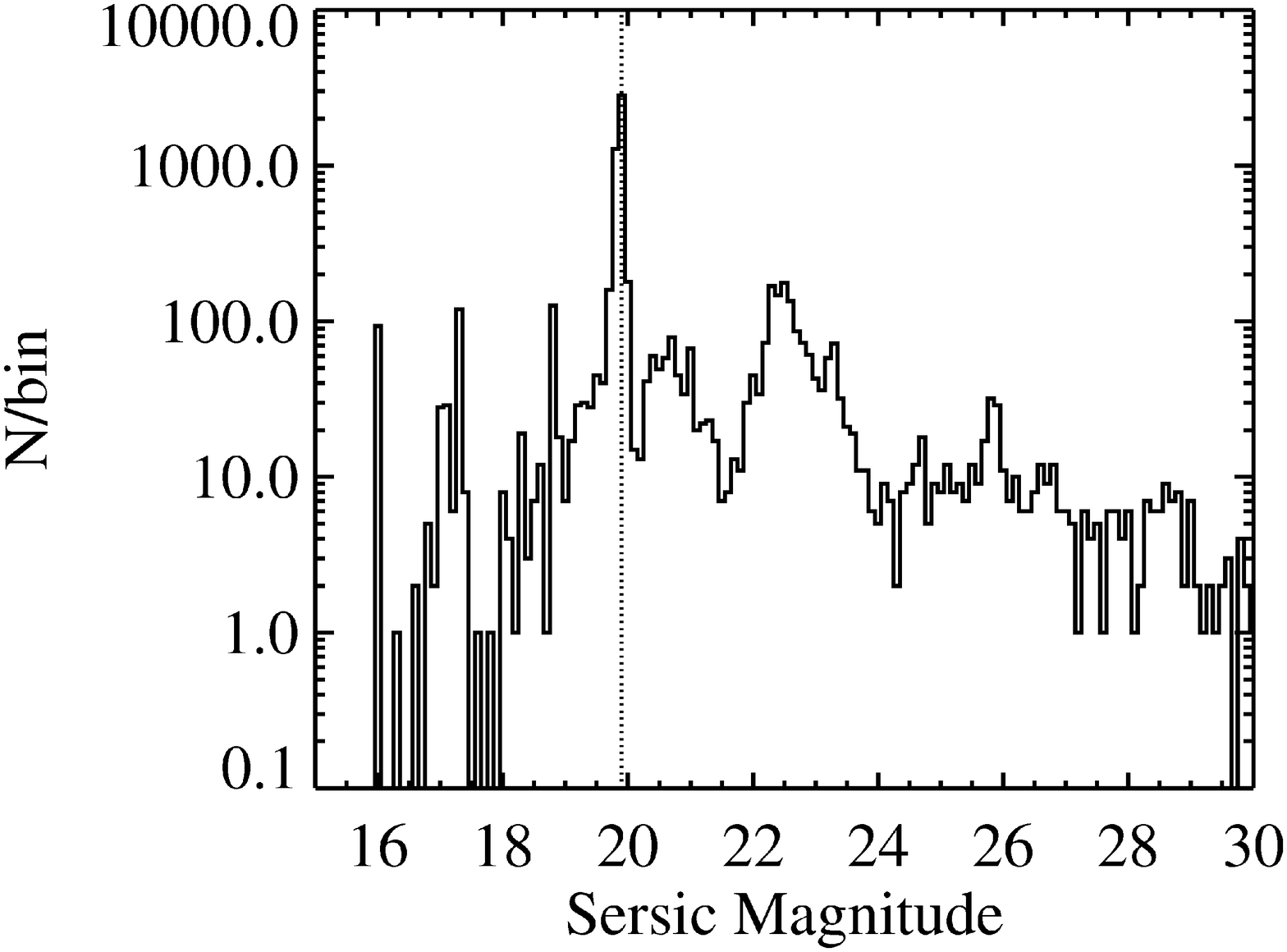} &
\includegraphics[width=0.48\textwidth]{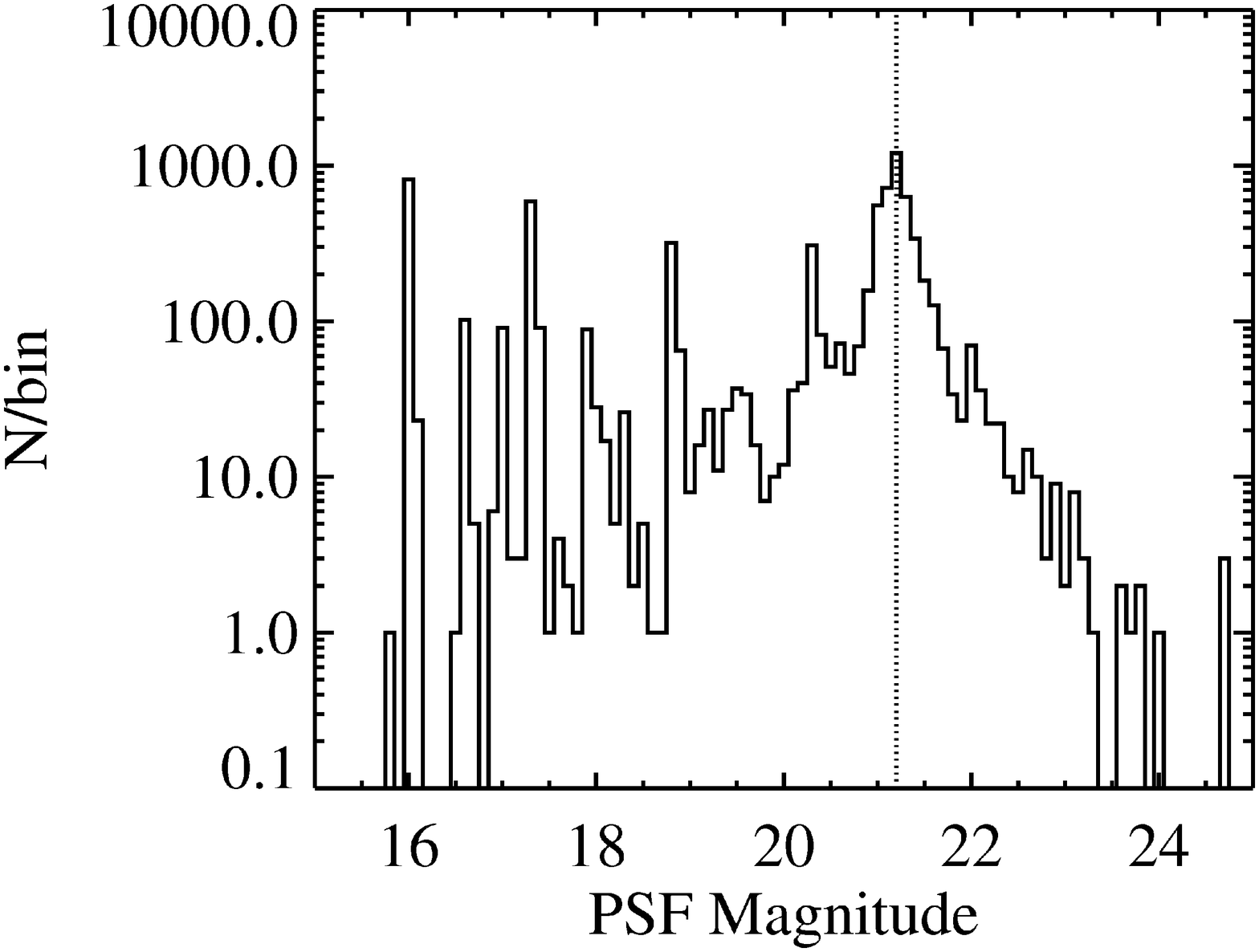}
\end{tabular}
\caption{The results of multiple GALFIT runs with different starting parameters
for J1713+28.  We chose this galaxy as a worst-case example because it
has two cores that are each fit with a S\'{e}rsic $+$ PSF model, and,
therefore, has the largest number of free parameters.  The first plot is for the
S\'{e}rsic only fit to this galaxy, and the remaining plots are for the S\'{e}rsic
$+$ PSF model.  The distribution of the S\'{e}rsic magnitudes for the S\'{e}rsic-only
model is singly and very tightly peaked like $r_e$.  All plots are for the primary
core.  The vertical dotted lines are at the best-fit values.}
\label{fig:grid_dist}
\end{figure*}

To explore the robustness of our derived model parameters, we began the fits
from fixed grids of a range of parameter starting values.  This also verifies
that GALFIT has found the global minimum and that the fit is not sensitive to
starting parameter values, which we found could be true for the more complicated
models (e.g., J1713+28:  a S\'{e}rsic and PSF for each core).  For each model
fit, we created a grid of hundreds to thousands of starting parameter
combinations (depending on the number of free parameters) and normally ran
GALFIT until it found a minimum or crashed.  We show some example distributions
for J1713+28 in Fig.~\ref{fig:grid_dist}.

This approach enables us to roughly quantify the degeneracies in our model
parameters.  The distributions in the parameter values are always at least as
large as the statistical uncertainties returned by GALFIT.  This approach is
particularly useful because it highlights the expected considerable increase in
uncertainty of the parameter values as more complicated models are introduced
(more degeneracies between model parameters).  For instance, note that, while
the effective radius is well-determined for a single S\'{e}rsic fit per core,
the uncertainty is almost always much larger when a PSF model is included.  In
this case for almost every galaxy, the addition of the PSF assumes a large
fraction of the core light, leading the S\'{e}rsic to fit more extended diffuse
emission.  However, because the underlying emission is so much fainter than the
core, its extent is not as well-determined as the extent of the predominant core
emission in a S\'{e}rsic-only model.  For instance, the spread in the effective
radii and S\'{e}rsic magnitudes are $\sim 10$~--~$100 \times$ larger when both a
PSF and a S\'{e}rsic model are simultaneously fit versus fitting only a single
S\'{e}rsic model.  This suggests that, while adding a PSF to the model to better
understand how PSF-like the galaxies are, the best-fit parameters should be
interpreted with caution.

\begin{figure*}[tbp]
\includegraphics[trim = 3.8in 0.1in 4.2in 0.9in, clip, scale=0.67, angle=90]{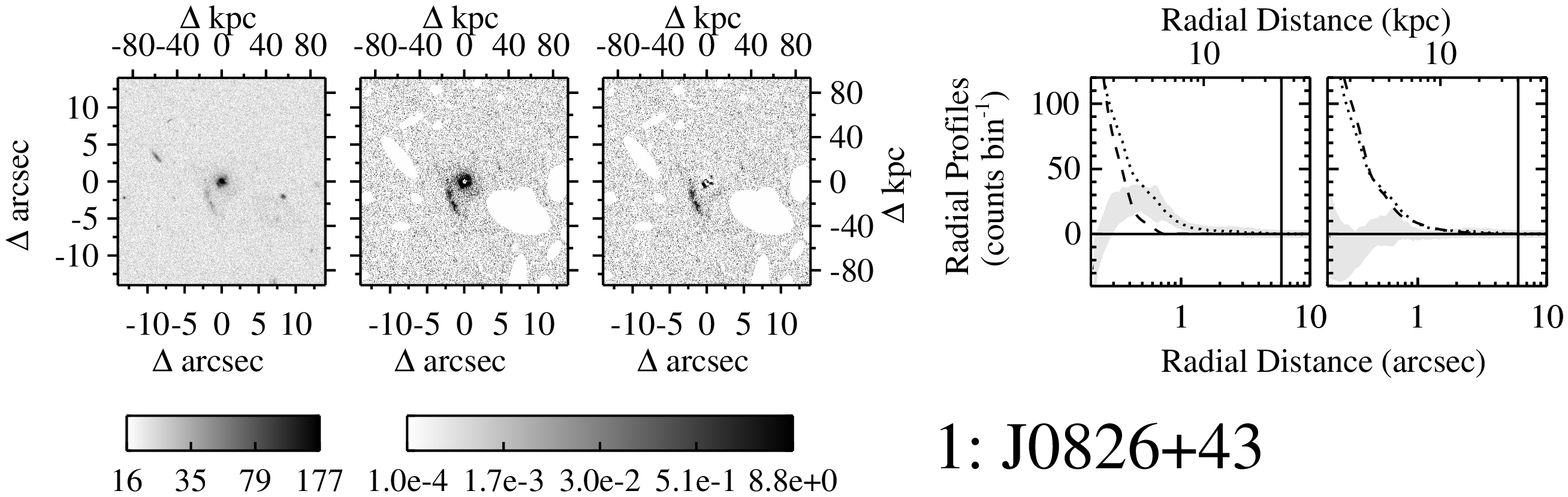}
\includegraphics[trim = 3.8in 0.1in 4.2in 0.9in, clip, scale=0.67, angle=90]{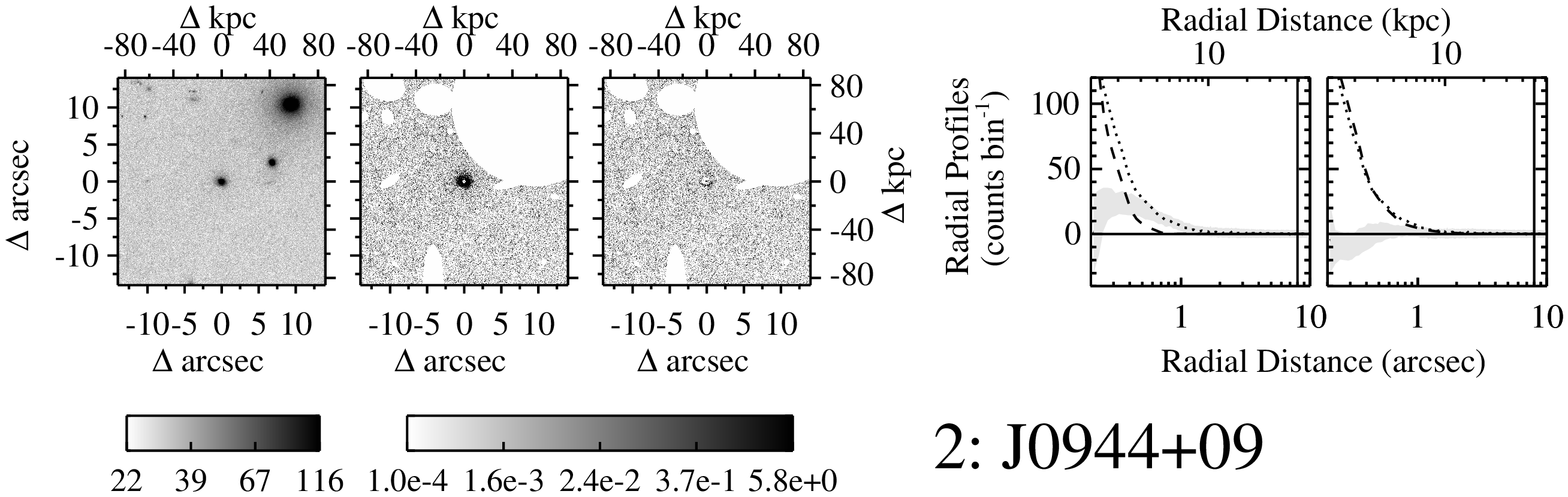}
\includegraphics[trim = 3.8in 0.1in 4.2in 0.9in, clip, scale=0.67, angle=90]{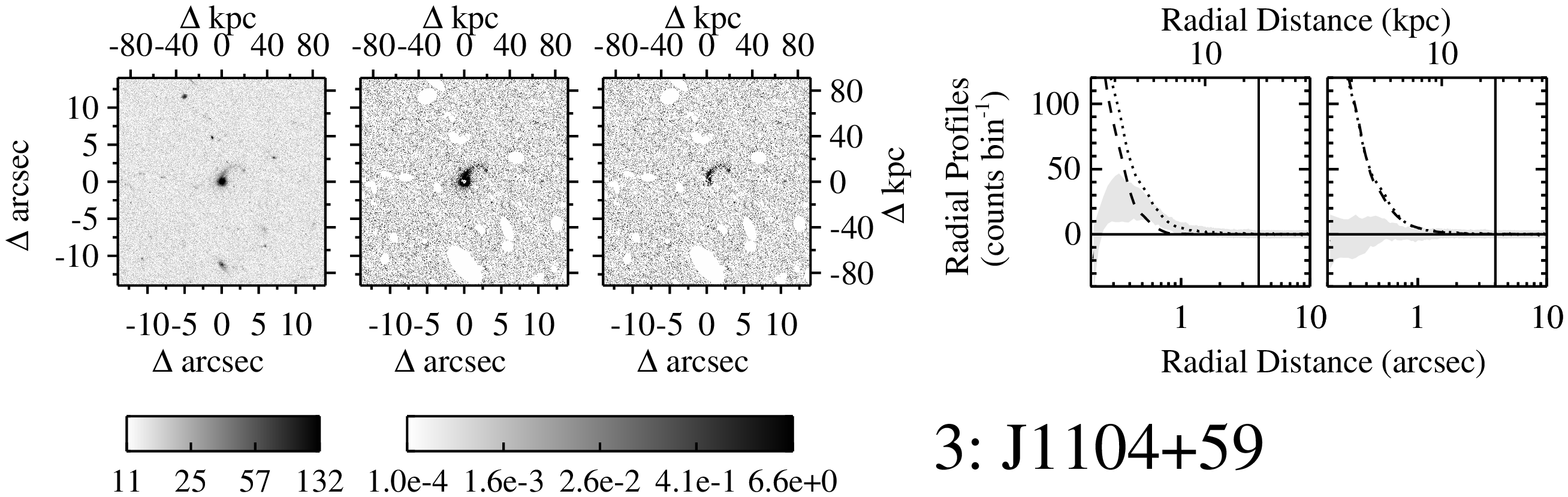}
\caption{From left to right for each of our 12 galaxies:  the 1400~pixel $\times$ 1400~pixel
cutout from the original image that we fit with GALFIT; the residual image from the
S\'{e}rsic model fit; the residual image from the S\'{e}rsic $+$ PSF model fit;
radial profiles of the galaxy, best-fit model, and residuals for the
S\'{e}rsic and S\'{e}rsic $+$ PSF fits.  The three images are logarithmically scaled
using the sky level and the Sersic magnitude to define the stretch; the
colorbars are in units of HST counts per pixel.  Regions that are masked
during the fit are shown as white in the residual images.  North is up
and east is left.
We plot the radial profiles for the masked, sky-subtracted original
image (dotted), the masked, sky-subtracted model (dashed), and the
masked residual image (shaded; the width corresponds to the standard
deviation of the pixel values in each radial bin).  We do not
show the innermost region of the radial profile ($r < 10$~pixels) because
these regions are dominated by noise from the imperfect PSFs.  If the inner
radius of the sky region is within 10~arcsec, it denoted by a solid vertical line.
Note that J1506+61 and J1713+28 are always fit with two n=4 S\'{e}rsic
models as they have two clearly distinct, associated, bright cores.  When
a PSF is also applied to J1506+61, only one PSF is used because the fit
to the fainter core does not favor a PSF, whereas, two PSFs are used for
J1713+28, one per core.}
\label{fig:data_fits}
\end{figure*}

\begin{figure*}
\centering
\ContinuedFloat
\includegraphics[trim = 3.8in 0.1in 4.2in 0.9in, clip, scale=0.67, angle=90]{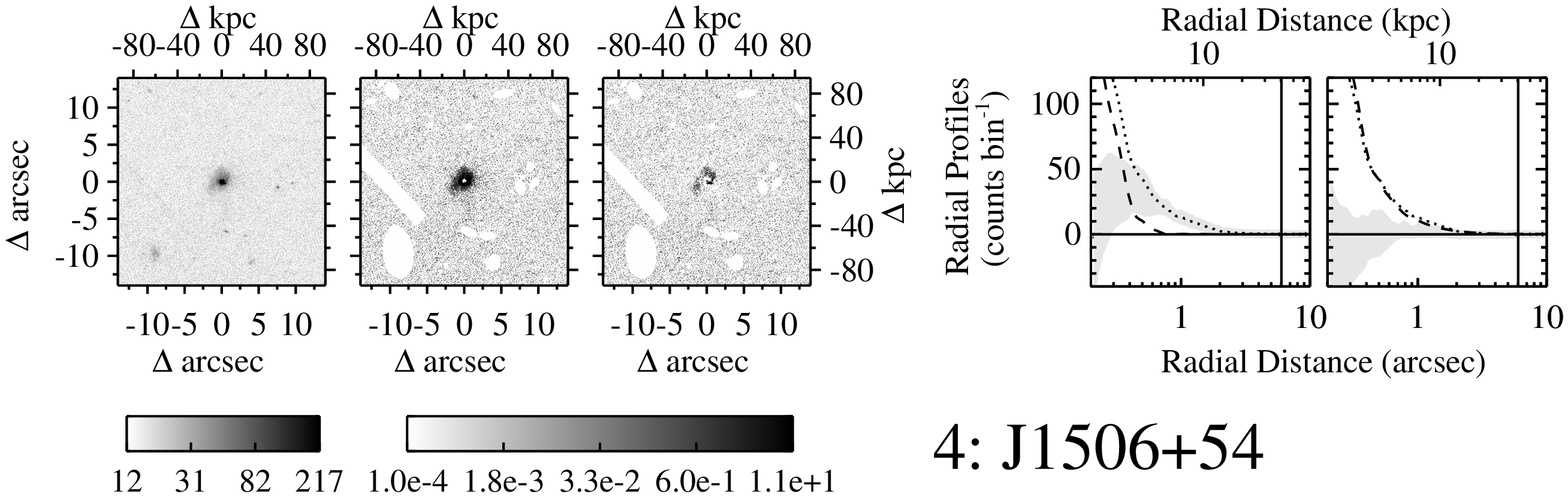}
\includegraphics[trim = 3.8in 0.1in 4.2in 0.9in, clip, scale=0.67, angle=90]{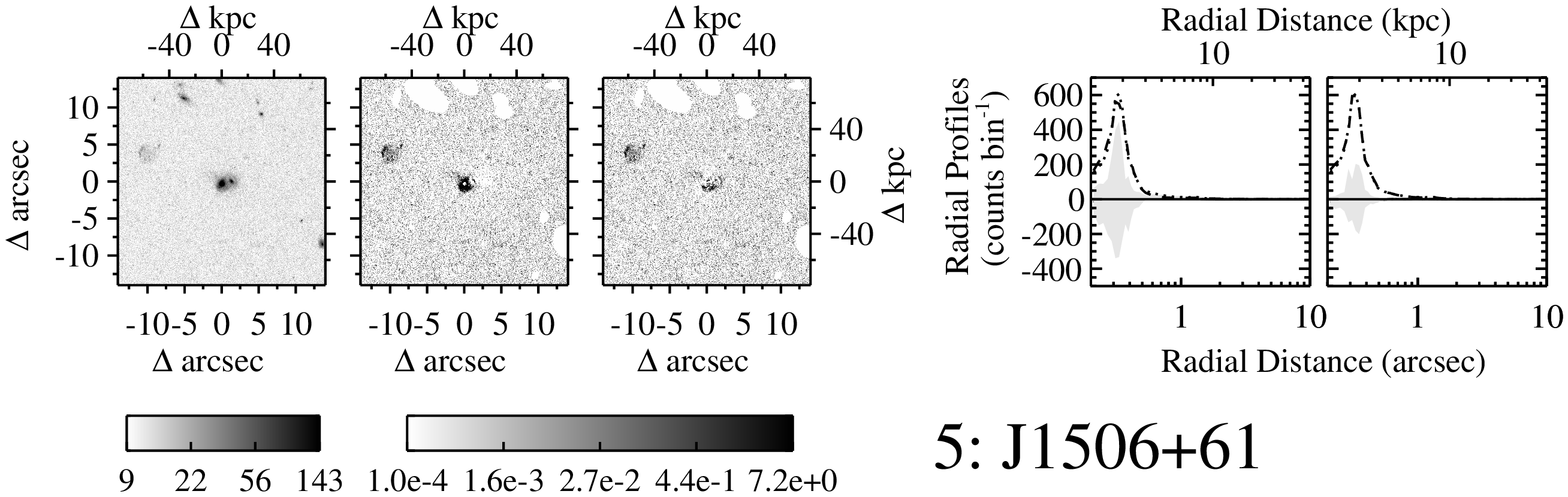}
\includegraphics[trim = 3.8in 0.1in 4.2in 0.9in, clip, scale=0.67, angle=90]{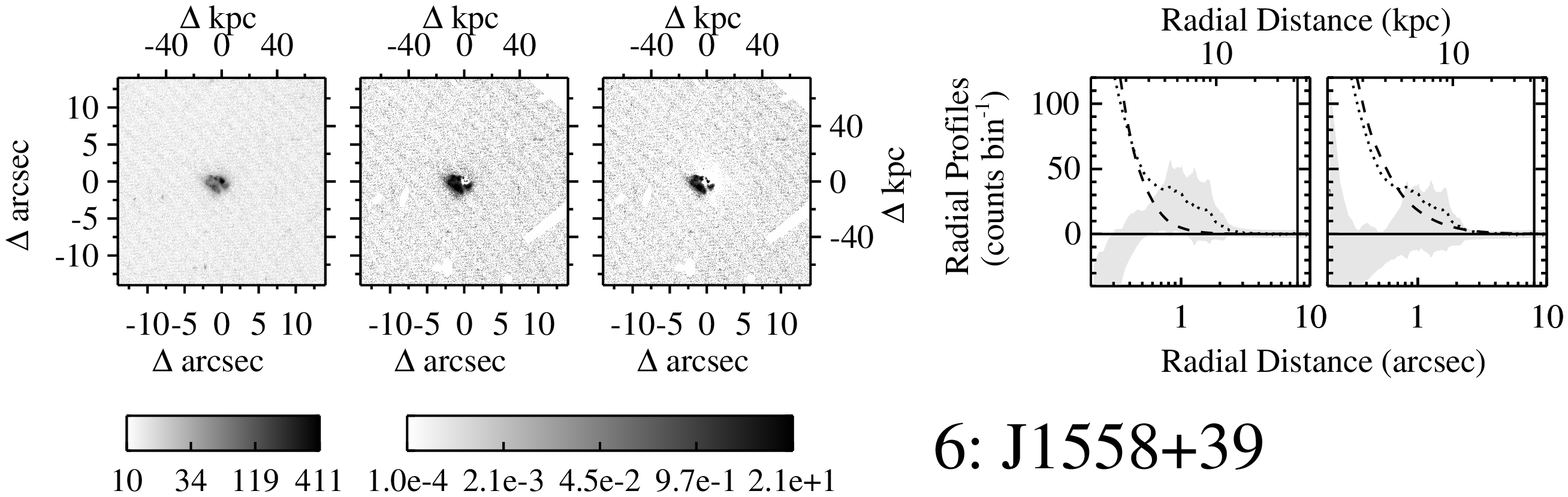}
\caption{Cont.}
\end{figure*}

\begin{figure*}
\centering
\ContinuedFloat
\includegraphics[trim = 3.8in 0.1in 4.2in 0.9in, clip, scale=0.67, angle=90]{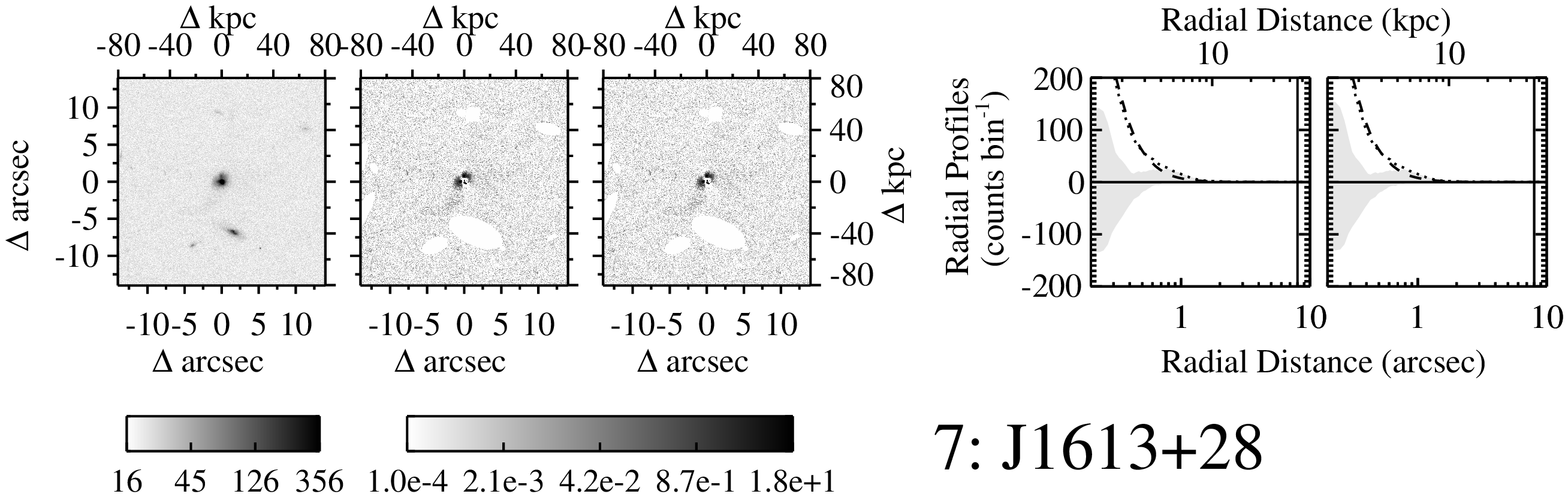}
\includegraphics[trim = 3.8in 0.1in 4.2in 0.9in, clip, scale=0.67, angle=90]{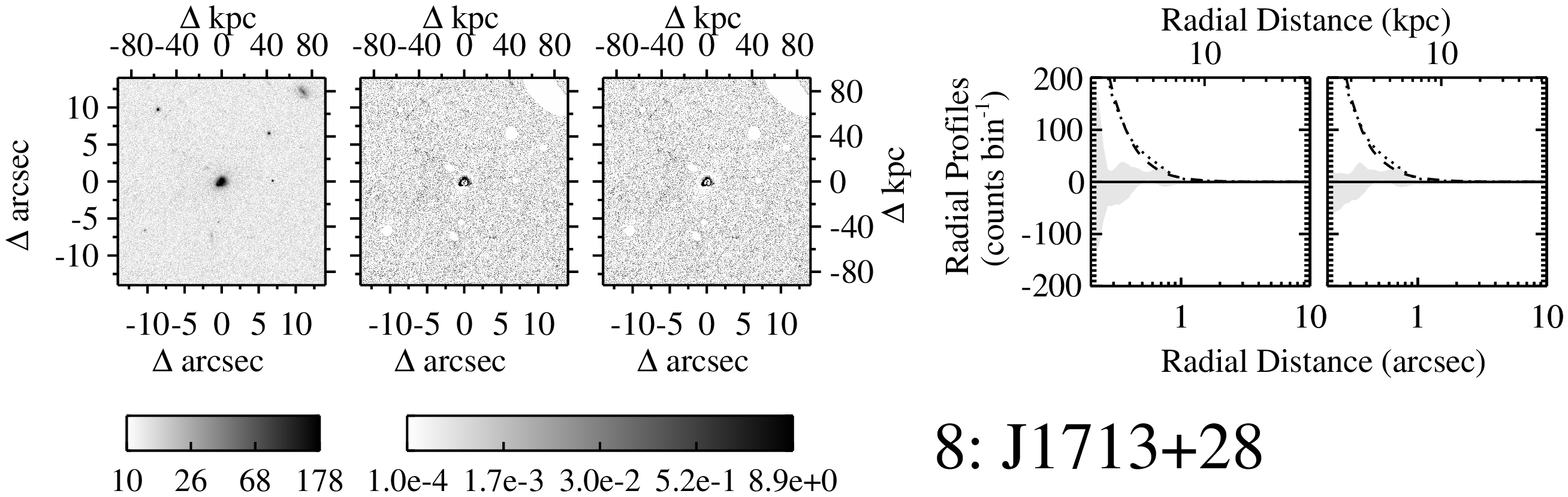}
\includegraphics[trim = 3.8in 0.1in 4.2in 0.9in, clip, scale=0.67, angle=90]{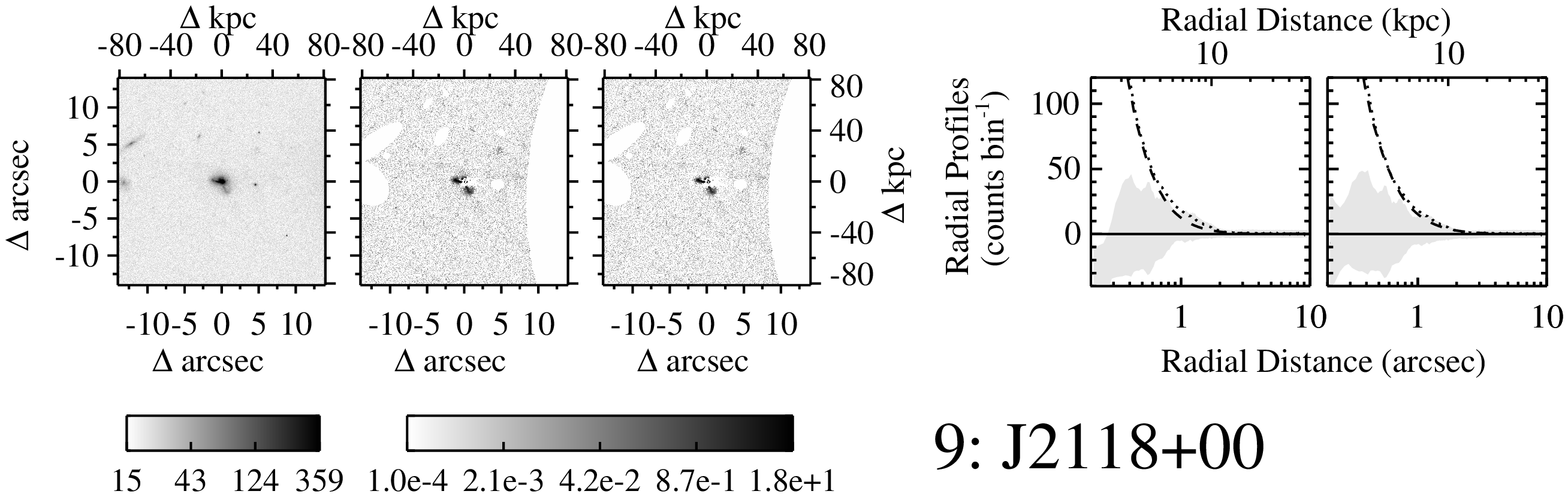}
\caption{Cont.}
\end{figure*}

\begin{figure*}
\ContinuedFloat
\centering
\includegraphics[trim = 3.8in 0.1in 4.2in 0.9in, clip, scale=0.67, angle=90]{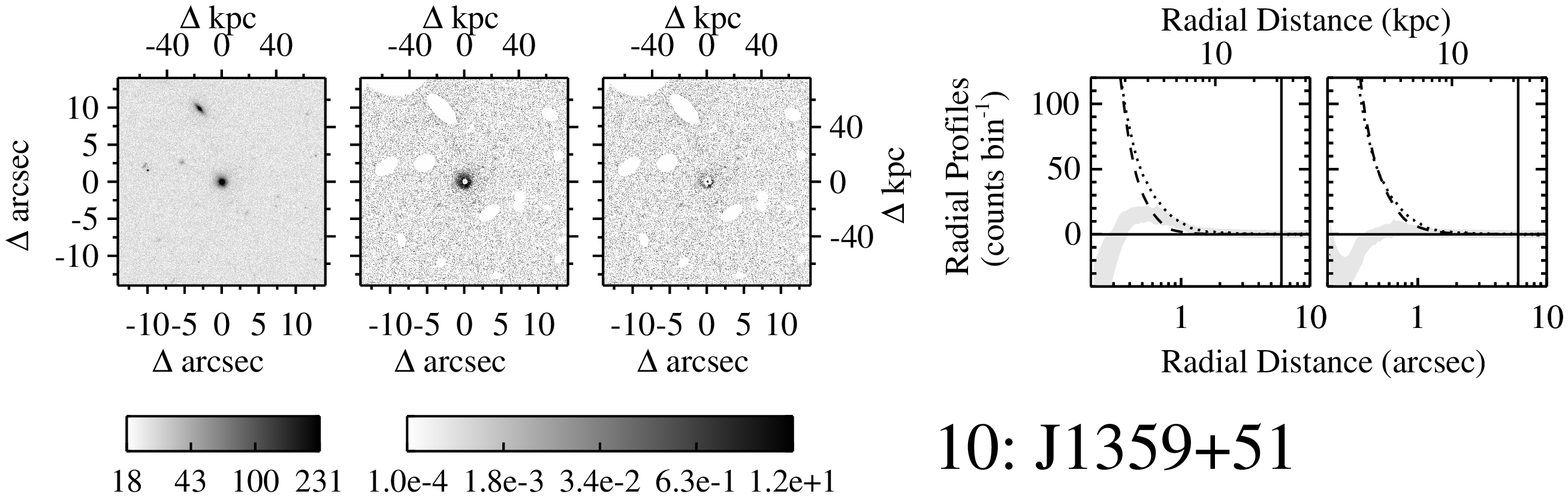}
\includegraphics[trim = 3.8in 0.1in 4.2in 0.9in, clip, scale=0.67, angle=90]{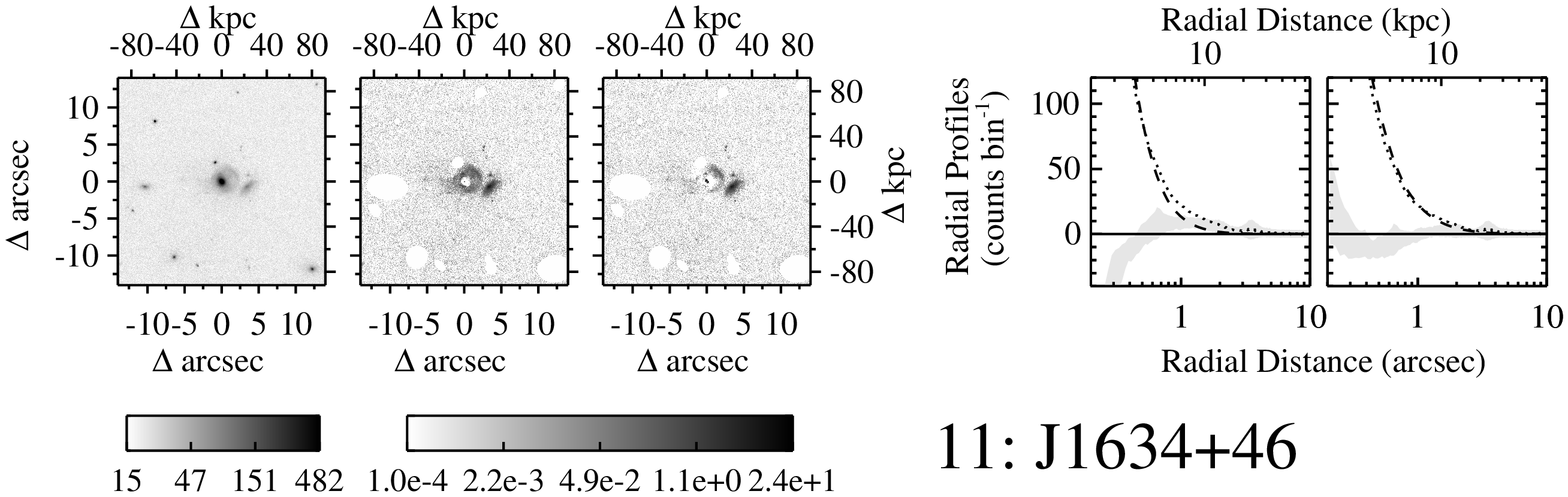}
\includegraphics[trim = 3.8in 0.1in 4.2in 0.9in, clip, scale=0.67, angle=90]{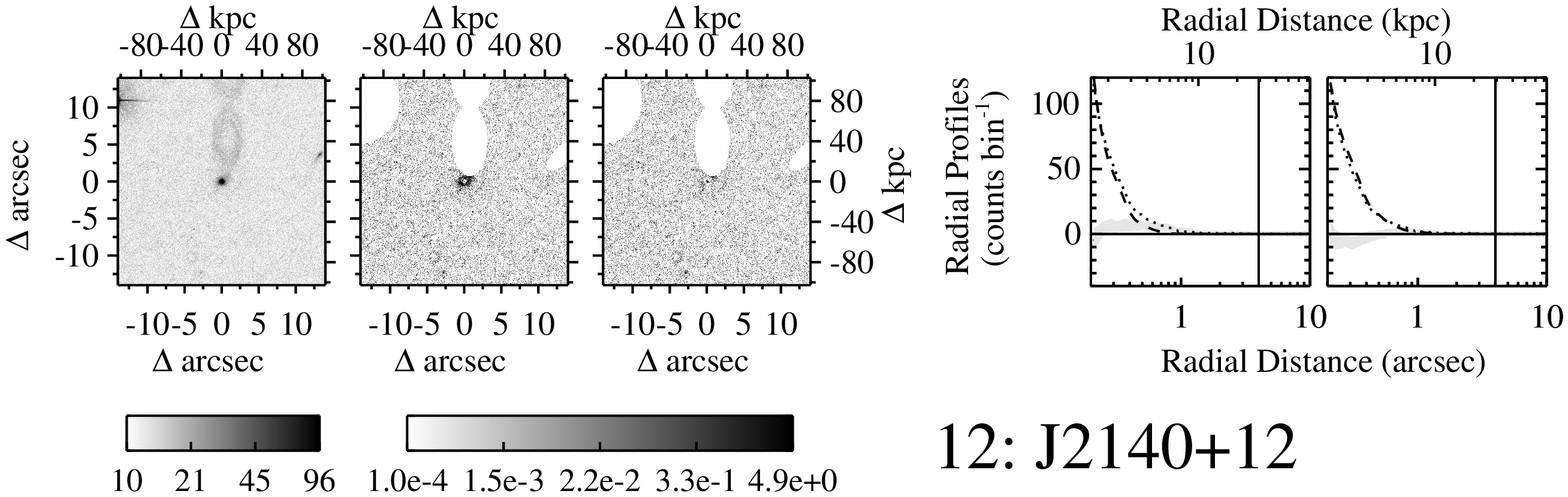}
\caption{Cont.}
\end{figure*}

\end{document}